\documentclass[11pt]{article}

\usepackage[preprint]{acl}

\usepackage{times}
\usepackage{latexsym}

\usepackage[T1]{fontenc}

\usepackage[utf8]{inputenc}

\usepackage{microtype}

\usepackage{inconsolata}

\usepackage{graphicx}

\usepackage{amsmath}
\usepackage{amssymb}
\usepackage{mathtools}
\usepackage{amsthm}
\usepackage{mathabx}
\usepackage{multirow}
\usepackage{color, colortbl}
\usepackage{enumitem}
\setitemize{noitemsep,topsep=0pt,parsep=0pt,partopsep=0pt}
\usepackage{booktabs}
\usepackage{subcaption}

\usepackage{xcolor}

\definecolor{LightCyan}{rgb}{0.88,1,1}

\usepackage[capitalize,noabbrev]{cleveref}

\theoremstyle{plain}
\newtheorem{theorem}{Theorem}[section]

\theoremstyle{definition}
\newtheorem{definition}[theorem]{Definition}

\theoremstyle{remark}

\title{Tacit Coordination of Large Language Models}

\author{
 \textbf{Ido Aharon\textsuperscript{1}},
 \textbf{Emanuele La Malfa\textsuperscript{2,3}},
 \textbf{Michael Wooldridge\textsuperscript{2}},
 \textbf{Sarit Kraus\textsuperscript{1}}
\\
\\
 \textsuperscript{1} Department of Computer Science, Bar-Ilan University \\
 \textsuperscript{2} Department of Computer Science, University of Oxford \\
 \textsuperscript{3} Institute for Decentralized AI (IDAI)
\\
 \small{
   \textbf{Correspondence:} \href{mailto:emanuele.lamalfa@cs.ox.ac.uk}{emanuele.lamalfa@cs.ox.ac.uk}
 }
}

\begin{document}
\maketitle
\begin{abstract}
Large Language Models (LLMs) are increasingly deployed in multi-agent settings that require coordination without communication, from human–AI interaction to safety-critical scenarios. Humans often overcome the absence of communication through focal points: salient solutions that naturally stand out to all participants.
We present the first large-scale evaluation of how, when, and why focal points emerge in LLMs, comparing their behaviour with humans across cooperative and competitive games, including realistic search \& rescue scenarios, demonstrating when focal points enable effective coordination.
Across more than 20 open- and closed-source models, we find that LLMs exhibit a remarkable ability to coordinate without communication, often matching or outperforming humans. However, the same models consistently fail in tasks requiring numerical common sense or culturally nuanced notions of salience. We additionally evaluate simple learning-free strategies that substantially improve coordination both among LLMs and between humans and LLMs.
Our results reveal striking coordination capabilities, as well as social limitations in modern LLMs, and offer new insight into the latent notions of salience encoded within them. Our findings caution against assuming that LLMs share humans' cultural and perceptual substrate when deployed in coordination settings.\footnote{The code to replicate \textbf{all} the results in the paper is available at \url{https://github.com/EmanueleLM/focal-points}.} 
\end{abstract}

\section{Introduction}
\begin{figure*}
    \centering
    \includegraphics[width=1\linewidth]{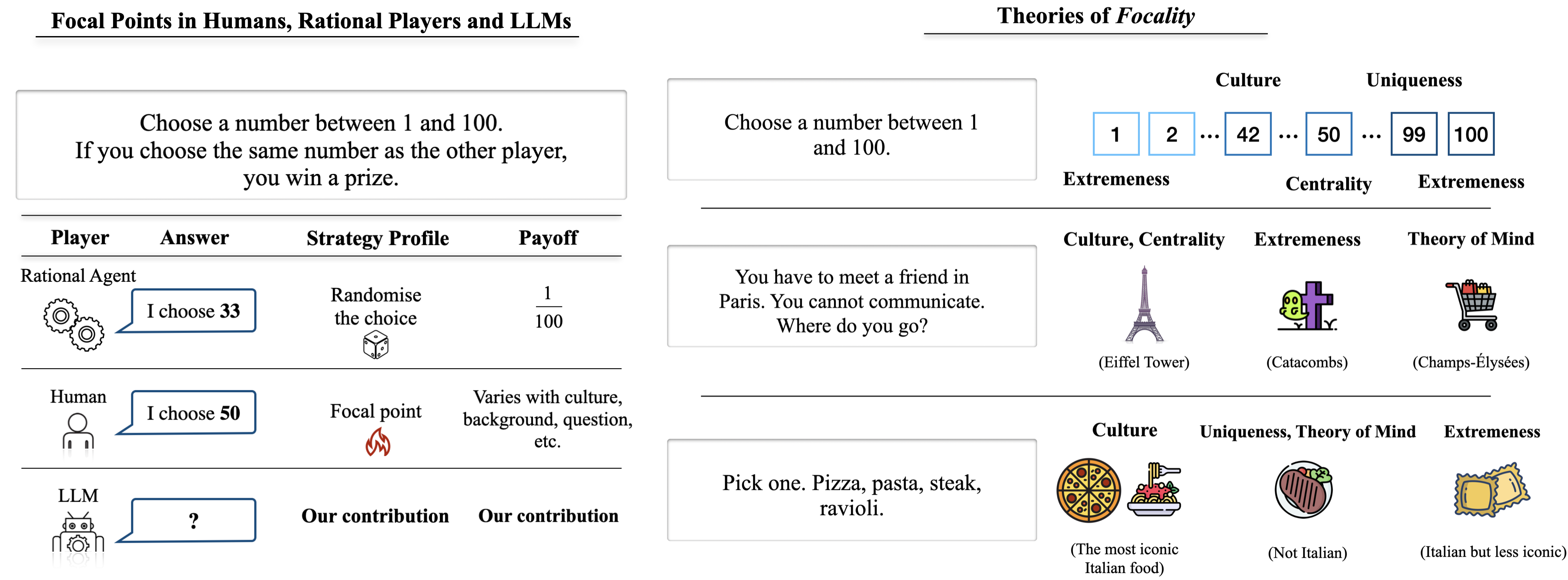}
    \caption{Left: This paper uses the theoretical framework of Schelling/focal points to study how, why, and when tacit coordination emerges in heterogeneous LLMs.
    Right: Examples of focality principles that humans leverage when they have to tacitly coordinate include \emph{extremeness}, \emph{centrality}, as well as cultural factors~\cite{kraus2000exploiting}.}
    \label{fig:intro-and-focality}
\end{figure*}
Humans routinely coordinate their behaviour without direct communication, often so seamlessly that the process goes unnoticed~\cite{tomasello2014natural}. Yet, this ease conceals a deeper problem: how do people converge on a preferred outcome when each must choose independently, without communication or binding commitments? Game theory characterises such settings through concepts such as Nash equilibria~\cite{nash1953two}, but these do not explain why humans coordinate more successfully than purely rational agents in cases of \emph{tacit coordination}.

Consider two agents who cannot communicate and must each choose a number between $1$ and $100$. They receive a prize only if their choices match. Of $100^2$ possible pairs, only $100$ are successful. Nevertheless, humans often converge on salient numbers such as $1$, $50$, or $100$, even though standard theory does not uniquely predict these choices. A distinctive feature of human coordination is precisely this ability to select strategies that are \emph{salient}: they stand out for reasons not reducible to payoff dominance~\cite{schelling1980strategy}.

Such salient strategies are known as \emph{focal points}~\cite{schelling1980strategy}. They arise when players rely on shared cues that go beyond the formal structure of payoffs. Supported by work in game theory, economics, and psychology~\cite{bardsley2010explaining,leland2018theory}, focal points explain why humans can break symmetry in tacit coordination games where many choices are formally equivalent.

Large Language Models (LLMs)~\cite{brown2020languagemodelsfewshotlearners} are increasingly used as agents in settings where coordination is essential~\cite{zhang2024proagent,ruan2025benchmarking}. In search \& rescue, for instance, a human commander, autonomous drones, and LLM-based planners may need to act under time pressure and incomplete information, converging on salient choices such as the last known location of a missing person or a natural rendezvous point. Similar dynamics arise in software engineering, automated customer support, and mixed human--AI scheduling. In each case, success depends not only on choosing a good action but on anticipating which action will also appear natural to others.

Focal points matter beyond human--AI interaction. Even LLM agents built from different models, prompts, roles, or partial task views may face many equally valid actions, with coordination succeeding only when they converge on the same one. This exposes two risks. First, if LLM salience diverges from human salience, human--AI coordination may fail when communication is sparse. Second, if different LLMs encode different notions of salience, AI--AI coordination may fail even when each agent is individually competent. Focal points, therefore, provide a lens for testing whether LLM agents act as rational players, deviate systematically from strict rationality, use higher-order reasoning~\cite{hu2020other,wu2025large}, or reproduce culturally shaped biases~\cite{tao2024cultural}.

This work investigates whether LLMs can coordinate tacitly and how focal points emerge in their behaviour. We compare more than $20$ open- and closed-source LLMs on tasks requiring coordination, cooperation, and competition, and present the first large-scale assessment of LLM tacit collaboration and comparison with human choices. Our analysis covers standard coordination games and real-world scenarios, including human search \& rescue. LLMs often coordinate as well as, or better than, humans; however, they fail on questions involving subtle cultural context, revealing biases that we mitigate using three learning-free techniques.

\begin{figure*}
    \centering
    \includegraphics[width=0.95\linewidth]{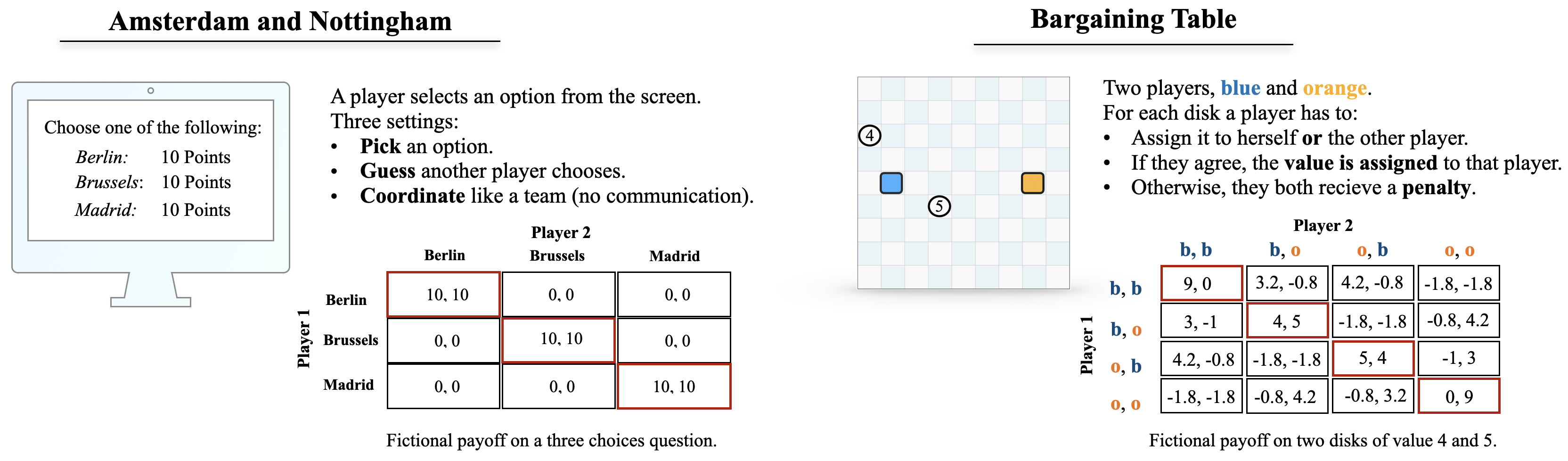}
    \caption{Left: Illustration of the rules and the \textcolor{purple}{pure Nash equilibria} of the Amsterdam and Nottingham coordination games~\cite{bardsley2010explaining}. 
    Right: Illustration of the Bargaining Table game~\cite{mizrahi2020using}, a semi-competitive game, and its \textcolor{purple}{pure Nash equilibria}.}
    \label{fig:tasks-picture}
\end{figure*}

\section{Related Work}

\noindent{\bf Focal points in game theory.}
Focal points originate with \citet{schelling1980strategy}, who argued that people often solve coordination problems by selecting outcomes that stand out beyond the payoff structure itself. Subsequent work studied focality in bargaining~\cite{murnighan1980effects}, two-player games~\cite{mehta1994nature}, and games with out-of-equilibrium outcomes~\cite{cooper1996cooperation}. Rather than deriving focal points from payoff/risk dominance or strategic sophistication, this literature models salience through cognitive and cooperative principles, including cognitive hierarchy theory~\cite{bardsley2010explaining,camerer2004cognitive}, $k$-level reasoning in Theory of Mind~\cite{sweller1994cognitive}, and team reasoning~\cite{sugden2003logic,bacharach1999interactive}. A broader line of work studies how focal points emerge in tacit cooperative settings~\cite{genesereth1988cooperation}, which saliency principles guide human choices~\cite{kraus2000exploiting}, and how such principles can support human--AI coordination~\cite{carroll2019utility}.

\noindent{\bf Tacit coordination and saliency in LLMs.}
Human--AI and AI--AI coordination is central to multi-agent systems~\cite{shoham2008multiagent,wooldridge2009introduction,kraus1997negotiation} and human--computer collaboration~\cite{terveen1995overview,frieder2012agent,gan2022defense,wang2020human}. Prior work has used focal points to enable coordination without explicit communication~\cite{kraus2000exploiting,zuckerman2011using,mizrahi2020using,mizrahi2023predicting}. With the rise of LLM agents~\cite{li2024more}, and building on reinforcement learning and multi-agent RL~\cite{guestrin2002coordinated}, recent studies examine tacit knowledge~\cite{budding2025large} and LLM coordination across cooperative, competitive, and hybrid settings~\cite{li2023theory,guo2024embodied,liu2024benchmark,zhu2025multiagentbench}. Relatedly, Machine Theory of Mind studies whether AI agents can model and coordinate with humans~\cite{rabinowitz2018machine}, including the debate over whether LLMs genuinely possess Theory of Mind or only mimic it from pre-training data~\cite{strachan2024testing,kosinski2023theory}. Complementary explanations focus on bias and culture~\cite{shaki2023cognitive,shaki2025out}: embeddings encode cultural and social stereotypes~\cite{bolukbasi2016mancomputerprogrammerwoman}, while LLM outputs can exhibit cultural incongruencies, inconsistencies, and stereotype reinforcement~\cite{prabhakaran2022cultural,tao2024cultural,kirk2021bias}.

For a systematic review of the topics mentioned above, we refer the reader to the following works~\cite{li2023theory,agashe2025llmcoordinationevaluatinganalyzingmultiagent,zhu2025multiagentbench}.

\section{Methodology}

\subsection{Nash Equilibria and Focal Points}
Let $G=(N,(\Sigma_i)_{i \in N},(u_i)_{i \in N})$ be a finite normal-form $n$-player game, where $N=\{1, \ldots, n\}$ is the set of players, $\Sigma_i$ the set of strategies of player $i$, and $u_i: \times_{i\in N} \Sigma_i \xrightarrow{} \mathbb{R}$ the utility function of player $i$, which maps a collection of strategies, one for each player, to the utility that player $i$ would receive if this collective choice was made.
We denote by $(\sigma_i, \sigma_{N \setminus i})$ the strategy profile in which $\sigma_i$ is the strategy adopted by player $i$ and $\sigma_{N \setminus i}$ the strategies adopted by any other player. 
A strategy profile $(\sigma_i^*, \sigma_{N \setminus i}^*)$ is a Nash equilibrium iff $\forall i \in N, \sigma_i^* \in \arg\max_{\sigma_i} u_i(\sigma_i, \sigma_{N \setminus i}^*)$. In other words, $(\sigma_i, \sigma_{N \setminus i})$ forms a Nash equilibrium if no player can benefit by unilaterally changing her strategy, assuming other players stay with theirs.
When games admit multiple Nash equilibria, the theory alone does not predict which will occur. A key question is then how players choose independently so as to coordinate on the same equilibrium. We call this \emph{tacit coordination}. To this end, we introduce a salience function that quantifies how \emph{focal} an equilibrium $e$ is for a player. Let $\mathcal{E}$ denote the set of (Nash) equilibria, and define
$S : \mathcal{E} \to \mathbb{R}_{\ge 0}$.
With this in mind, it is possible to define, over a set of Nash equilibria, a focal point equilibrium $e^*$ with respect to $\mathcal{E}$ is one satisfying $e^* \in \arg\max_{e \in \mathcal{E}} S(e)$.
One can interpret the raw scores of such a function to be a probability distribution over the set $\mathcal{E}$ of equilibria, namely $P: \mathcal{E} \xrightarrow{} [0, 1], \ s.t. \ \sum_{e \ \in \mathcal{E}} P(e) = 1$.  
The economics and multi-agent systems literature discusses tie-breaking conventions humans rely on to commit to focal points (e.g., \emph{uniqueness}, \emph{uniqueness complement}, \emph{centrality}, and \emph{extremeness}~\cite{kraus2000exploiting}, as shown in Figures~\ref{fig:intro-and-focality} (right)).
In other words, they suggest that we need to make strong assumptions (i.e., same salience function, common monotone transform, common tie-breaking conventions) to guarantee the existence of a unique focal point in a game. 
On the other hand, less studied is the problem of how a focal point emerges in tacit coordination games: we rigorously account for that in Appendix~\ref{sec:orbits}.

\begin{figure*}
    \centering
    \includegraphics[width=.98\linewidth]{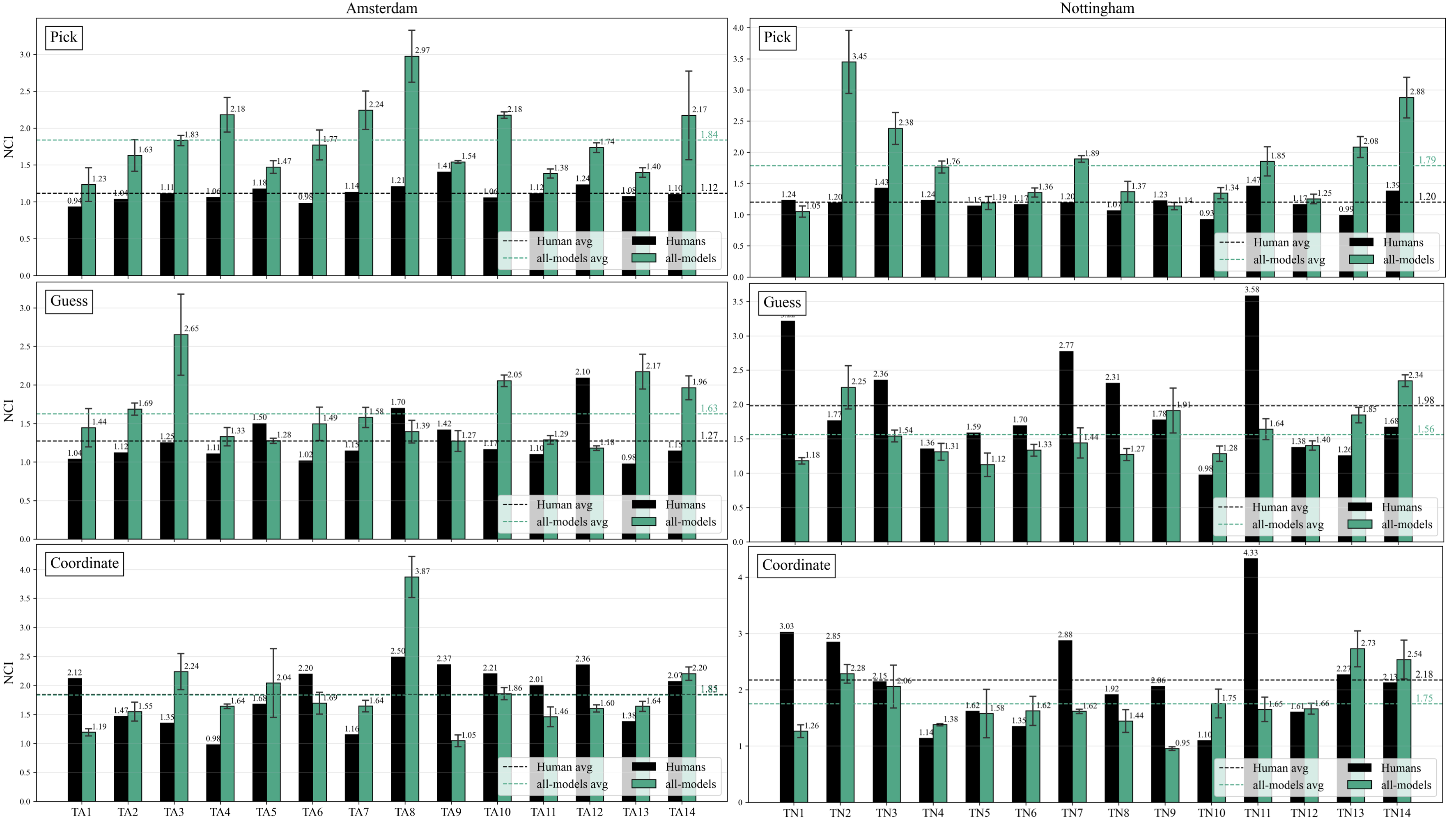}
    \caption{Normalised Coordination Index (NCI) of humans and LLMs (Llama-3, 3.1, and 3.3 70B; Qwen 2 and 2.5 72B; GPT-oss 20B and 120B) on the Amsterdam and Nottingham datasets.}
    \label{fig:all-models-vanilla-AN}
\end{figure*}

\subsection{Quantifying Tacit Coordination}
Measuring coordination with focal points requires performing controlled experiments on humans or machines, as the salience function is unknown and highly contextual.
The individual payoff and the social welfare (i.e., the sum of the player payoffs) are standard metrics for coordination in normal-form competitive games. 
On the other hand, the Coordination Index (CI)~\cite{bardsley2010explaining} provides a better measure for games that are purely cooperative. 
Let $G$ be an $n$-players game where each players share the same set of $m \ge 2$ strategies, $\{s_1, ..., s_m\}$. Let $m_j$ be the number of players who choose the strategy $s_j$. The coordination index is defined as:
\begin{equation}
    \text{CI} = \sum_{j=1}^{n}\frac{m_j(m_j - 1)}{n(n-1)}
\end{equation}
The CI is the probability that two randomly-chosen individuals choose the same strategy (more details in Appendix~\ref{a:ci}); the Normalised Coordination Index (NCI), computed as $\text{NCI} = m CI$, scales the Coordination Index with the number of strategies available to the players, and provides a measure of the concentration/dispersion of the strategy choices.

\section{Experimental Evaluation}
\subsection{The Amsterdam and Nottingham Human Evaluation}

To study when and why focal points emerge in LLMs, we build on the human experiments of \citet{bardsley2010explaining}. These experiments tested around $50$ participants in Amsterdam and Nottingham on multiple-answer coordination questions, reporting both the CI and NCI under three instructions: ``pick'', where participants choose an answer without further information; ``guess'', where they guess what another randomly paired participant would choose; and ``coordinate'', where they are explicitly asked to choose as if tacitly coordinating. Figure~\ref{fig:tasks-picture} (left) illustrates the game. The setting is purely cooperative and symmetric, as shown by the illustrative payoff matrix.

The human benchmark contains $14$ questions for Amsterdam, labelled \{TA1,\dots,TA14\}, and $14$ for Nottingham, labelled \{TN1,\dots,TN14\}. We evaluate LLMs on the same answer sets. To control for order effects, each question is tested $30$ times under three random answer permutations, giving $90$ runs per question. The main paper reports a high-performing open-weight subset: Llama-3-70B, Llama-3.1-70B, Llama-3.3-70B~\cite{grattafiori2024llama3herdmodels}, Qwen-2-72B, Qwen-2.5-72B~\cite{yang2024qwen2technicalreport,qwen2025qwen25technicalreport}, and GPT-oss-20B and GPT-oss-120B.\footnote{\href{https://openai.com/index/introducing-gpt-oss/}{openai.com/index/introducing-gpt-oss/}} The full evaluation, covering more than $20$ models, is reported in the Appendix and released code. For methodological consistency, the main analysis focuses on open-source or open-weight models that expose their full Chain of Thought~\cite{wei2023chainofthoughtpromptingelicitsreasoning}, excluding API-only models. Appendix~\ref{a:additional-res-AN} reports results for GPT 5.4 and Gemini 2.5 Pro and 3 Pro.

\begin{figure}
    \centering
    \includegraphics[width=1\linewidth]{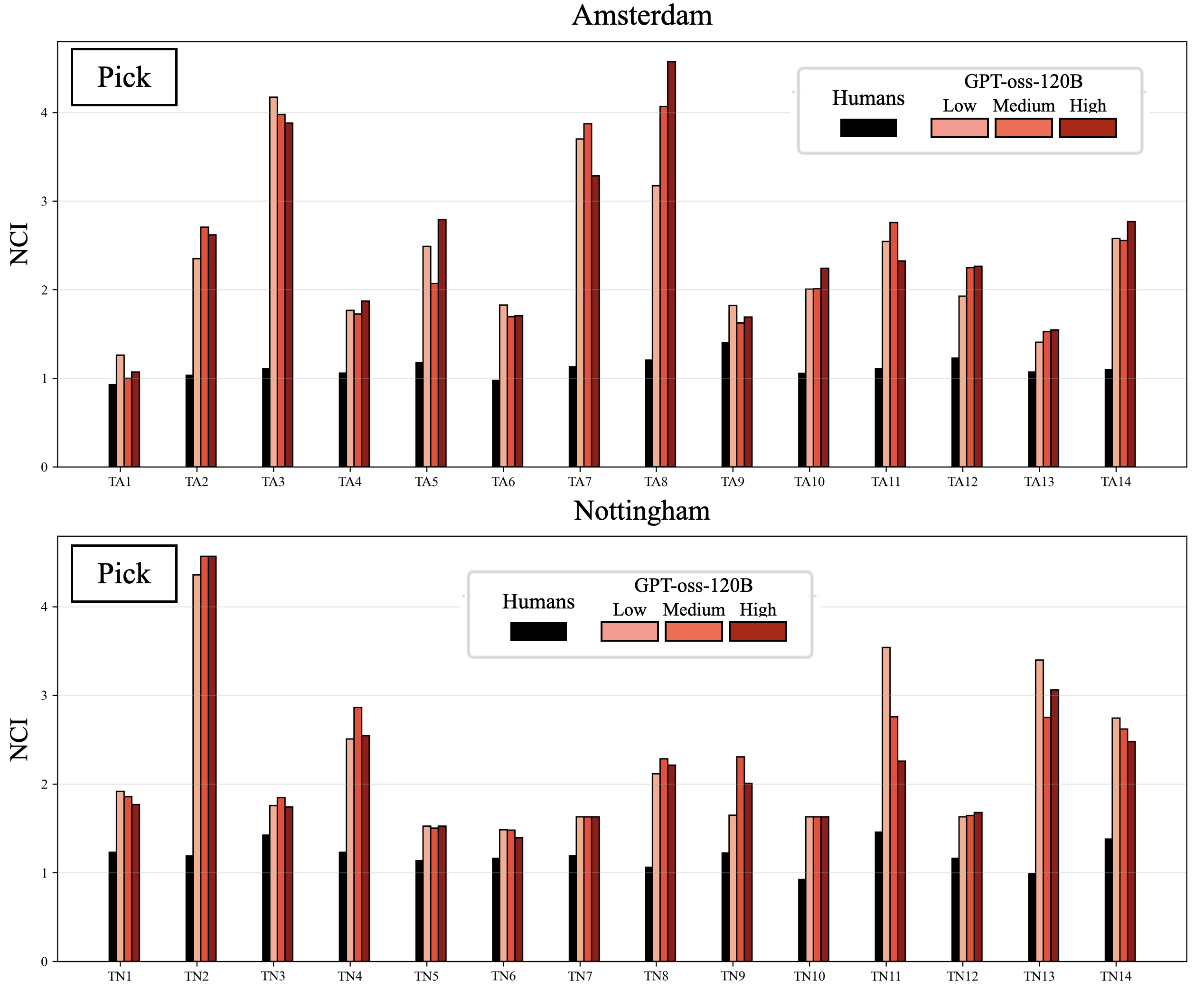}
    \caption{The effect of reasoning (low, medium, and high: the darker the red, the higher the reasoning) on the coordination of GPT-oss-120B in Amsterdam and Nottingham. There is no clear evidence that reasoning improves the NCI of LLMs. Same results hold for other settings and prompting techniques (full results in Appendix~\ref{a:reasoning-AN}).}
    \label{fig:gpt-120b-reasoning-AN}
\end{figure}

\subsubsection{The \textit{Vanilla} Amsterdam and Nottingham}
\label{sec:results-vanilla-AN}

In the \emph{vanilla} setting, LLMs receive the same questions given to humans, with only minimal formatting changes; prompts are reported in Appendix~\ref{a:prompting-AN}. Figure~\ref{fig:all-models-vanilla-AN} shows a striking pattern. LLMs outperform humans under ``pick'', but the advantage narrows under ``guess'' and ``coordinate''. They remain competitive in Amsterdam, whereas humans clearly outperform them in Nottingham. The largest gaps occur on TN1, TN7, and TN11.

These failures are informative. Humans often coordinate through culturally salient or personally meaningful options, which can produce strong agreement in some settings but dispersion in others, especially under ``pick''. LLMs instead rely more heavily on systematic cues such as option order, structural regularity, or surface salience. This makes them robust in some cases, but brittle when the focal point depends on subtle cultural context. We analyse these question-level differences in Appendix~\ref{a:cultural-AN}.

Prior work suggests that humans coordinate better on focal points when they can reason more about the task and possible strategies~\cite{laufer2022electrophysiological}. We test whether the same holds for LLMs using GPT-oss-20B and GPT-oss-120B, whose reasoning level can be varied and is associated with output length and benchmark performance~\cite{meincke2025promptingsciencereport2}.\footnote{\href{https://openai.com/index/learning-to-reason-with-llms/}{openai.com/index/learning-to-reason-with-llms/}} Surprisingly, more reasoning does not improve tacit coordination. As shown in Figure~\ref{fig:gpt-120b-reasoning-AN}, and in full in Appendix~\ref{a:reasoning-AN}, stronger reasoning settings can even degrade performance. Log analysis shows that GPT-oss-120B often produces increasingly elaborate rationales that ultimately collapse to arbitrary positional heuristics, such as selecting the first option in a randomly permuted list. This suggests that tacit coordination is not simply a reasoning-depth problem, and may belong to the class of abilities that frontier LLMs struggle to improve through more deliberation alone~\cite{malek2025frontierllmsstrugglesimple}.

\begin{figure}
    \centering
    \includegraphics[width=1\linewidth]{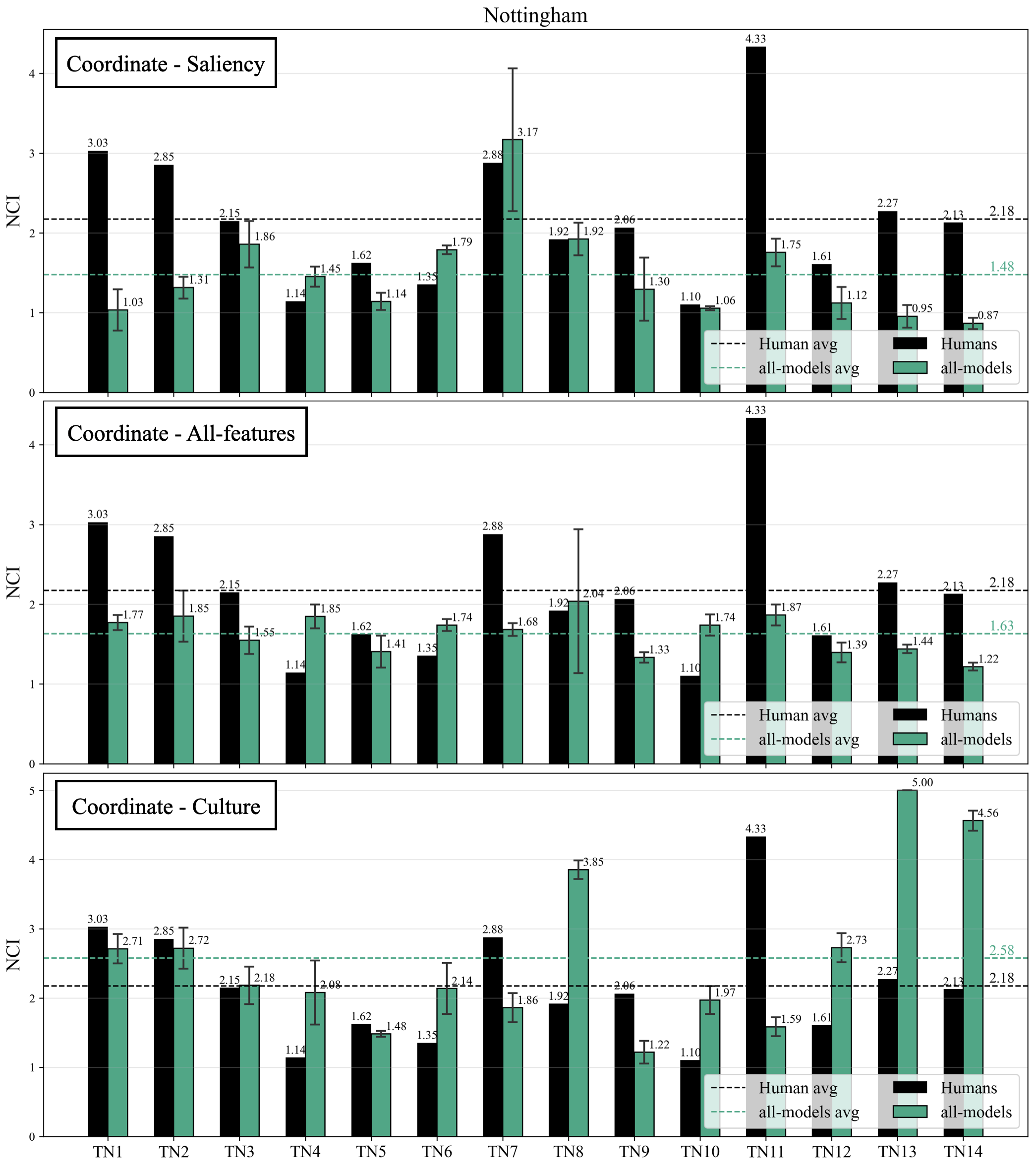}
    \caption{Improving tacit coordination of several LLMs (Llama-3, 3.1, 3.3 70B, Qwen-2, 2.5 72B, and GPT-oss-20B, 120B, with low, medium, and high reasoning), on Nottingham. The prompting technique ``culture'' improves the performance of the models and surpasses humans, while the others do not.}
    \label{fig:improve-AN}
\end{figure}

\subsubsection{Improving Tacit Coordination}

Using models from the same family substantially improves NCI by reducing output diversity, as shown in Appendix~\ref{a:same-models-AN}. Appendix~\ref{a:other-models-AN} also reports scaling results for seven Llama models (1B, 3B, 8B, and 70B) and $13$ Qwen models (0.5B--72B). However, real deployments often involve heterogeneous agents, where the other models are unknown or uncontrolled. We therefore test three learning-free prompting techniques on the models from Section~\ref{sec:results-vanilla-AN}; prompts are reported in Appendix~\ref{a:prompting-AN}.

The first prompt, ``saliency'', asks the model to choose the option most salient to humans, approximating theory-theory accounts of Theory of Mind~\cite{sweller1994cognitive,leiberg2006multiple}. The second, ``all-features'', asks the model to reason over the focality principles of \citet{kraus2000exploiting}. The third, ``culture'', asks for the option most culturally relevant to humans, motivated by work on cultural representations in LLMs~\cite{tao2024cultural}. Figure~\ref{fig:improve-AN} shows that ``saliency'' and ``all-features'' have little effect, whereas ``culture'' markedly improves NCI, matching or surpassing human performance in both Amsterdam and Nottingham. Thus, when culture is part of focality, a simple cultural prompt can make heterogeneous LLMs coordinate more human-like behaviourally; further analysis appears in Appendices~\ref{a:cultural-AN} and~\ref{a:cultural-AN-salient}.

\begin{figure}
    \centering
    \includegraphics[width=1\linewidth]{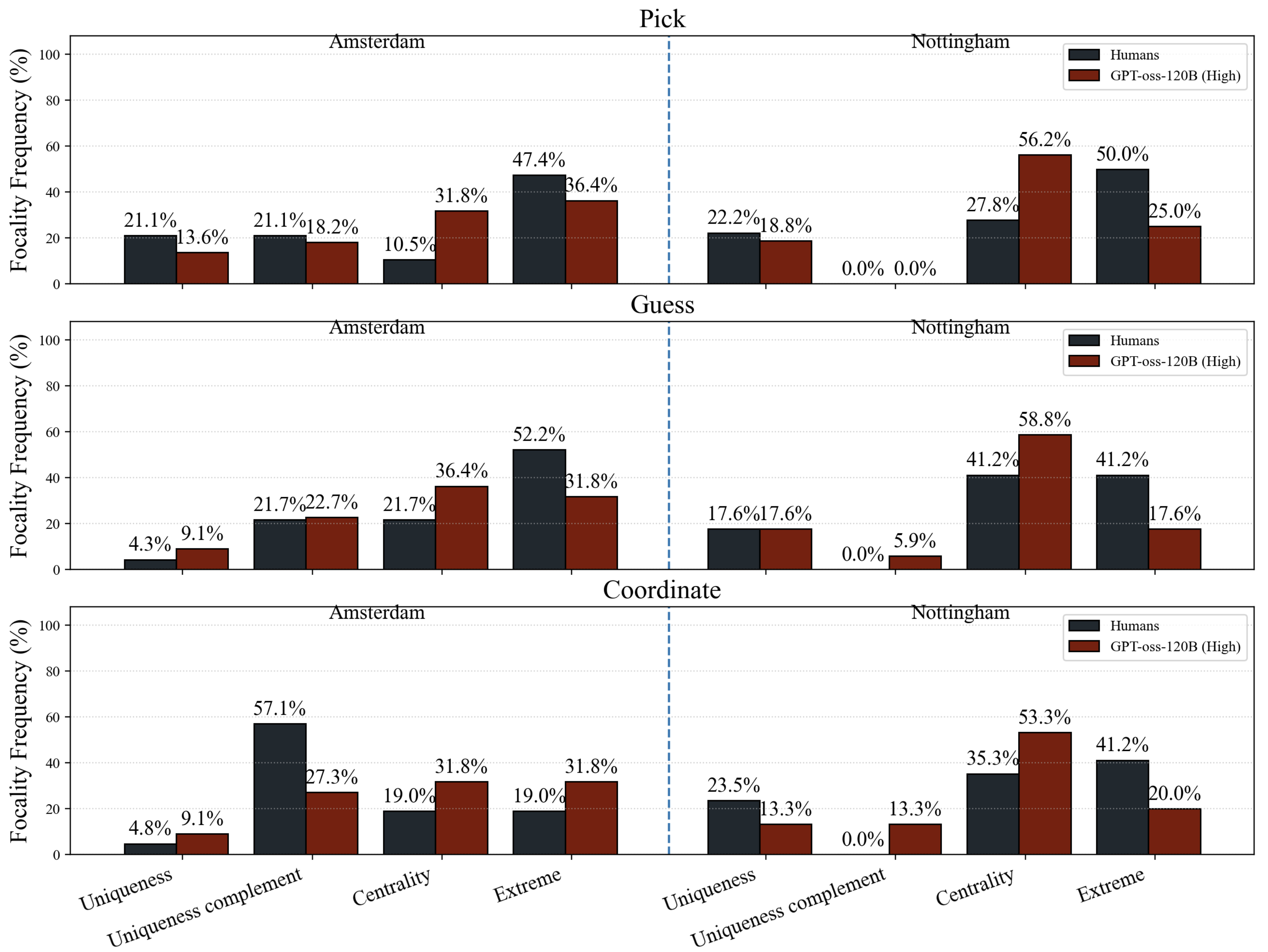}
    \caption{Analysis of which of the four principles of focality (as described in~\citet{kraus2000exploiting}) are picked up by humans and LLMs for Amsterdam and Nottingham. It clearly emerges that while humans leverage, although to varying degrees, all the principles of focality in both the experiments, in Nottingham LLMs tend to privilege centrality.}
    \label{fig:focality-AN}
\end{figure}

\subsubsection{Unique Focal Points and LLMs Saliency}

We further analyse the Amsterdam and Nottingham results through the focality framework of \citet{kraus2000exploiting}. Using GPT-5.2 with high reasoning, we assign each answer to one or more saliency groups: ``uniqueness'', ``uniqueness complement'', ``centrality'', and ``extremeness''. We then cluster the choices of humans and GPT-oss-120B with high reasoning.
Figure~\ref{fig:focality-AN} reveals different focality profiles. In Amsterdam, humans rely more on ``extremeness'' for ``pick'' and ``guess'', but shift towards ``uniqueness complement'' under ``coordinate'', where their performance improves. LLMs distribute choices more evenly across clusters, except for weaker use of ``uniqueness''. In Nottingham, humans use most focality dimensions except ``uniqueness complement'', whereas LLMs concentrate on ``centrality''. This centrality bias helps explain their weaker Nottingham performance: the models select structurally central options even when human focality is driven by other cues.

\begin{table}[t]
\centering

\begin{subtable}[t]{0.98\linewidth}
\centering
\resizebox{\linewidth}{!}{%
\begin{tabular}{ll|cccc}
\hline
 &  & \multicolumn{2}{c}{Orange Player (Human)} & \multicolumn{2}{c}{Blue Player (GPT-oss)} \\
\cline{3-4}\cline{5-6}
Variant & Model Size & Mean & Median & Mean & Median \\
\hline
\multirow{2}{*}{All-features}
 & 20B  &  9.62 & 10.60 & 26.90 & 30.07 \\
 & 120B & \textbf{30.18} & \textbf{31.33} & \textbf{28.74} & \textbf{31.33} \\
\hline
\multirow{2}{*}{Cooperative}
 & 20B  & 18.10 & 18.30 & \textbf{29.93}$^*$ & \textbf{32.70} \\
 & 120B & \textbf{32.89} & \textbf{33.30} & 29.10 & 29.67 \\
\hline
\multirow{2}{*}{Greedy}
 & 20B  & 15.00 & 15.57 & 25.25 & 28.60 \\
 & 120B & \textbf{31.51} & \textbf{32.03} & \textbf{27.32} & \textbf{28.57} \\
\hline
\multirow{2}{*}{Saliency}
 & 20B  & 26.44 & 27.47 & 24.27 & 26.20 \\
 & 120B & \cellcolor{LightCyan}\textbf{33.26}$^*$ & \cellcolor{LightCyan}\textbf{33.70}$^*$ & \cellcolor{LightCyan}\textbf{28.38} & \cellcolor{LightCyan}\textbf{29.00} \\
\hline
\multirow{2}{*}{Vanilla}
 & 20B  & 18.81 & 18.77 & 28.47 & \textbf{32.57}$^*$ \\
 & 120B & \textbf{32.03} & \textbf{32.10} & \textbf{28.53} & 29.93 \\
\hline
\end{tabular}%
}
\caption{Grouped by model size.}
\label{tab:bargaining-by-size}
\end{subtable}

\vspace{0.75em}

\begin{subtable}[t]{0.98\linewidth}
\centering
\resizebox{\linewidth}{!}{%
\begin{tabular}{ll|cccc}
\hline
 &  & \multicolumn{2}{c}{Orange Player (Human)} & \multicolumn{2}{c}{Blue Player (GPT-oss)} \\
\cline{3-4}\cline{5-6}
Variant & Reasoning & Mean & Median & Mean & Median \\
\hline
\multirow{3}{*}{All-features}
 & low    & \textbf{21.19} & 22.00 & 26.23 & 28.90 \\
 & medium & 19.96 & \textbf{21.15} & \textbf{28.93} & 31.55 \\
 & high   & 18.55 & 19.35 & 28.31 & \textbf{32.05} \\
\hline
\multirow{3}{*}{Cooperative}
 & low    & 24.91 & 25.70 & 29.36 & 31.30 \\
 & medium & \textbf{25.88} & \textbf{26.30} & \textbf{29.72} & \textbf{31.80} \\
 & high   & 25.69 & 26.05 & 29.46 & 31.10 \\
\hline
\multirow{3}{*}{Greedy}
 & low    & 21.05 & 21.90 & 25.59 & 26.85 \\
 & medium & 24.25 & \textbf{25.65} & \textbf{26.75} & \textbf{29.20} \\
 & high   & \textbf{24.48} & 25.10 & 26.52 & 27.95 \\
\hline
\multirow{3}{*}{Saliency}
 & low    & \cellcolor{LightCyan}\textbf{30.56}$^*$ & \cellcolor{LightCyan}\textbf{31.75}$^*$ & \cellcolor{LightCyan}26.06 & \cellcolor{LightCyan}27.15 \\
 & medium & 29.69 & 31.20 & 26.06 & \textbf{28.55} \\
 & high   & 29.30 & 31.50 & \textbf{26.85} & 28.15 \\
\hline
\multirow{3}{*}{Vanilla}
 & low    & 24.88 & 25.30 & 28.20 & 31.45 \\
 & medium & \textbf{26.75} & \textbf{27.25} & 28.32 & 30.35 \\
 & high   & 24.61 & 24.50 & \textbf{28.98}$^*$ & \textbf{32.35}$^*$ \\
\hline
\end{tabular}%
}
\caption{Grouped by reasoning level.}
\label{tab:bargaining-by-reasoning}
\end{subtable}

\caption{Mean and median Bargaining Table payoffs for Orange (Human) and Blue (GPT-oss) players across prompting variants. Top: results grouped by model size, averaging over reasoning levels. Bottom: results grouped by reasoning level, averaging over GPT-oss-20B and GPT-oss-120B. Best individual payoffs within each prompting technique are highlighted in \textbf{bold}; the overall best-performing entry is marked with $^*$; and the configuration with the highest social welfare is highlighted in \colorbox{LightCyan}{LightCyan}.}
\label{tab:bargaining_oss_summary}
\end{table}

\begin{figure*}
    \centering
    \includegraphics[width=1\linewidth]{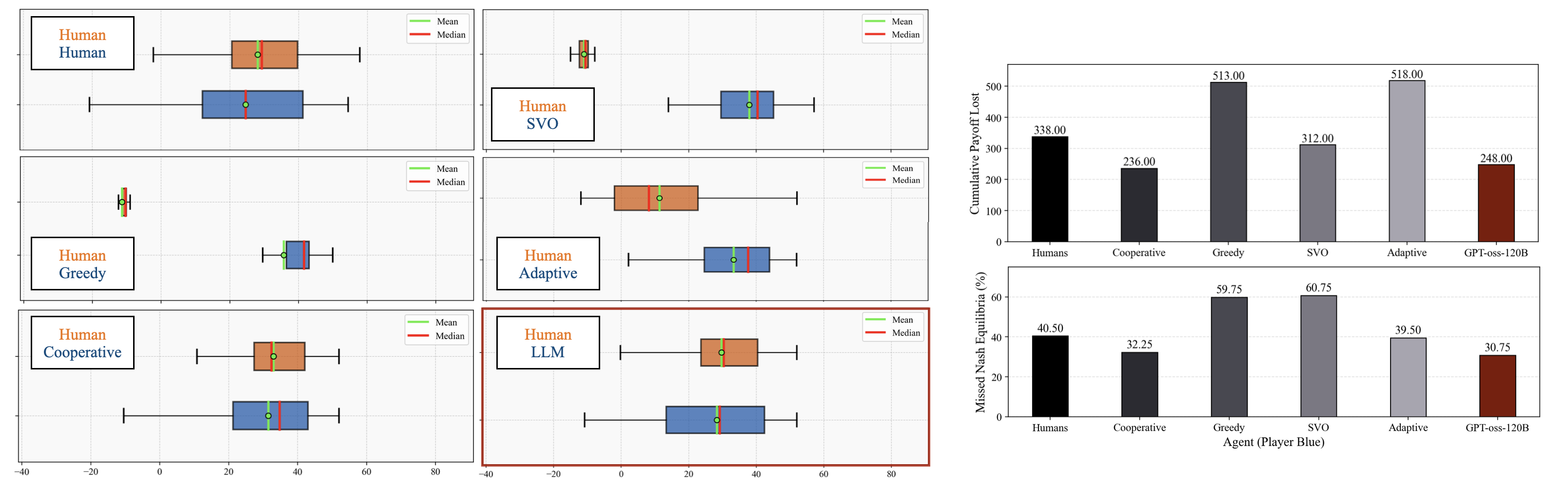}
    \caption{Left: Each barplot reports the mean and median payoff of the \emph{blue} and \emph{orange} players in $100$ iterations, per typology of game, of the Bargaining Table. While the blue agent changes her strategy, the data for the orange player is that of humans who played the game and comes from~\cite{mizrahi2020using}. The bottom-right player is the \emph{blue player} powered by GPT-oss-120B.
    Right: Missed Nashed equilibria and cumulative payoff lost (the lower, the better) in $100$ iterations of the ten games of the Bargaining Table per category of \emph{blue agent}. GPT-oss-120B, as a \emph{blue agent}, has performance comparable to a cooperative player.}
    \label{fig:bargaining-blue}
\end{figure*}

\subsection{The Bargaining Table}

The Bargaining Table is a mixed cooperative--competitive game in which two players, blue and orange, occupy positions on a $9 \times 9$ board and simultaneously assign each disk either to themselves or to the other player. If both agree on the same assignment, the assigned player receives the disk value; otherwise, both players receive a penalty equal to $20\%$ of that value. Final payoffs are the sum over all disk assignments.

Unlike the Amsterdam and Nottingham tasks, this game is semi-competitive: players must coordinate while maximising their own utility, as illustrated in Figure~\ref{fig:tasks-picture} (right). With $k$ disks, the game has $2^{k+1}$ joint strategies, of which $2^k$ are Nash equilibria: precisely those with no contested disk, since any disagreement creates a negative payoff that either player can avoid by deviating.

\citet{mizrahi2020using} collected data from $93$ university students playing both blue and orange roles, and compared them with three strategies: ``cooperative'', which assigns each disk to the closest player; ``greedy'', which assigns every disk to itself; and ``SVO'', which uses the other player's inferred \emph{social value orientation}~\cite{griesinger1973toward}. They also collected data from $38$ additional students to train an ``adaptive'' agent that maximises utility using features such as SVO, distance to disks, and opponent behaviour.

\begin{figure*}
    \centering
    \includegraphics[width=0.32\textwidth]{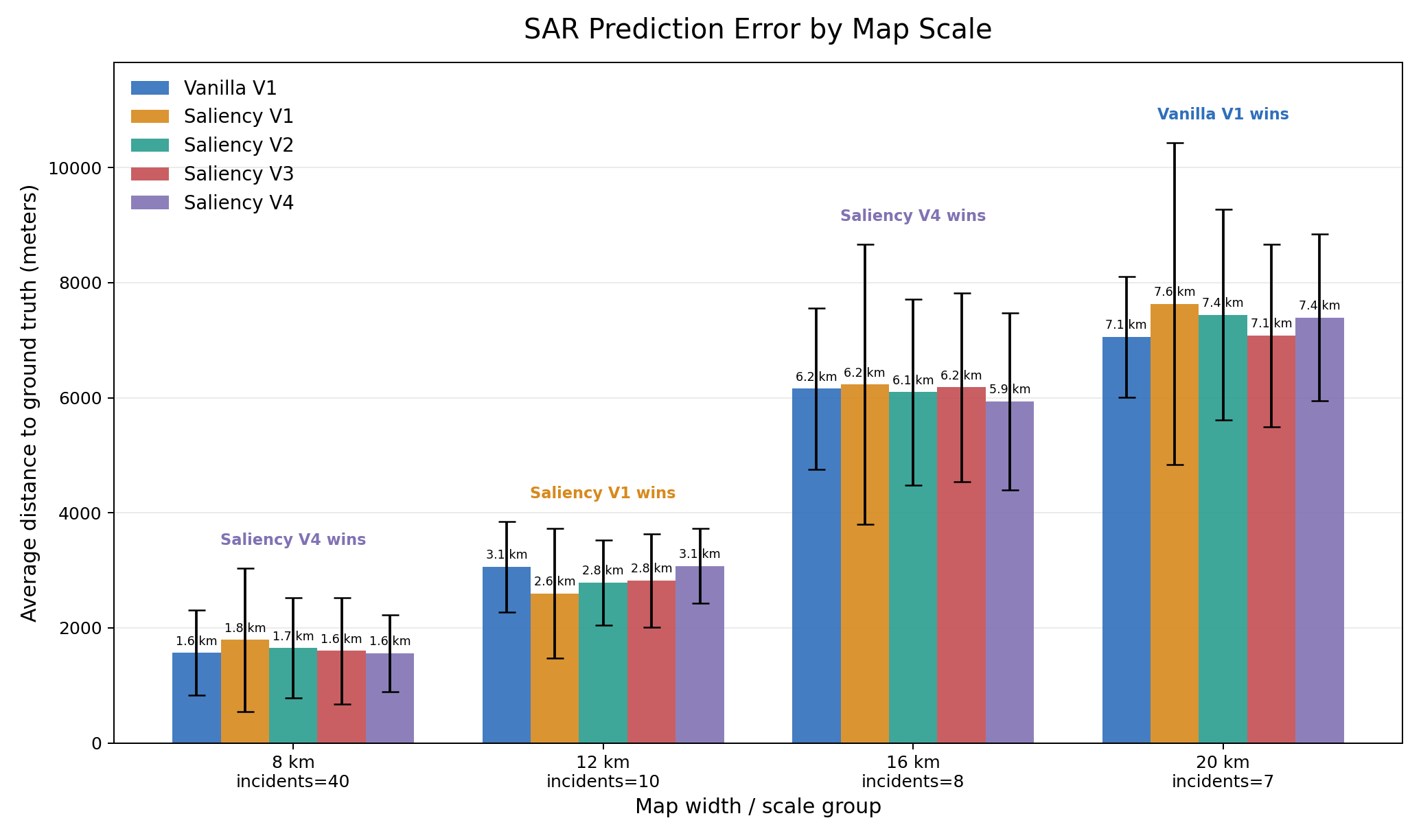}
    \hfill
    \includegraphics[width=0.32\textwidth]{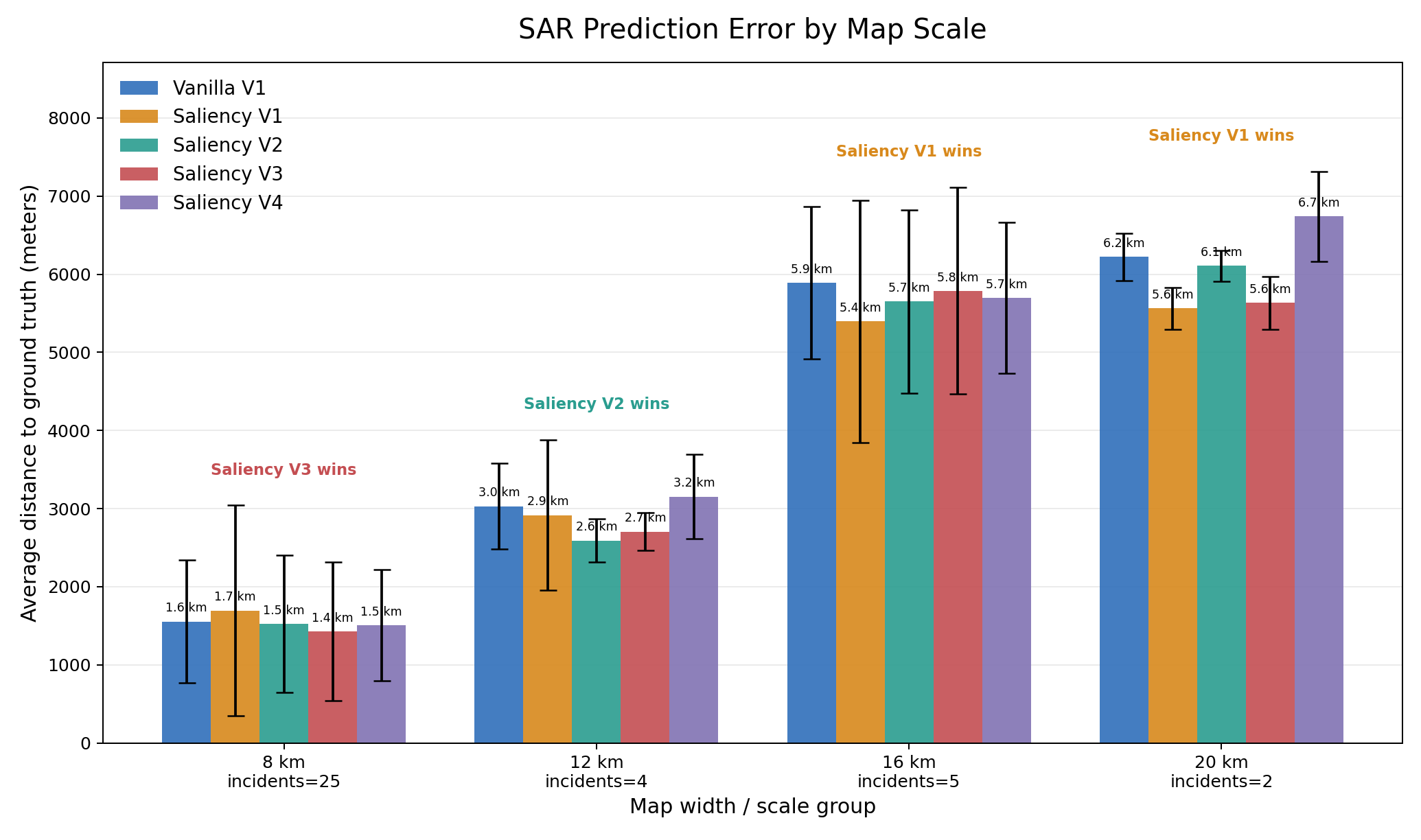}
    \hfill
    \includegraphics[width=0.32\textwidth]{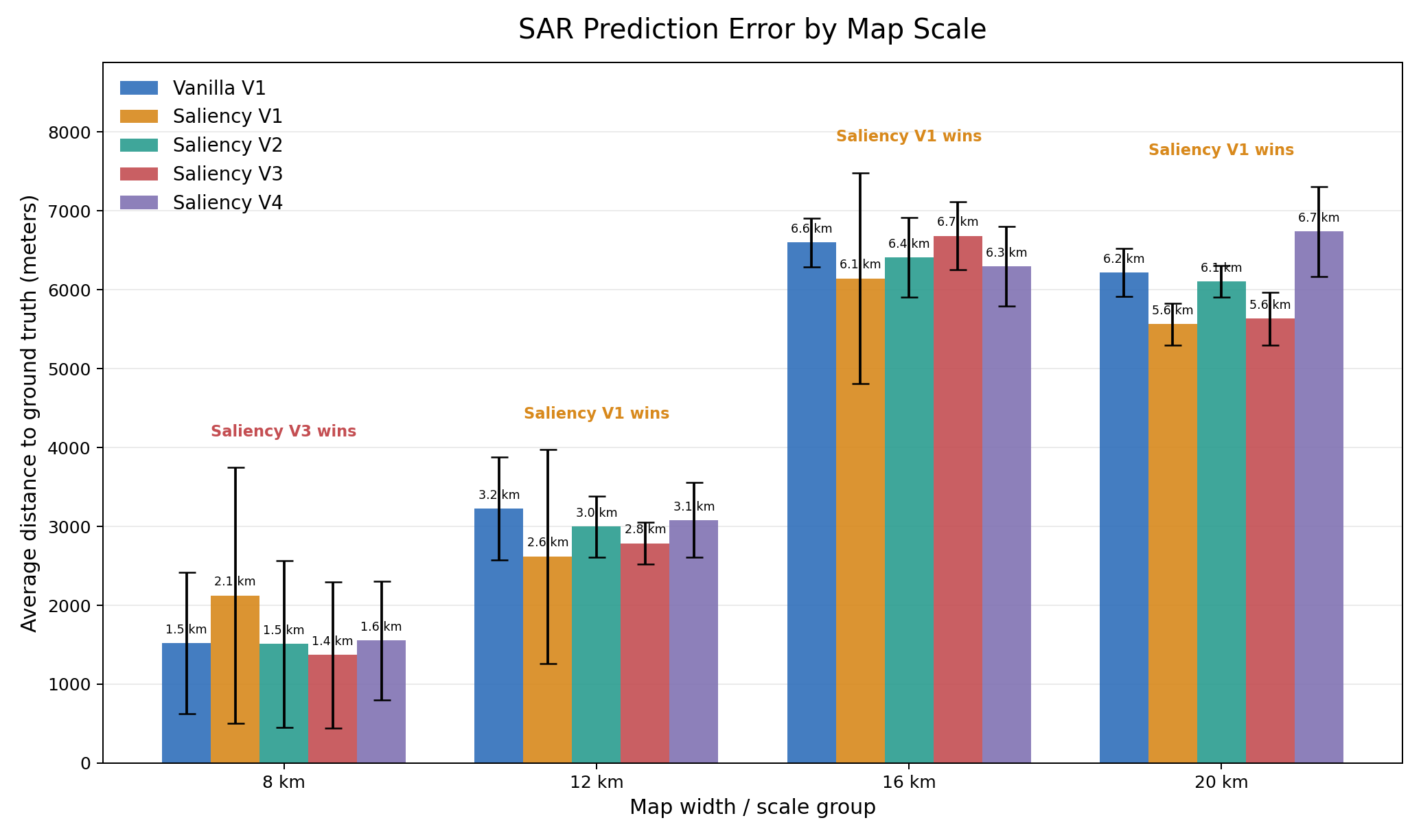}
    \caption{SAR prediction error by map scale. Bars show average Euclidean distance to the ground-truth find location for vanilla and the four behaviour-aware prompts, shown as Saliency V1--V4. Left: all 65 incidents. Middle: incidents classified as focal by GPT-5.5. Right: incidents classified as focal by Gemini 3.1 Pro Preview. Error bars show variation across predictions.}
    \label{fig:sar-error-by-scale}
\end{figure*}

\subsubsection{Results of the Bargaining Table}

We adapt the Bargaining Table into a one-shot LLM prompt and evaluate GPT-oss-20B and GPT-oss-120B against humans and the agents from \citet{mizrahi2020using}; prompts appear in Appendix~\ref{a:prompting-bargaining}. Since the original results report only the utility of the focal strategy player, we reconstruct the experiments from their data to compute both players' utilities and NCI.

Figure~\ref{fig:bargaining-blue} shows that LLMs behave closest to cooperative agents: they achieve high payoff and high social welfare for both players. GPT-oss-120B, in particular, minimises contested assignments and hence negative payoffs when playing as the blue agent.
Scaling improves performance. Table~\ref{tab:bargaining-by-reasoning} (top) shows that GPT-oss-120B substantially outperforms GPT-oss-20B in payoff, welfare, and coordination, suggesting that model size benefits both competition and coordination. We observe the same trend under prompts adapted from the Amsterdam/Nottingham experiments (``saliency'' and ``all-features'') and under explicit ``greedy'' and ``cooperative'' prompts. By contrast, increasing reasoning level does not improve payoff or social welfare, as shown in Table~\ref{tab:bargaining-by-reasoning} (bottom). Additional reasoning-scale, orange-player, and model-family results are reported in Appendices~\ref{a:reasoning-scale-bargaining},~\ref{a:yellow-bargaining}, and~\ref{a:other-models-bargaining}.

\begin{figure}
    \centering
    \includegraphics[width=0.46\textwidth]{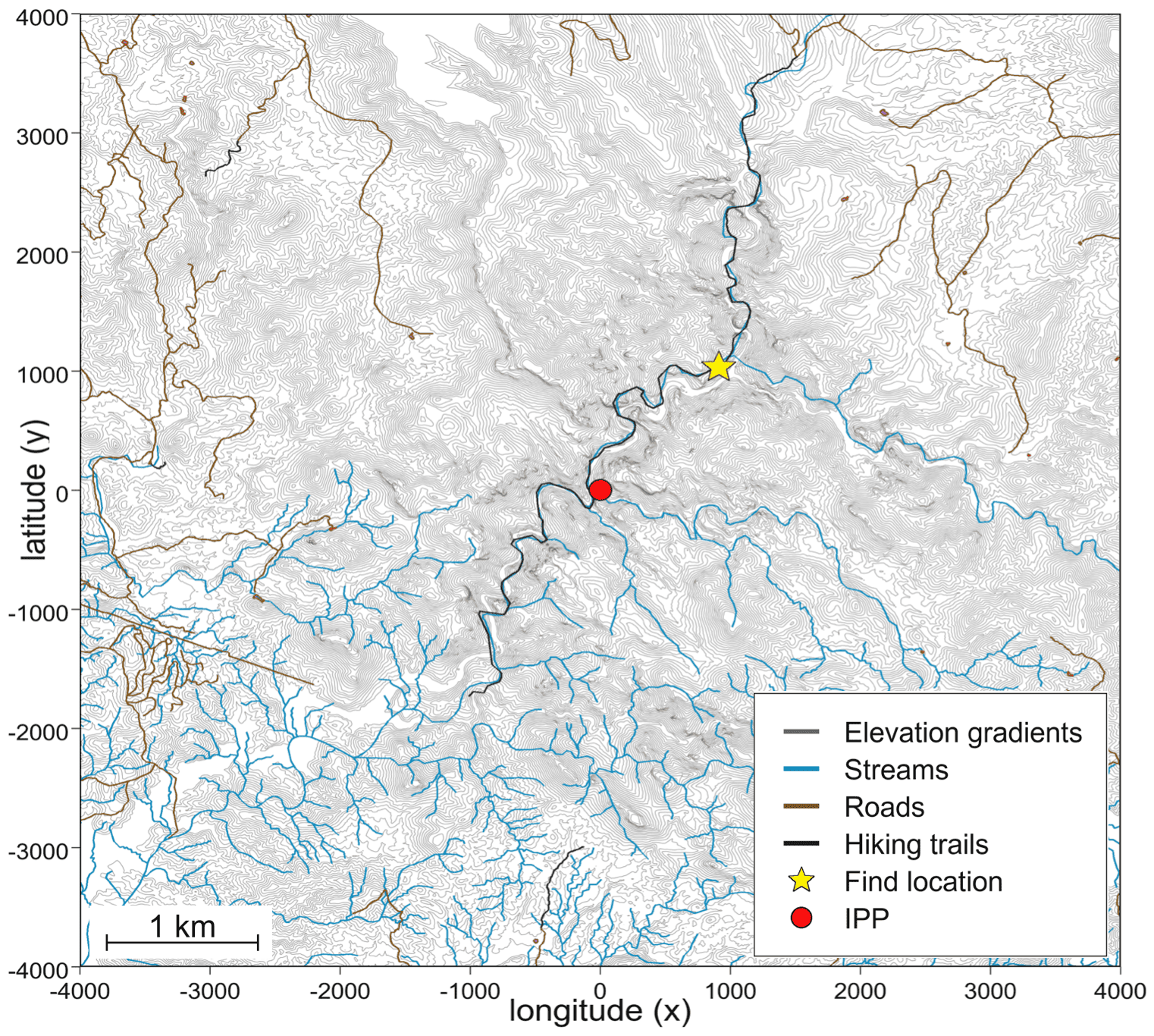}
    \caption{A rendered version of a SAR incident with the target-revealing map, which additionally overlays the yellow ground-truth find marker. The target-revealing location is removed at test-time.}
    \label{fig:sar-map-versions}
\end{figure}

\subsection{Focal Points in Search \& Rescue Missions}
\label{sec:sar-main}

We use wilderness search \& rescue (SAR) as a spatial testbed for focal-point reasoning. Given a terrain map centered on the initial planning point (IPP), an LLM predicts a single local coordinate where a missing hiker is likely to be found. The true find location is withheld. We use the 65 missing-hiker incidents from \citet{hashimoto2022agent}, projecting each IPP and finding the location into an IPP-centered metric frame and scoring predictions by Euclidean distance to the ground truth. Full task, map-generation, and evaluation details are reported in Appendix~\ref{app:sar-details}.

For each incident, we render two maps. The prediction map, shown to the model, contains terrain, OpenStreetMap vector layers, and a red IPP marker. The target-revealing map additionally overlays the true find location in yellow and is used only for classification, evaluation, and visualisation (see Figure~\ref{fig:sar-map-versions}).
Before prediction, we classify whether each incident is spatially \emph{focal}: that is, whether the realised find location lies near a salient movement or search affordance, such as a trail junction, road, stream crossing, shoreline, drainage corridor, settlement, or distinctive landmark. GPT-5.5 classified 36 of 65 incidents as focal, while Gemini 3.1 Pro Preview classified 19 of 65 as focal. Appendix~\ref{app:sar-task} illustrates both classes.

Behaviour-aware prompting (i.e., LLMs choose, in four variants S1-S4, a rescue location that is focal; expanded in Appendix~\ref{app:sar-prediction-eval}) does not improve SAR prediction uniformly. Across all 65 incidents, the best prompt, S3, improves over the vanilla baseline by only 11\,m on average. The effect appears specifically in focal cases. On GPT-5.5 focal incidents, S3 reduces mean error from 2580\,m to 2410\,m; on Gemini focal incidents, from 3086\,m to 2882\,m. These improvements are statistically significant: 170\,m under the GPT-5.5 split, with bootstrap 95\% CI [48, 292]\,m and permutation $p=0.0048$; and 203\,m under the Gemini split, with bootstrap 95\% CI [13, 388]\,m and permutation $p=0.0257$. No behaviour-aware prompt reliably improves the full set or the non-focal splits. Thus, focal-point prompting helps specifically when the missing person's endpoint is plausibly tied to a salient terrain cue.
Figure~\ref{fig:sar-error-by-scale} confirms the same pattern descriptively: larger maps produce larger absolute errors, while the S3 gain over vanilla is concentrated in the focal splits.

\section{Discussion, Findings and Conclusions}

This work studies tacit coordination in LLMs through Schelling focal points across cooperative coordination games, a semi-competitive bargaining task, and a spatial search \& rescue setting. We test whether LLMs converge on the same salient choices as humans, when they diverge, and whether prompting can improve this alignment. Overall, LLMs coordinate reliably, often matching or exceeding human baselines, but their success depends on whether model salience aligns with the structure of the task.

Across Amsterdam/Nottingham and the Bargaining Table, coordination is shaped by different constraints: shared salience in the former, and payoff asymmetry in the latter. Human and LLM failures, therefore, differ. Humans rely more on cultural and contextual cues, while LLMs favour structural heuristics such as centrality, explaining their weaker Nottingham performance and the gains from cultural prompting. More chain-of-thought reasoning does not reliably improve coordination, payoff, or welfare; instead, it often behaves like arbitrary tie-breaking. Model scale helps more consistently, reducing missed equilibria, improving payoffs, and producing more cooperative bargaining.
The SAR experiments extend this pattern to spatial focality. Behaviour-aware prompting improves localisation only when the true endpoint is itself focal, suggesting that LLM coordination succeeds when model salience matches human salience. 

Overall, LLM coordination is best understood as implicit social and spatial modelling. For mixed human--AI systems, the central challenge is not simply making agents reason more, but aligning what they find salient.

\section*{Limitations}
Our human baselines come from established datasets~\citep{bardsley2010explaining,mizrahi2020using}, which sample specific populations at specific times. This aids reproducibility and direct comparison, but the human salience we benchmark against reflects those groups; the cultural-prompting gains should be read as alignment with a given population's salience, not a universal human one. When we describe LLMs as exhibiting a centrality ``bias'', we mean a descriptive, behavioural tendency to over-select structurally central options relative to the human choice distribution on the same task; we do not make claims about social bias or representational harm.

Our coverage spans more than $20$ models, with the main analysis focused on open-weight models that expose their full chain of thought. Results for frontier API LLMs (GPT-5.5, Gemini) are reported as a complement rather than at the same scale. The focality classification relies on proprietary LLMs, and the two we use apply different thresholds for what counts as focal; we therefore report both splits throughout and treat focality as a model-relative judgement rather than a fixed property of each incident.

The SAR study uses 65 incidents from a single dataset~\citep{hashimoto2022agent} and scores a single predicted coordinate by distance to the realised find location. This controlled setup isolates focal-point reasoning, and the behaviour-aware gains it reveals are correspondingly specific: they appear on focal incidents, where the endpoint is tied to a salient terrain cue, and not on the full set. We treat this conditional effect as the finding, and leave richer search formulations (uncertainty, multiple search points, cost asymmetry) to future work.

\section*{Ethical Considerations}
This work uses only previously published, de-identified datasets of game choices and missing-person incidents, and conducts no new experiments with human participants. Our SAR analysis is a controlled research probe, not a deployable search tool, and we want to be explicit about the risk of misreading it as one. A model that outputs a confident single coordinate can be wrong, and in a real search, a wrong prediction may divert finite teams away from a missing person, with potential cost to life. Our own results underline this: behaviour-aware prompting helps only when the endpoint is focal and can degrade predictions when it is not, so the method is not safe to apply uniformly. We therefore frame LLM output in SAR as decision support that must remain subordinate to trained human judgement and established search-theoretic practice, never as an autonomous recommender. More broadly, the cultural-prompting results show that LLM salience can be steered toward a target population, which is useful for coordination but could equally be used to manipulate or homogenise choices; we report it to make the mechanism transparent rather than to encourage that use.

\section*{Acknowledgments}
ELM is affiliated with the Institute for Decentralized AI, which he thanks for its support. MW is supported by an AI 2050 Senior Fellowship from the Schmidt Sciences Foundation. The team thanks Dor Mizrahi (and his team) for his support in sharing the data and useful insights to replicate the experiments in his paper~\cite{mizrahi2020using}.

\clearpage
\bibliography{custom}

\clearpage
\appendix
\onecolumn
\section*{Appendix - Table of Content}
\begin{itemize}
    \item[] \textbf{\ref{a:methodology}. Methodology - Theory and Proofs}
        \begin{itemize}
            \item[] \textbf{\ref{a:symmetry}. Symmetry Groups}
            \item[] \textbf{\ref{a:ci}. Coordination Index}
        \end{itemize}
    \item[] \textbf{\ref{a:prompting}. Prompting Techniques}
        \begin{itemize}
            \item[] \textbf{\ref{a:prompting-AN}. Amsterdam and Nottingham - Prompting Protocol}
            \item[] \textbf{\ref{a:prompting-bargaining}. Bargaining Table - Prompting Protocol}
        \end{itemize}
    \item[] \textbf{\ref{a:additional-results}. Experimental Evaluation and Additional Results}
        \begin{itemize}
            \item[] \textbf{\ref{a:models-list}. List of Models and Experimental Details}
            \item[] \textbf{\ref{a:AN-performance-across-tasks}. Amsterdam and Nottingham - Performance Across Tasks}
            \item[] \textbf{\ref{a:additional-res-AN}. Additional Results for Amsterdam and Nottingham Human Experiments}
            \item[] \textbf{\ref{a:additional-res-bargaining}. Additional Results for the Bargaining Table}
            \item[] \textbf{\ref{a:additional-res-SAR}. Search \& Rescue: Details and Additional Experimental Results}
        \end{itemize}
\end{itemize}

\clearpage
\twocolumn
\section{Methodology - Theory and Proofs}~\label{a:methodology}

\subsection{Existence and Uniqueness of Focal Point Equilibria}\label{sec:orbits}
This section first discusses under which conditions a unique focal point equilibrium exists and is chosen by all the players.
It then mathematically characterises two cases where a unique focal point equilibrium emerges.

The existence of a unique focal point, despite being observed in several aspects of everyday decision making, stands on strong assumptions about the salience function of each player.
To show that, we assume two cases: when players share the same salience function, and when players do not.

\begin{figure}
    \centering
    \includegraphics[width=1\linewidth]{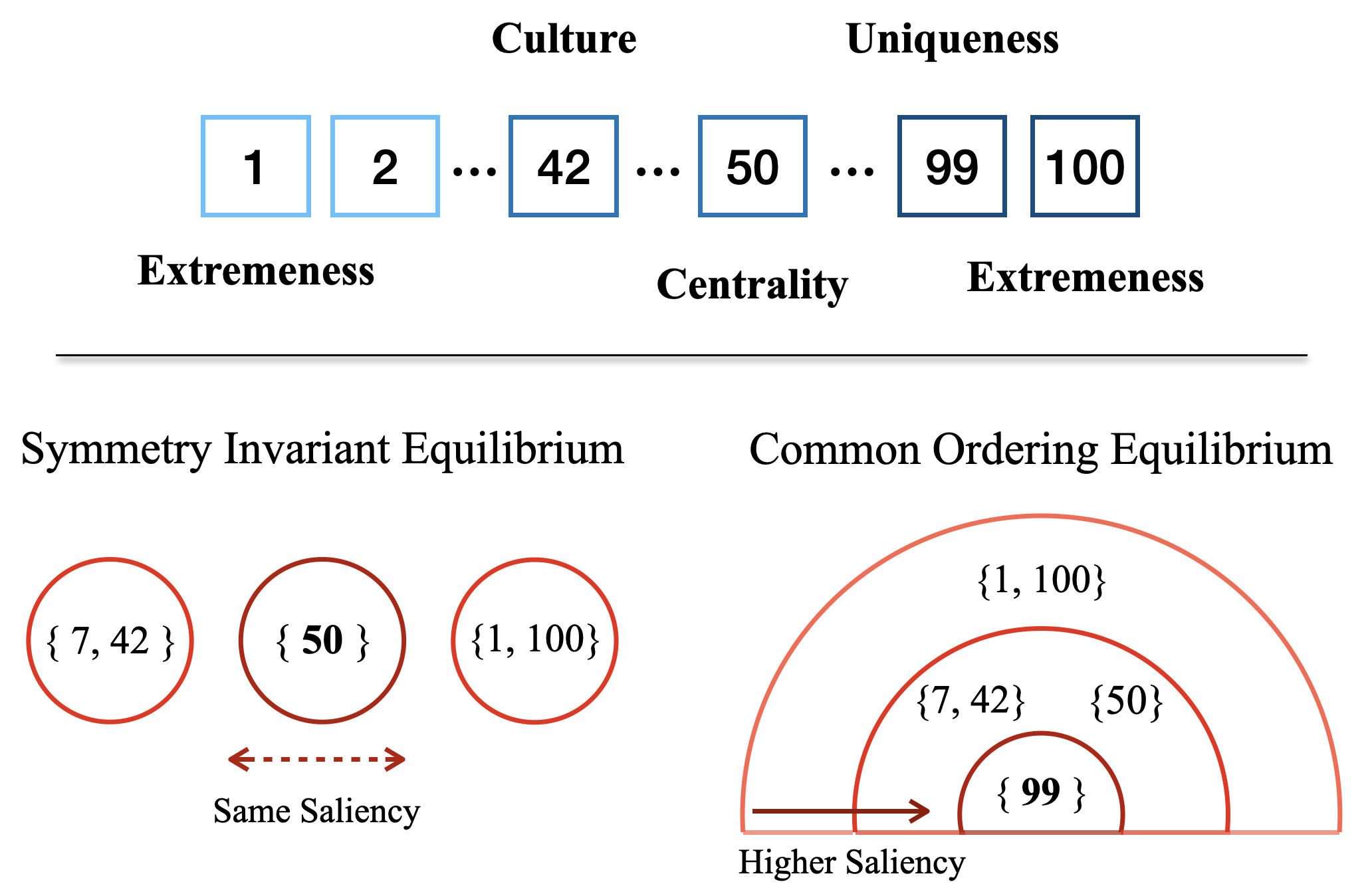}
    \caption{Example of symmetry invariant (left) and common ordering (right) orbits that lead to a unique focal point equilibrium (in \textbf{bold}) on the question ``Pick a number between $1$ and $100$''.}
    \label{fig:orbits-example}
\end{figure}

First, consider the case where players share the same salience function. Let $G$ be a normal-form two-player game with a finite nonempty set of Nash equilibria $\mathcal{E}$. 
\begin{itemize}
\item When the players share the same salience function, it is sufficient to introduce a realised noise on its output (e.g., i.i.d. continuous noise across equilibria), i.e., $\bar{S}(e) = S(e) + \eta(e)$, to induce, with probability $1$, the existence of a unique focal equilibrium $e^* \in \arg\max_{e \in \mathcal{E}} \bar{S}(e)$. 
For the existence, $\mathcal{E}$ is finite, so a maximiser of $\bar{S}$ exists. For uniqueness, suppose there are two distinct equilibria $(i \neq j)$ s.t., $\bar{S}(e_i) = \bar{S}(e_j)$. That implies that $S(e_i) - S(e_j) = \eta(e_j) - \eta(e_i)$. Because $\eta$ has a density, the probability for a tie to exactly lie on this hyperplane is $0$. On the other hand, the previous case assumes there is a mechanism (noise) to break ties and introduces some sort of coordination; otherwise, no equilibrium exists.
\item It is trivial to prove that, under different salience functions of players $1$ and $2$, namely $S_1$ and $S_2$, a unique focal point does not exist (even in the presence of realised noise). Consider the case where $\mathcal{E}=\{e_1, e_2\}$, s.t., $S_1(e_1) \gg S_1(e_2) \wedge S_2(e_2) \gg S_2(e_1)$.
\end{itemize}
When players have different salience functions, it is interesting to characterise when the outcome is still a unique focal point equilibrium.
We denote two cases: (i) symmetry-invariant equilibrium and (ii) common ordering induced by symmetry classes: the intuition behind them is depicted in Figure~\ref{fig:orbits-example}.

We distinguish the salience functions of each player, namely $S_1$ and $S_2$, and introduce a symmetry group~$\circ: \Gamma \times \mathcal{E} \xrightarrow{} \mathcal{E}$ that maps the equilibria into ``orbits'' of the same saliency for both players, i.e., $\mathcal{E} = \bigcup_k \operatorname{Orb}_k \ . \ \forall (e_i, e_j) \in \operatorname{Orb}_k, \ S_p(e_i) = S_p(e_j), \ p \in \{1, 2\}$. By the definition of a symmetry group, orbits are non-empty, pairwise disjoint, and their union forms the original set of Nash equilibria: Section~\ref{a:symmetry} provides a formal definition of a symmetry group and its properties.

For symmetry-invariant equilibrium (i), $\Gamma$ identifies orbits of which only one is a singleton and thus the focal point equilibrium, i.e., $\exists! \ \operatorname{Orb}^*=\{e^*\}$. This case captures when players only agree on the unique focal point equilibrium, which stands out among the others.
For the common ordering induced by symmetry classes (ii), similarly to case (i), a symmetry group $\Gamma$ partitions equilibria into orbits, of which more than one can be a singleton, yet players still agree on which orbit is the most salient, and thus the unique focal point equilibrium. Namely, $\exists! \ \operatorname{Orb}^*=\{e^*\} \ . \ e^* \in \arg\max_{e \in \operatorname{Orb}_k} S_p(e), \ p \in \{1, 2\}$.
This case tightens up case (i) and captures the case when players may identify different unique focal points that stand out, yet one is still chosen.

\subsection{Symmetry Groups}\label{a:symmetry}
\begin{definition}[Symmetry group and induced partition]
Let $\mathcal{E}$ be a set and let $\Gamma$ be a group with identity element $x$.
A \emph{symmetry group} of $\mathcal{E}$ is a group action
\[
\circ : \Gamma \times \mathcal{E} \to \mathcal{E}
\]
satisfying
\[
x \circ e = e \quad \forall e \in \mathcal{E},
\qquad
(\gamma_1 \gamma_2) \circ e
= \gamma_1 \circ (\gamma_2 \circ e)
\]
\[
\forall \gamma_1,\gamma_2 \in \Gamma,\; e \in \mathcal{E}.
\]

\noindent For each $e \in \mathcal{E}$, define the orbit
\[
\operatorname{Orb}_\Gamma(e)
:= \{ \gamma \circ e \mid \gamma \in \Gamma \}.
\]

\noindent The collection of orbits
\[
\mathcal{E} / \Gamma
:= \{ \operatorname{Orb}_\Gamma(e) \mid e \in \mathcal{E} \}
\]
forms a partition of $\mathcal{E}$ into pairwise disjoint subsets.
\end{definition}

\subsection{Coordination Index}\label{a:ci}
Let $G=(N,(\Sigma_i)_{i\in N},(u_i)_{i\in N})$ be a two-player normal-form symmetric coordination game, where both players share the same finite set of pure strategies
$\{\hat{\sigma}_1,\ldots,\hat{\sigma}_m\}$ and have symmetric, non-negative payoffs, i.e.,
$u_1(\sigma)=u_2(\sigma)\ge 0$ for all pure strategy profiles $\sigma$.
With a slight abuse of notation, we identify the pure strategy sets of the two players.

Consider a sample of $n$ independent plays of the game. Let $m_j$ denote the number of times strategy $\hat{\sigma}_j$ is chosen in the sample, with
$\sum_{j=1}^m m_j = n$.
Select two distinct players uniformly at random from the sample (sampling without replacement), and let $\sigma_1$ and $\sigma_2$ denote their chosen strategies.

The probability that the two selected players choose the same strategy is
\[
\mathbb{P}(\sigma_1=\sigma_2)
= \sum_{j=1}^m
\mathbb{P}(\sigma_2=\hat{\sigma}_j)\,
\mathbb{P}(\sigma_1=\hat{\sigma}_j \mid \sigma_2=\hat{\sigma}_j)
\]
\[
= \sum_{j=1}^m \frac{m_j}{n}\frac{m_j-1}{n-1},
\]

which is the definition of Coordination Index (CI).

\clearpage
\section{Prompting Techniques}~\label{a:prompting}
\noindent This section introduces the prompting techniques used in the Amsterdam and Nottingham experiments, as well as the Bargaining Table.

\subsection{Amsterdam and Nottingham - Prompting Protocol}\label{a:prompting-AN}

For the Nottingham and Amsterdam experiments, we instantiated three task templates, corresponding to standard variants of coordination and prediction problems. In all cases, the placeholder \texttt{@objects@} was replaced by the task-specific list of options and associated payoffs (e.g., \{\texttt{Friday lunchtime: 10, Monday morning: 10, Saturday night: 10, Sunday night: 10, Wednesday evening: 10}\}).

\paragraph{Task templates.}
We used the following base instructions:
\begin{itemize}
    \item \textbf{Problem-Pick:} \emph{``In this task, you have to choose one of the following options. This is the list of options and their score: \texttt{@objects@}. You will earn the number of points shown on the option you choose. Your earnings are determined by your decisions alone. Return the option you chose, without the score, between \texttt{<answer></answer>} tags.''}
    \item \textbf{Problem-Guess:} \emph{``In this task, you have to choose one of the following options. This is the list of options and their score: \texttt{@objects@}. You will earn the number of points shown on the option you choose. You have been paired with another LLM and you have to guess what it chooses. If you guess correctly, you will receive the number of points indicated on that option. If not, you will receive nothing for that task. Return the option you chose, without the score, between \texttt{<answer></answer>} tags.''}
    \item \textbf{Problem-Coordinate:} \emph{``In this task, you have to choose one of the following options. This is the list of options and their score: \texttt{@objects@}. You have been paired with another LLM. If you choose the same object as your partner, you will both receive the number of points indicated on that option. If not, neither will receive anything. Return the option you chose, without the score, between \texttt{<answer></answer>} tags.''}
\end{itemize}

\paragraph{Prompt variants.}
For each task template, we evaluated four prompting variants:
\begin{itemize}
    \item \textbf{Vanilla:} the base instruction as written above, with no additional guidance.

    \item \textbf{Saliency:} we encouraged the selection of an option that is maximally distinct from the others:
    \begin{quote}\small
    \emph{``Choosing an option that is entirely different from the others is highly recommended (for example, in a different category).''}
    \end{quote}
    
    \item \textbf{All-features:} we appended an explicit definition of four focal-point properties and instructed the model to prioritise options satisfying these properties:
    \begin{quote}\small
    \emph{``There are four intuitive properties that make a choice desirable:
    (i) uniqueness: it is the only object with a given property;
    (ii) uniqueness complement: it is the only object without a given property;
    (iii) centrality: it is a central point around which a domain is symmetric;
    (iv) extremeness: it is an object that has the largest or the smallest feature among all the others.
    Now, you have to prioritize the selection of the only object that satisfies one of the above-mentioned properties, i.e., it is the only object with that property in the list. If there are multiple objects that satisfy one or more of the above-mentioned properties, select the one that is unique, then extreme, then central, and eventually a unique complement.''}
    \end{quote}

    \item \textbf{Culture:} we conditioned the model on sharing the same cultural background as the human participants from the relevant location. For Nottingham (analogously for Amsterdam), we used task-specific phrasing:
    \begin{itemize}
        \item \emph{Pick:} \emph{``You are in Nottingham, so make your decision based on the activity or object that you would like to do or obtain as a person from that place.''}
        \item \emph{Guess:} \emph{``You have been paired with another human from Nottingham. You have to guess what he/she chooses: remember that you and your partner are both in Nottingham, so make your decision based on the activity or object you think your partner would like to do or obtain as a person from that place.''}
        \item \emph{Coordinate:} \emph{``You are from Nottingham and you have been paired with another human from Nottingham. If you choose the same object as your partner, you will both receive the number of points indicated on that option. If not, neither will receive anything. Make your decision based on the activity or object that you and your partner would like to do or obtain as a person from Nottingham.''}
    \end{itemize}

\end{itemize}

\subsection{Bargaining Table - Prompting Protocol}\label{a:prompting-bargaining}
\paragraph{Task template.}
For the Bargaining Table experiment, we used a single task template with two placeholders: \texttt{@state@}, describing the concrete board configuration, and \texttt{@variant@}, specifying the behavioural objective. The complete prompt was introduced as follows:
\begin{quote}\small
\emph{"Bargaining Table is a tacit coordination game played on a 9$\times$9 board with two special squares, one representing each player (e.g., a blue square and a yellow square). Several discs are scattered on the board, and each disc has a numeric value. Without communicating and without knowing the other player's choices, each player must decide, for every disc, which of the two player-squares the disc should be assigned to. A disc's value is awarded only if both players assign that disc to the same player: if both assign it to Blue, Blue receives the disc's value (and Yellow receives nothing for that disc), and if both assign it to Yellow, Yellow receives the disc's value (and Blue receives nothing for that disc). If the two players assign the same disc to different players, then both players are penalized by losing 20\% of that disc's value. The total score for each player is the sum of outcomes over all discs.\\
The coordinates are written as (row, column) on a 9$\times$9 grid, where row 1 is the top row and column 1 is the leftmost column. You will have to play that game in the following state:\\
\texttt{@state@}\\
Inside \texttt{<answer></answer>} tags, output only a JSON object where each key is a disc coordinate in the form \texttt{"(row,col)"} and each value is either \texttt{"blue"} or \texttt{"yellow"}, indicating which player you assign that disc to. Include every disc exactly once and no extra text. For example:}
\texttt{<answer>{"(row,col)":"blue","(row,col)":"yellow"}</answer>}
\emph{\texttt{@variant@} Start the solution below."}
\end{quote}

\paragraph{Game instance encoding.}
The placeholder \texttt{@state@} was replaced by a natural-language description of the concrete board configuration, including (i) the model's player identity (Blue or Yellow), (ii) the locations of both player-squares, and (iii) the set of discs with their values and coordinates. For example, when the model played as Blue in the first game:
\begin{quote}\small
\emph{You are the Blue player, and your square is located at (6, 2). The other player's square (Yellow) is located at (6, 9). There are 5 discs on the board: a value-3 disc at (8, 1), a value-3 disc at (4, 4), a value-3 disc at (1, 7), a value-1 disc at (1, 8), and a value-2 disc at (9, 8).}
\end{quote}

\paragraph{Prompt variants.}
We instantiated \texttt{@variant@} using one of the following directives:
\begin{itemize}
    \item \textbf{Vanilla:} no additional text.
    \item \textbf{Greedy:} \emph{``Try to maximize your own payoff.''}
    \item \textbf{Cooperative:} \emph{``Try to be cooperative: aim to maximize the total payoff of both players (joint outcome), not just your own.''}
    \item \textbf{All-features:} we appended an explicit definition of four focal-point properties and instructed the model to prioritize assignments accordingly:
    \begin{quote}\small
    \emph{``There are four intuitive properties that make a choice desirable:
    - uniqueness: it is the only object with a given property.
    - uniqueness complement: it is the only object *without* a given property.
    - centrality: it is a central point around which a domain is symmetric.
    - extremeness: it is an object that has the largest or the smallest property among all the others.
    Now, you have to prioritise the selection of your discs based on the mentioned properties.''}
    \end{quote}
    \item \textbf{Saliency:} \emph{``Anticipate the other player's moves and prefer discs he is unlikely to pick for himself.''}
\end{itemize}
\clearpage
\section{Experimental Evaluation and Additional Results}~\label{a:additional-results}
\subsection{List of Models and Experimental Details}~\label{a:models-list}
For the results in the main paper and the Appendix, we relied on the HuggingFace Python library and their off-the-shelf configuration and temperature.
For the GPT-oss reasoning models, we tuned their ``reasoning effort'' as suggested by OpenAI here: \href{https://openai.com/index/introducing-gpt-oss/}{https://openai.com/index/introducing-gpt-oss/}. For Gemini and ChatGPT, we relied on their API endpoints.

While in the main paper we report the results on Llama-3, 3.1, 3.3 70B, Qwen-2, 2.5 72B, and GPT-oss 20B and 120B, our analysis encompasses many more LLMs.
We report additional experiments in the next sections; yet, all the results are available in the code material.

We hereby report the full list of models we used:
\begin{itemize}

  \item \textbf{Meta}
  \begin{itemize}
    \item \textbf{LLaMA 3.0}~\cite{grattafiori2024llama3herdmodels}
    \begin{itemize}
      \item Meta-Llama-3-70B-Instruct
      \item Meta-Llama-3-8B-Instruct
    \end{itemize}

    \item \textbf{LLaMA 3.1} (\href{https://ai.meta.com/blog/meta-llama-3-1/}{ai.meta.com/blog/meta-llama-3-1/})
    \begin{itemize}
      \item Llama-3.1-70B-Instruct
      \item Llama-3.1-8B-Instruct
    \end{itemize}

    \item \textbf{LLaMA 3.2} (\href{https://ai.meta.com/blog/llama-3-2-connect-2024-vision-edge-mobile-devices/}{ai.meta.com/blog/llama-3-2-connect-2024-vision-edge-mobile-devices/})
    \begin{itemize}
      \item Llama-3.2-3B-Instruct
      \item Llama-3.2-1B-Instruct
    \end{itemize}

    \item \textbf{LLaMA 3.3} (\href{https://www.llama.com/docs/model-cards-and-prompt-formats/llama3_3/}{www.llama.com/docs/model-cards-and-prompt-formats/llama3\_3/})
    \begin{itemize}
      \item Llama-3.3-70B-Instruct
    \end{itemize}
  \end{itemize}
  
  \item \textbf{Qwen}
  \begin{itemize}
    \item \textbf{Qwen 2.5}~\cite{yang2024qwen2technicalreport}
    \begin{itemize}
      \item Qwen/Qwen2-72B-Instruct
      \item Qwen/Qwen2-7B-Instruct
      \item Qwen/Qwen2-0.5B-Instruct
    \end{itemize}

    \item \textbf{Qwen 2.5}~\cite{qwen2025qwen25technicalreport}
    \begin{itemize}
      \item Qwen2.5-72B-Instruct
      \item Qwen2.5-32B-Instruct
      \item Qwen2.5-14B-Instruct
      \item Qwen2.5-14B-Instruct-1M
      \item Qwen2.5-7B-Instruct
      \item Qwen2.5-7B-Instruct-1M
      \item Qwen2.5-3B-Instruct
      \item Qwen2.5-1.5B-Instruct
      \item Qwen2.5-0.5B-Instruct
    \end{itemize}
    
  \end{itemize}

  \item \textbf{OpenAI}
  \begin{itemize}
    \item \textbf{GPT-OSS} (\href{https://openai.com/index/introducing-gpt-oss/}{openai.com/index/introducing-gpt-oss/})
    \begin{itemize}
      \item GPT-oss-20B
      \item GPT-oss-120B
    \end{itemize}
    \item \textbf{ChatGPT} (\href{https://developers.openai.com/api/docs/models/gpt-5.4}{developers.openai.com/api/docs/models/gpt-5.4})
    \begin{itemize}
      \item GPT 5.4
    \end{itemize}
  \end{itemize}

  \item \textbf{Google}
  \begin{itemize}
    \item \textbf{Gemini} (\href{https://openai.com/index/introducing-gpt-oss/}{openai.com/index/introducing-gpt-oss/})
    \begin{itemize}
      \item Gemini2.5-pro (\href{https://blog.google/innovation-and-ai/models-and-research/google-deepmind/gemini-model-thinking-updates-march-2025/}{Gemini2.5})
      \item Gemini3-pro (\href{https://blog.google/products-and-platforms/products/gemini/gemini-3/}{Gemini3})
    \end{itemize}
  \end{itemize}

\end{itemize}

\subsection{Amsterdam and Nottingham - Performance Across Tasks}\label{a:AN-performance-across-tasks}

\paragraph{Model results.}
\textbf{Qwen} reports results aggregated across \texttt{Qwen2-72B-Instruct} and \texttt{Qwen2.5-72B-Instruct}.
\textbf{Meta-Llama} reports results aggregated across \texttt{Llama-3.1-70B-Instruct}, \texttt{Llama-3.3-70B-Instruct}, and \texttt{Meta-Llama-3-70B-Instruct}.
For each family, we first merge model outputs within the family and then compute the normalized coordination index on the merged data (rather than averaging indices computed per model). Afterwards, the reported scores are averages across all questions in the experiment.

\subsubsection{Amsterdam}

\begin{table*}[h!]
\centering
\caption{Amsterdam: performance by method and task.}
\begin{tabular}{llccc}
\toprule
Method & Task & Humans & Meta-Llama & Qwen \\
\midrule
\multirow{3}{*}{Vanilla}
  & coordinate & 1.85 & 2.24 & 1.92 \\
  & guess      & 1.27 & 1.79 & 1.82 \\
  & pick       & 1.12 & 2.02 & 1.77 \\
\midrule
\multirow{3}{*}{Culture}
  & coordinate & 1.85 & 2.59 & 2.80 \\
  & guess      & 1.27 & 2.58 & 3.00 \\
  & pick       & 1.12 & 2.34 & 2.98 \\
\midrule
\multirow{3}{*}{Saliency}
  & coordinate & 1.85 & 2.39 & 2.17 \\
  & guess      & 1.27 & 2.56 & 2.08 \\
  & pick       & 1.12 & 2.83 & 2.30 \\
\midrule
\multirow{3}{*}{All-Features}
  & coordinate & 1.85 & 2.17 & 2.08 \\
  & guess      & 1.27 & 2.06 & 1.98 \\
  & pick       & 1.12 & 2.23 & 1.90 \\
\bottomrule
\end{tabular}
\end{table*}

In Amsterdam, both model families consistently outperform humans across all methods and tasks. Overall, \textit{Culture} yields the strongest results for Qwen on every task (\textit{coordinate}: 2.80, \textit{guess}: 3.00, \textit{pick}: 2.98), while Meta-Llama performs best under \textit{Culture} on \textit{coordinate} (2.59) and \textit{guess} (2.58). For Meta-Llama, the main exception is \textit{pick}, where \textit{Saliency} is clearly strongest (2.83). In contrast, \textit{All-Features} is generally weaker than \textit{Culture} and \textit{Saliency} for both models, and \textit{Vanilla} typically yields the lowest model scores in this setting.

\subsubsection{Nottingham}

\begin{table*}[h!]
\centering
\caption{Nottingham: performance by method and task.}
\begin{tabular}{llccc}
\toprule
Method & Task & Humans & Meta-Llama & Qwen \\
\midrule
\multirow{3}{*}{Vanilla}
  & coordinate & 2.18 & 2.50 & 2.14 \\
  & guess      & 1.98 & 2.01 & 2.11 \\
  & pick       & 1.20 & 2.21 & 1.99 \\
\midrule
\multirow{3}{*}{Culture}
  & coordinate & 2.18 & 2.66 & 2.31 \\
  & guess      & 1.98 & 2.47 & 2.58 \\
  & pick       & 1.20 & 2.48 & 2.47 \\
\midrule
\multirow{3}{*}{Saliency}
  & coordinate & 2.18 & 2.29 & 1.69 \\
  & guess      & 1.98 & 1.94 & 2.00 \\
  & pick       & 1.20 & 2.24 & 2.02 \\
\midrule
\multirow{3}{*}{All-Features}
  & coordinate & 2.18 & 2.05 & 1.77 \\
  & guess      & 1.98 & 2.07 & 1.78 \\
  & pick       & 1.20 & 1.87 & 1.74 \\
\bottomrule
\end{tabular}
\end{table*}

In Nottingham, performance is more method-dependent, but \textit{Culture} is the strongest overall condition for both model families across all three tasks. Under \textit{Culture}, Meta-Llama and Qwen exceed human performance on \textit{coordinate} (2.66/2.31 vs.\ 2.18), \textit{guess} (2.47/2.58 vs.\ 1.98), and \textit{pick} (2.48/2.47 vs.\ 1.20). By contrast, \textit{All-Features} is the weakest method overall and includes the notable reversal on \textit{coordinate}, where humans slightly lead (2.18 vs.\ 2.05 and 1.77). \textit{Saliency} produces mixed outcomes, with Meta-Llama generally ahead of Qwen and modest gains over humans, while \textit{Vanilla} is typically intermediate between \textit{Culture} and \textit{All-Features}.

\subsection{Additional Results for Amsterdam and Nottingham Human Experiments}~\label{a:additional-res-AN}

\subsubsection{Cultural Elements in Amsterdam and Nottingham}\label{a:cultural-AN}
Human choices and preferences are often strongly shaped by cultural conventions that determine which options are perceived as salient, appropriate, or mutually expected~\cite{schelling1980strategy,bardsley2010explaining,kraus2000exploiting}. Such culturally grounded priors can lead to highly concentrated behaviour when a norm is widely shared within a population (e.g., preferences regarding food, weather, or leisure activities). In contrast, large language models do not intrinsically participate in culture; rather, they reflect cultural regularities only insofar as these are encoded in their training data and elicited by prompts. This distinction motivates our analysis of TN1 and TN7, in which we examine whether human–LLM gaps in coordination can be attributed to the presence or absence of culturally structured preferences and expectations. In the Amsterdam and Nottingham experiments, we observed that, in many cases, human choices are driven by cultural preferences. In some instances, we were able to substantiate this hypothesis using dedicated culture prompting, which explicitly instructs the LLM to base its choice on cultural aspects. We found that, when the cultural aspect is salient, the LLM can identify it and thereby substantially improve its coordination performance relative to an unstructured prompt. By contrast, when the cultural aspect underlying human behaviour is less transparent and more complex, particularly in cases involving highly specific cultural preferences, the LLM sometimes still struggles to reliably identify them. In such cases, its performance can be improved through alternative strategies, such as saliency prompting, which directs the model to choose the option that is most distinctive relative to the others.

In TN1, the participants have to choose between Friday lunchtime, Monday morning, Saturday night, Sunday night, and Wednesday evening. In the guess and coordinate question types, humans overwhelmingly choose Saturday night. This is a natural choice for humans, since humans are strongly influenced by culture [cite], and in most cultures Saturday night is considered a time for rest and social gatherings. In contrast, across most variations, the LLM splits between two different answers, yet still manages to coordinate relatively well. In the pick question, however, humans split between two main options but also show substantial presence across all other options, which leads them to fail completely at this question type. The LLM, on the other hand, remains consistent with only two options across all variations, and therefore achieves better results in pick. 

To substantiate this hypothesis, we evaluated LLM performance under a culture-oriented prompt and obtained normalized coordination scores of 4.74 for coordinate, 4.78 for guess, and 4.67 for pick, all substantially exceeding the human normalized coordination index. Consistent with our hypothesis, when explicitly instructed to reason in terms of culture, the LLM selects Saturday night in nearly all cases, exhibiting a more decisive and concentrated choice pattern than that observed among human participants. 

In TN7, participants are asked to choose among Colorado, Florida, Louisiana, Nevada, and Ontario. In the guess and coordinate settings, human participants select Florida by a large margin. Florida exhibits a strong degree of cultural salience for human participants. Its cultural prominence surpasses that of all alternative options, owing to its status as a major tourist destination-encompassing beaches, theme parks, and cruise ports—its frequent representation in social media and popular culture, and its widespread recognition even among individuals with limited knowledge of U.S. geography. In contrast, LLMs exhibit difficulty in resolving among the alternatives and consequently tend to select the first option almost systematically, a behaviour that is characteristic of LLMs and consistent with prior findings showing that option ordering can significantly influence model choices [cite]. For the pick question type, human responses are more widely dispersed, likely reflecting individual preferences, which results in a lower normalized coordination index compared to that of the LLMs. Unlike the previous case, prompting the model to reason in terms of culture does not improve LLM performance, suggesting that the LLM struggles to identify the underlying cultural aspect. Instead, the coordination appears to be governed by a distinctive attribute of one option that renders it uniquely salient. 

To validate this hypothesis, we evaluated LLM responses under a saliency-oriented prompt, instructing the model to select the option that is most different from the others. Under this prompt, the LLM almost exclusively selects Ontario, which is a Canadian province, whereas the remaining options are all located in the United States. Consequently, the LLM attains a normalized coordination index of 2.86 in coordinate (nearly identical to the human score of 2.88), 4.75 in guess, and 4.96 in pick. The latter two scores are substantially higher than the corresponding human scores (2.77 and 1.20, respectively), indicating a more pronounced sensitivity to structural differences among the options and resulting in more decisive and consistent coordination behaviour than that observed in humans.

\subsubsection{Salient Elements in Amsterdam and Nottingham}\label{a:cultural-AN-salient}
The selection of salient points may differ substantially between humans and large language models (LLMs). This divergence arises from the fact that human choices are shaped by dynamic factors such as environmental context, cultural background, individual preferences, and personal experience. In contrast, LLMs learn exclusively from human-generated data and therefore typically acquire such preferences only indirectly, through latent patterns and subcontextual signals present in their training data. In some relatively unambiguous cases, LLMs select options that are also salient to humans; however, in other settings, pronounced differences emerge in the underlying logic governing salient point selection. Our findings indicate that the ordering of available options plays a particularly significant role for LLMs, which often report selecting the first option under the assumption that this is likely the choice made by another LLM-a heuristic that is largely uninformative and rarely employed by humans. In this section, we further examine these discrepancies in salient point selection logic between humans and LLMs, drawing on experimental evidence from studies conducted in Amsterdam and Nottingham.

In TN9, participants are asked to choose among the years 1978, 1979, 1980, 1981, and 2000. Because the options correspond to calendar years, two salient points are particularly prominent for human participants. First, the year 2000 stands out as the beginning of a new millennium, in contrast to the other options, which all fall within the twentieth century. A second salient point is a year associated with significant events in Nottingham, either in the city itself or within one of its sports teams. With respect to the first salient point, LLMs are expected to identify it autonomously and select it accordingly. By contrast, the second salient point is more complex and is strongly shaped by factors such as the background and cultural context of the human participants. Across both the pick and guess conditions, human behaviour largely aligns with these expectations: participants predominantly selected the year 2000, followed by 1981, which is plausibly linked to salient events in Nottingham, and then 1980, which may also reflect the influence of notable local events. This interpretation is further supported by the observation that, in the pick condition, most participants favored 1981 and exhibited an increased tendency to choose 1980, suggesting a personal or culturally grounded preference for these years. In contrast, many LLMs rely on purely mathematical considerations in their decision-making. For instance, they frequently select 1979 because it is the only prime number in the set, reflecting a mode of reasoning that differs fundamentally from that of humans, most of whom are unlikely to be aware of this property. When prompted to make their choice based on cultural considerations, however, LLMs explicitly assign greater importance to years associated with significant events in Nottingham, indicating an understanding that, when coordinating with a human from Nottingham, such culturally salient choices are more likely to align with human preferences.

In TN11, participants are asked to choose among the options: win champagne, win chocolate, win money, win nothing, and win a trophy. In both the guess and coordinate question types, human participants overwhelmingly select win money, whereas LLMs most frequently select win chocolate in the coordinate question type, and exhibit a more evenly distributed set of responses in the guess question type. In the pick question type, humans tend to incorporate personal preferences and therefore select a broader range of options beyond win money. By contrast, LLMs most often choose either win chocolate or win money, which results in better performance than humans. Overall, humans demonstrate very strong coordination on this question, achieving a normalized coordination index of 4.33 in coordinate, 3.58 in guess and 1.46 in pick. For LLMs, the corresponding values are 2.71, 1.89, and 2.8, respectively, indicating that their coordination performance is in fact higher when they are not explicitly instructed to coordinate. As in the previous question (TN7), the preference for money in this setting does not appear to stem from cultural factors, but rather from a distinctive attribute of a single option that renders it uniquely salient. To investigate this hypothesis, we re-evaluated the LLMs using a saliency-oriented prompt that explicitly instructs the model to select the option that is most distinctive relative to the others. Under this prompt, the model achieved scores of 3.25, 3.25, and 4.13 in the coordinate, guess, and pick question types, respectively - representing a substantial improvement over the baseline and approaching the performance observed in human participants. Notably, the model again performs best in the pick question type, reinforcing the observation that LLMs coordinate more effectively on this question when they are not explicitly asked to do so. The most salient focal point selected by the models, in contrast to humans, is win nothing: unlike the other options, which all involve receiving some reward, this option uniquely entails receiving nothing at all, highlighting a form of salience-driven reasoning that differs qualitatively from human decision-making.

\begin{table*}[h]
\centering
\begin{tabular}{llcccc}
\hline
 & & Vanilla & Saliency & All-features & Culture \\
\hline
\multirow{3}{*}{Amsterdam} 
 & Pick & 0.09 & 0.52 & 0.16 & 0.24 \\
 & Guess & -0.17 & 0.34 & 0.16 & 0.31 \\
 & Coordinate & -0.54 & 0.01 & -0.06 & -0.09 \\
\hline
\multirow{3}{*}{Nottingham} 
 & Pick & -0.19 & 0.08 & -0.02 & 0.15 \\
 & Guess & -0.68 & -0.40 & -0.65 & -0.45 \\
 & Coordinate & -0.87 & -1.04 & -0.66 & -0.49 \\
\hline
\end{tabular}
\caption{Improvement of the NCI when humans and GPT-oss-120B are combined. We sampled $50$ results from humans, $50$ from GPT-oss, and $50$ from their joint distribution, and measured the difference between the NCI of humans and that of humans with GPT.}
\end{table*}

\subsubsection{Full Results for Reasoning}\label{a:reasoning-AN}
\begin{figure*}
    \centering
    \includegraphics[width=1.1\linewidth]{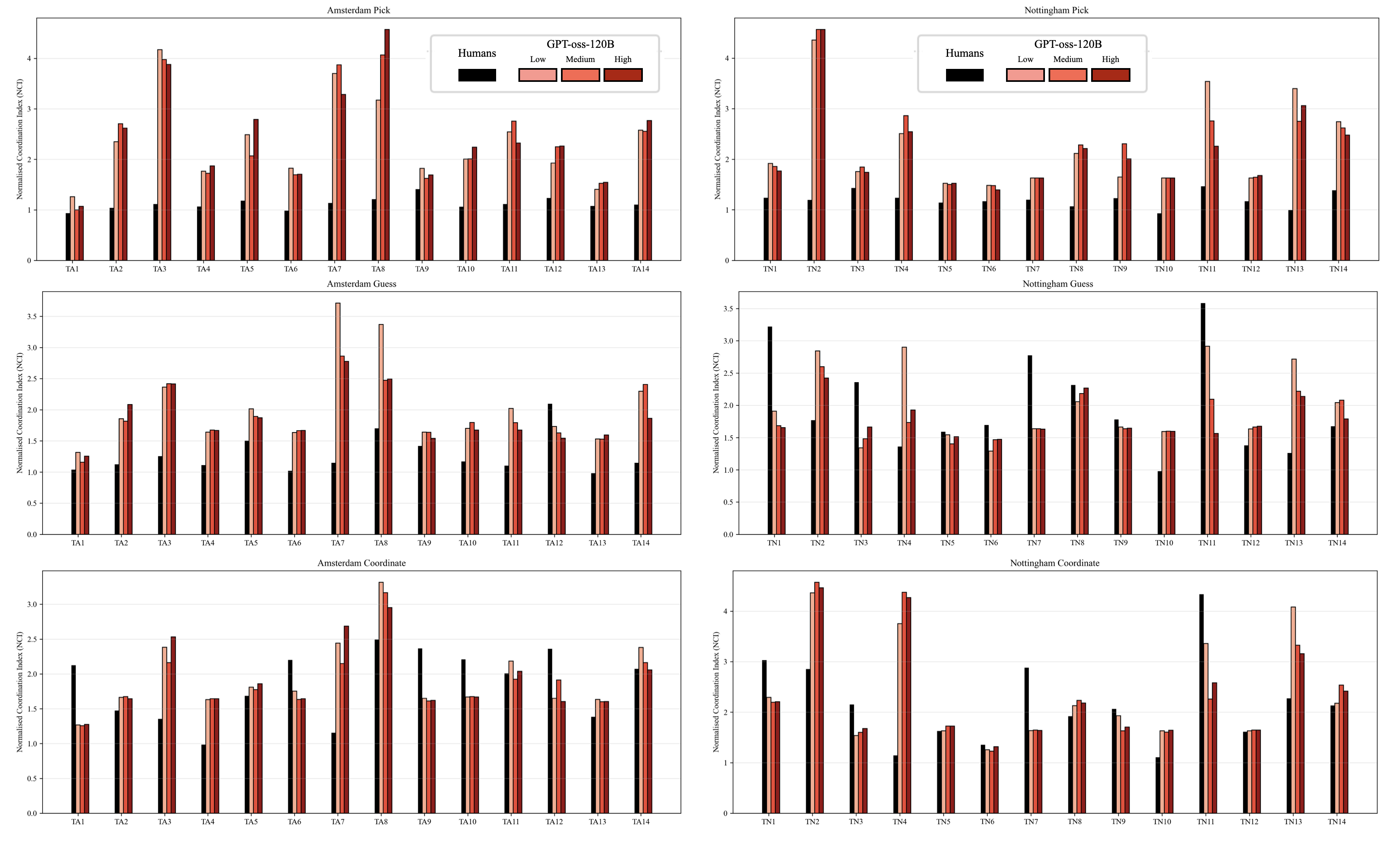}
    \caption{The effect of reasoning (low, medium, and high: the darker the red, the higher the reasoning) on the coordination of GPT-oss-120B in Amsterdam and Nottingham. There is no clear evidence that reasoning improves the NCI of LLMs.}
    \label{fig:appendix-gpt-120b-reasoning-AN}
\end{figure*}

Figure~\ref{fig:appendix-gpt-120b-reasoning-AN} shows that more reasoning does not help GPT-oss to perform better in Amsterdam and Nottingham.

\paragraph{When LLMs overthink.}

A number of interesting answers in Amsterdam and Nottingham reveal that LLMs tend to overthink, when they are asked to do so.
An interesting example is when we use the ``culture'' prompting technique in Nottingham. Consider the question TN9, which asks the model to choose between the following options: \{1987, 1988, 1989, 2000\}. While it seems natural for humans to choose 2000 (centrality and uniqueness), GPT-oss-120B with high reasoning often chooses 1978 as it is the year when the Nottingham Forest Club (a UK football club), won the league. Similarly, GPT-oss-120B with high reasoning chooses the color red (the choices are \{red, blue, orange, yellow, purple\} in TN12 as it is the official color of the football club.
Several other examples occur when models like GPT-oss are prompted with high reasoning, and makes us conclude that in coordination, LLMs can degrade their performance when they think too much.

\subsubsection{Using the Same Family of Models}\label{a:same-models-AN}
\begin{figure*}
    \centering
    \includegraphics[width=1.1\linewidth]{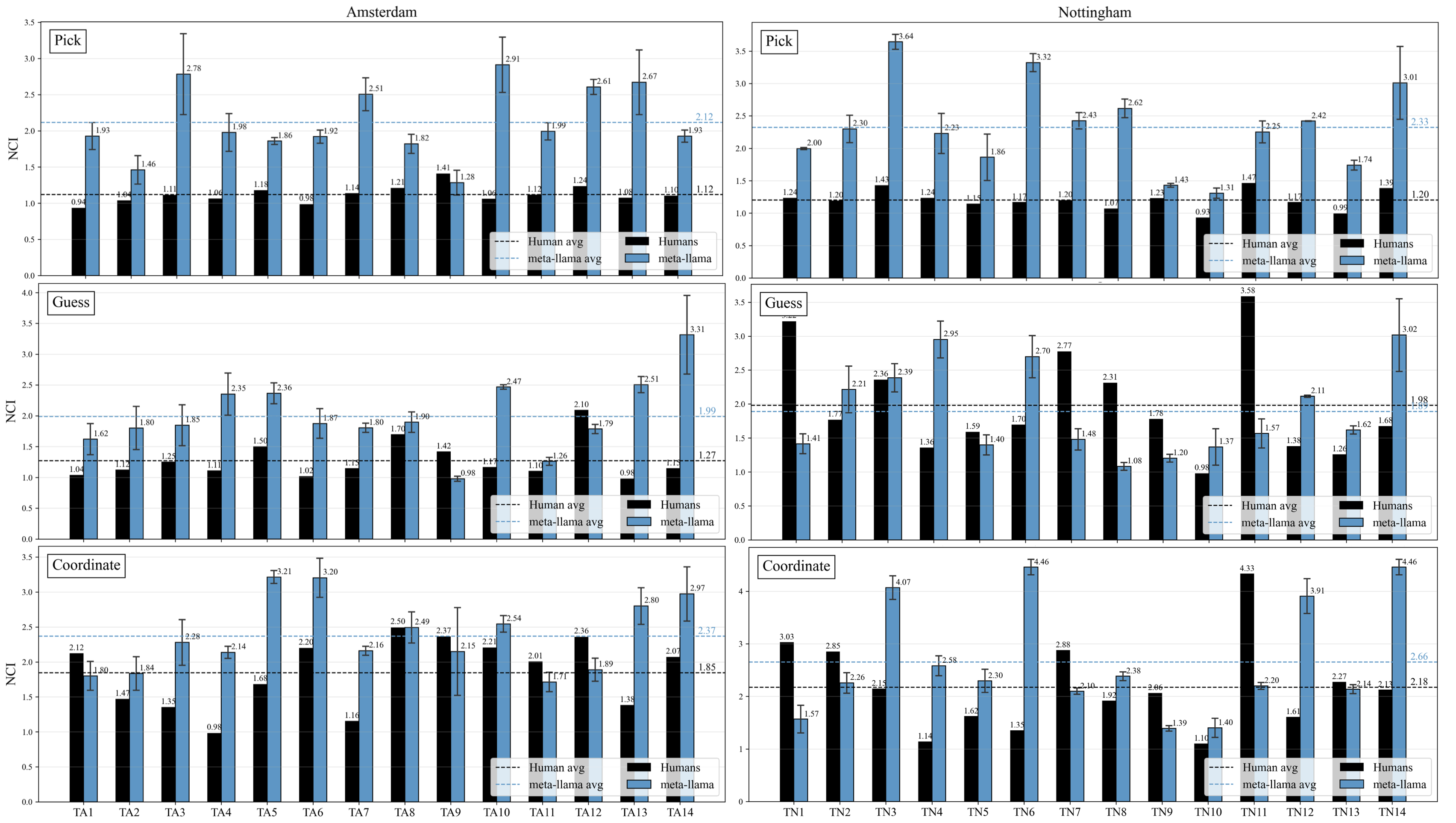}
    \caption{Normalised Coordination Index (NCI) of humans and LLMs (Llama-3, 3.1, and 3.3 70B) on the Amsterdam and Nottingham datasets.}
    \label{fig:appendix-llamas-AN}
\end{figure*}

\begin{figure*}
    \centering
    \includegraphics[width=1.1\linewidth]{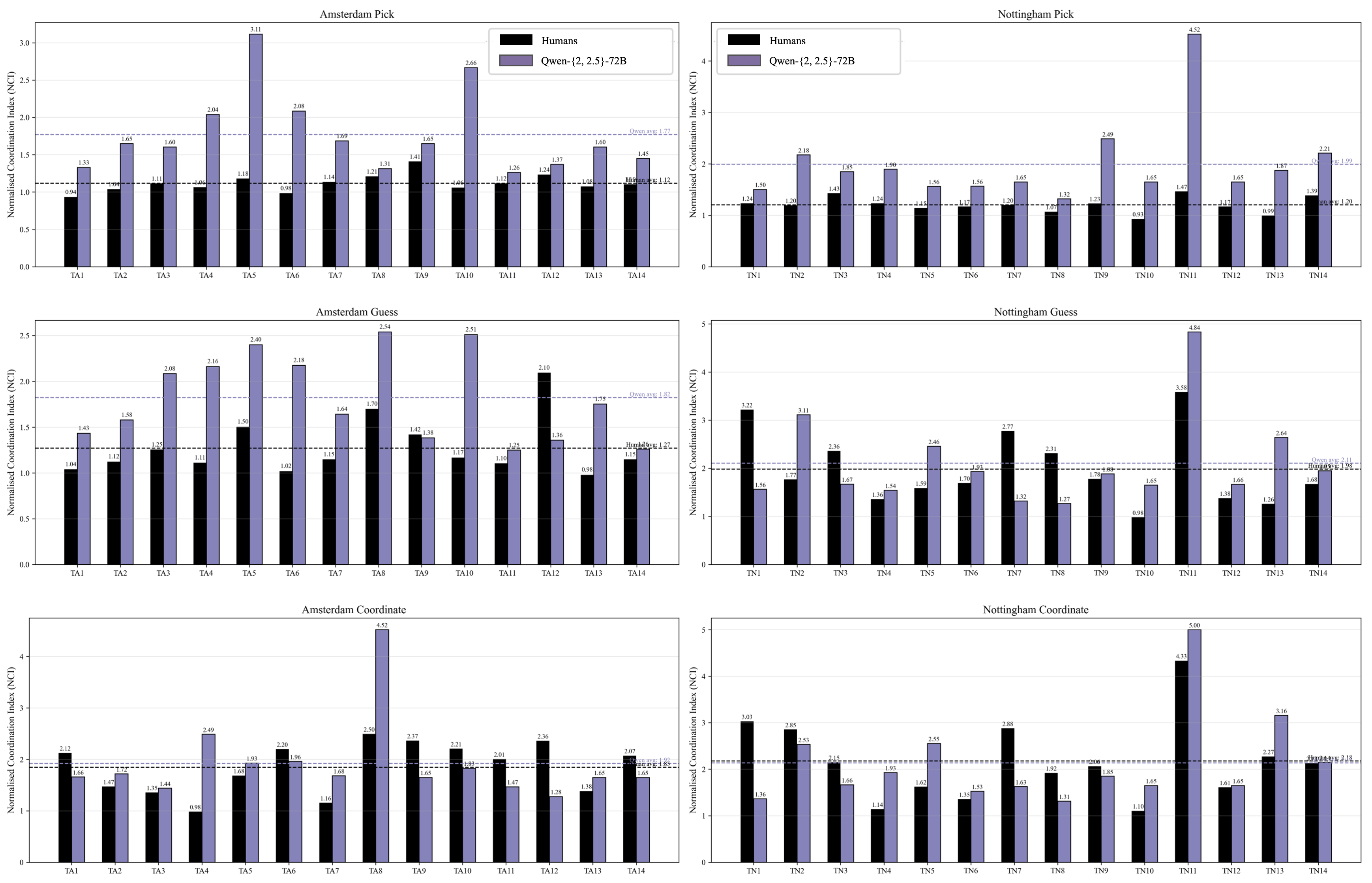}
    \caption{Normalised Coordination Index (NCI) of humans and LLMs (Qwen 2 and 2.5 72B) on the Amsterdam and Nottingham datasets.}
    \label{fig:appendix-qwen-AN}
\end{figure*}

\begin{figure*}
    \centering
    \includegraphics[width=1.1\linewidth]{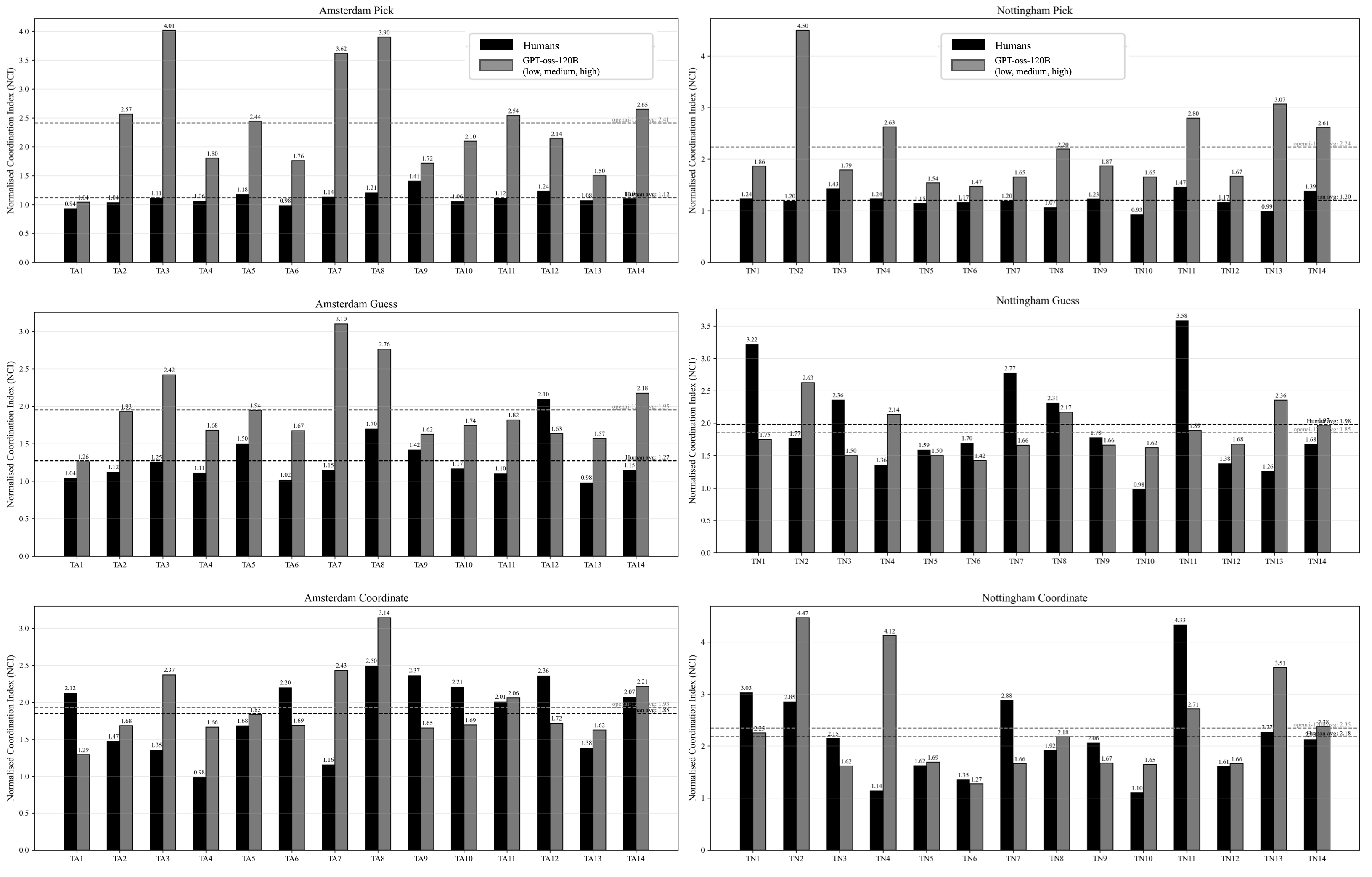}
    \caption{Normalised Coordination Index (NCI) of humans and LLMs (GPT-oss 20B and 120B) on the Amsterdam and Nottingham datasets.}
    \label{fig:appendix-gpt-AN}
\end{figure*}

Figures~\ref{fig:appendix-llamas-AN},~\ref{fig:appendix-qwen-AN}, and~\ref{fig:appendix-gpt-AN} report the results of Amsterdam and Nottingham for the same family of models, namely Llama-3, 3.1, and 3.3 70B, Qwen-2, and 2.5-72B, and GPT-oss-20B and 120B. Results evidence how using the same family of models improves the coordination in each task, despite Nottingham remaining a task where humans, for ``guess'' and ``coordinate'', still outperform them.

\subsubsection{Results with Other Models}\label{a:other-models-AN}
\begin{figure*}
    \centering
    \includegraphics[width=1.1\linewidth]{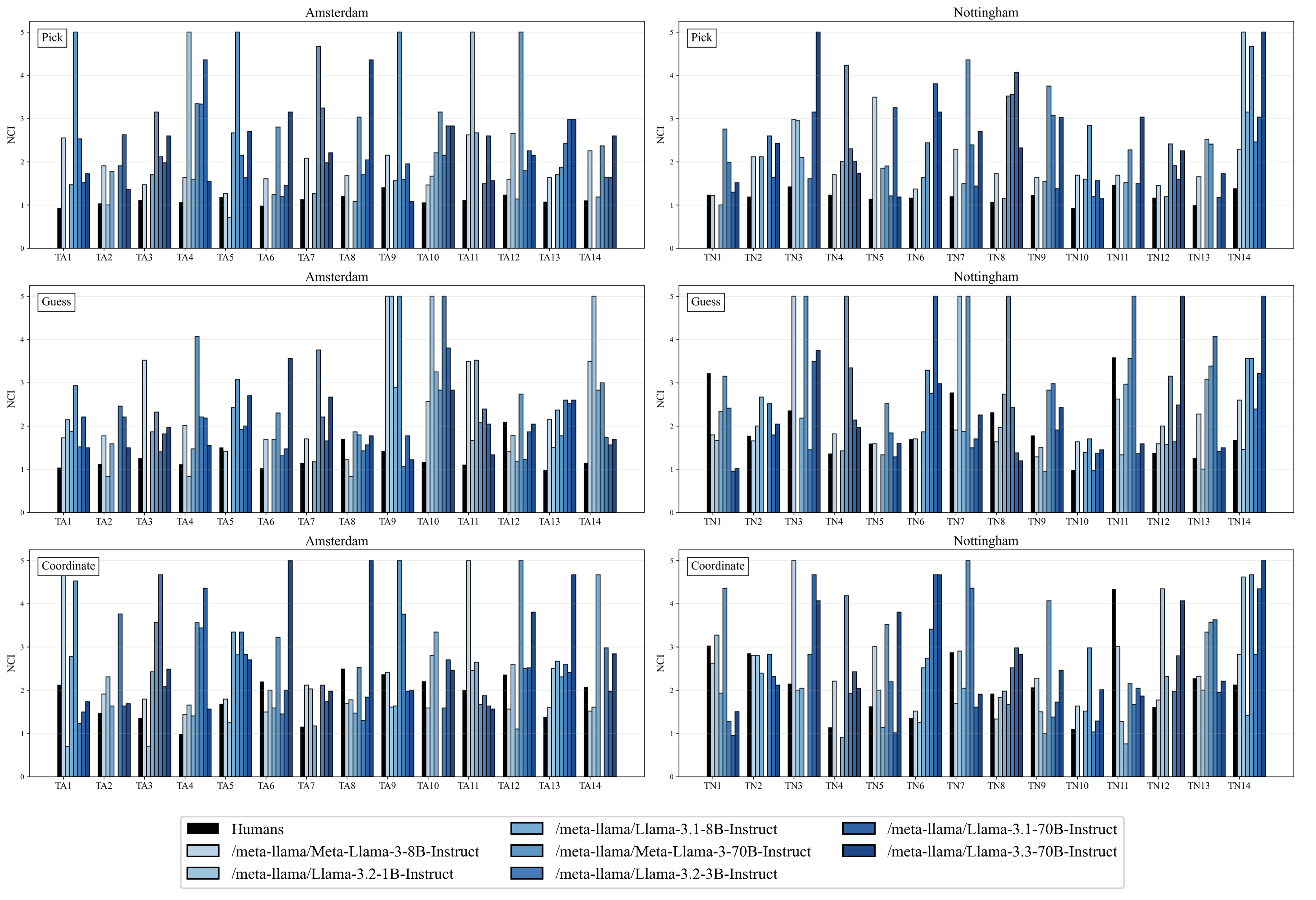}
    \caption{Normalised Coordination Index (NCI) of humans and all the Llama models on the Amsterdam and Nottingham datasets.}
    \label{fig:appendix-llama-all-AN}
\end{figure*}

\begin{figure*}
    \centering
    \includegraphics[width=1.1\linewidth]{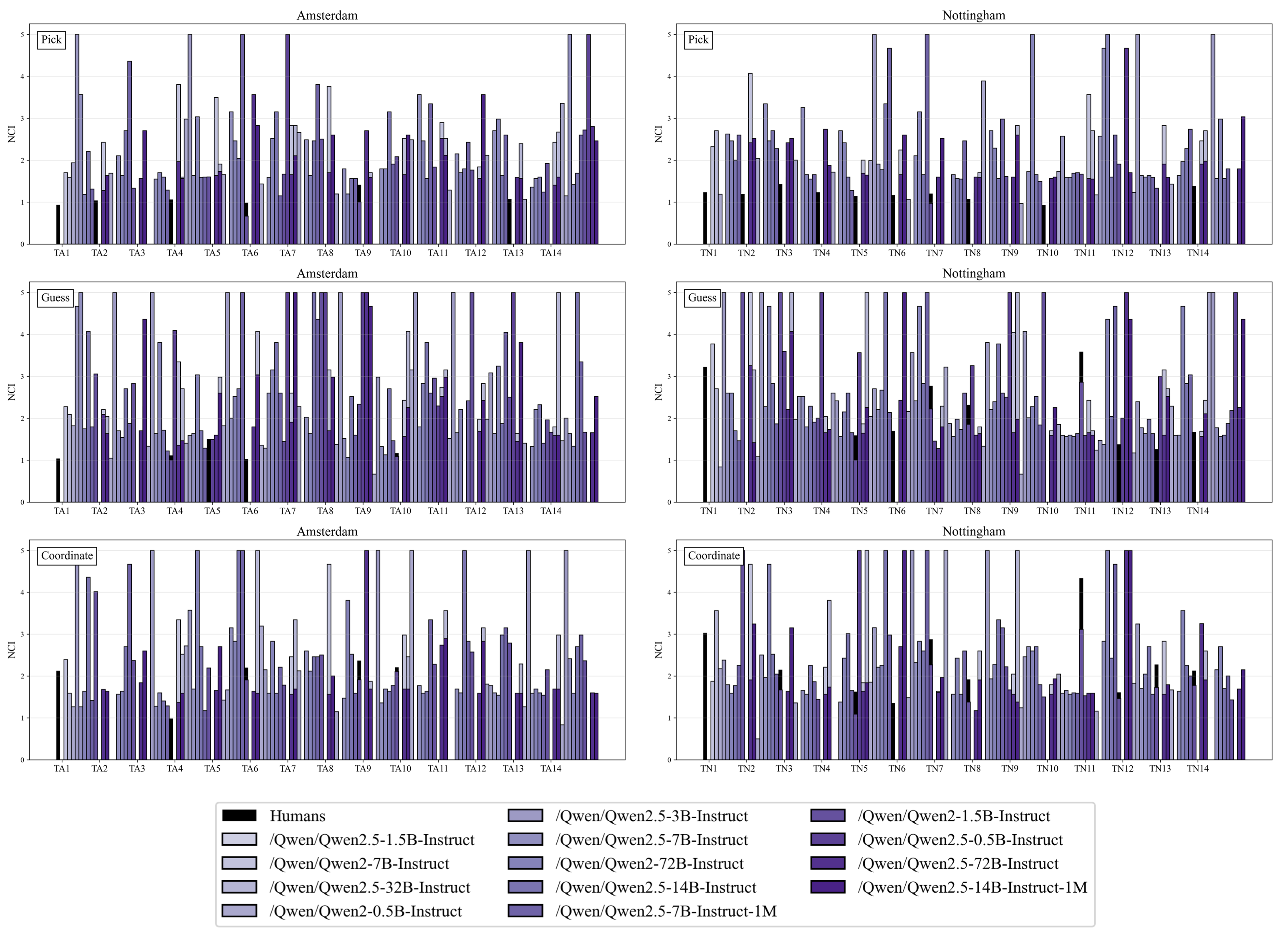}
    \caption{Normalised Coordination Index (NCI) of humans and all the Qwen models on the Amsterdam and Nottingham datasets.}
    \label{fig:appendix-qwen-all-AN}
\end{figure*}

\begin{figure*}
    \centering
    \includegraphics[width=1.1\linewidth]{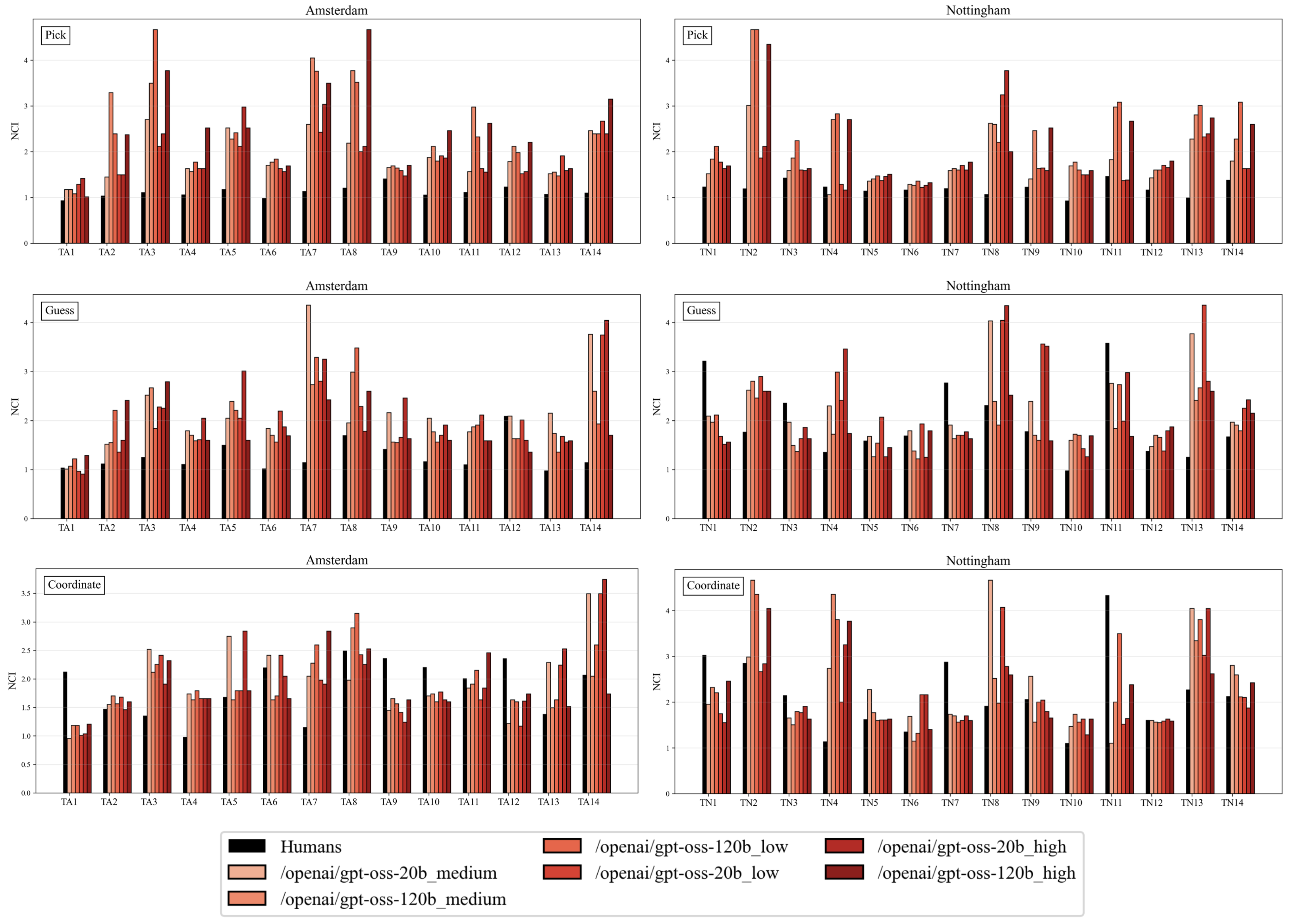}
    \caption{Normalised Coordination Index (NCI) of humans and all the GPT models on the Amsterdam and Nottingham datasets.}
    \label{fig:appendix-gpt-all-AN}
\end{figure*}

Figures~\ref{fig:appendix-llama-all-AN},~\ref{fig:appendix-qwen-all-AN}, and~\ref{fig:appendix-gpt-all-AN} report the results with all the Llama, Qwen, and GPT-oss models for Amsterdam and Nottingham. The prompting technique is \emph{vanilla}. The results for the other techniques, i.e., saliency, all-features, and culture, are reported, for reasons of space, in the code extension (folder results).

\subsection{Additional Results for the Bargaining Table}~\label{a:additional-res-bargaining}
\begin{figure*}
    \centering
    \includegraphics[width=0.9\linewidth]{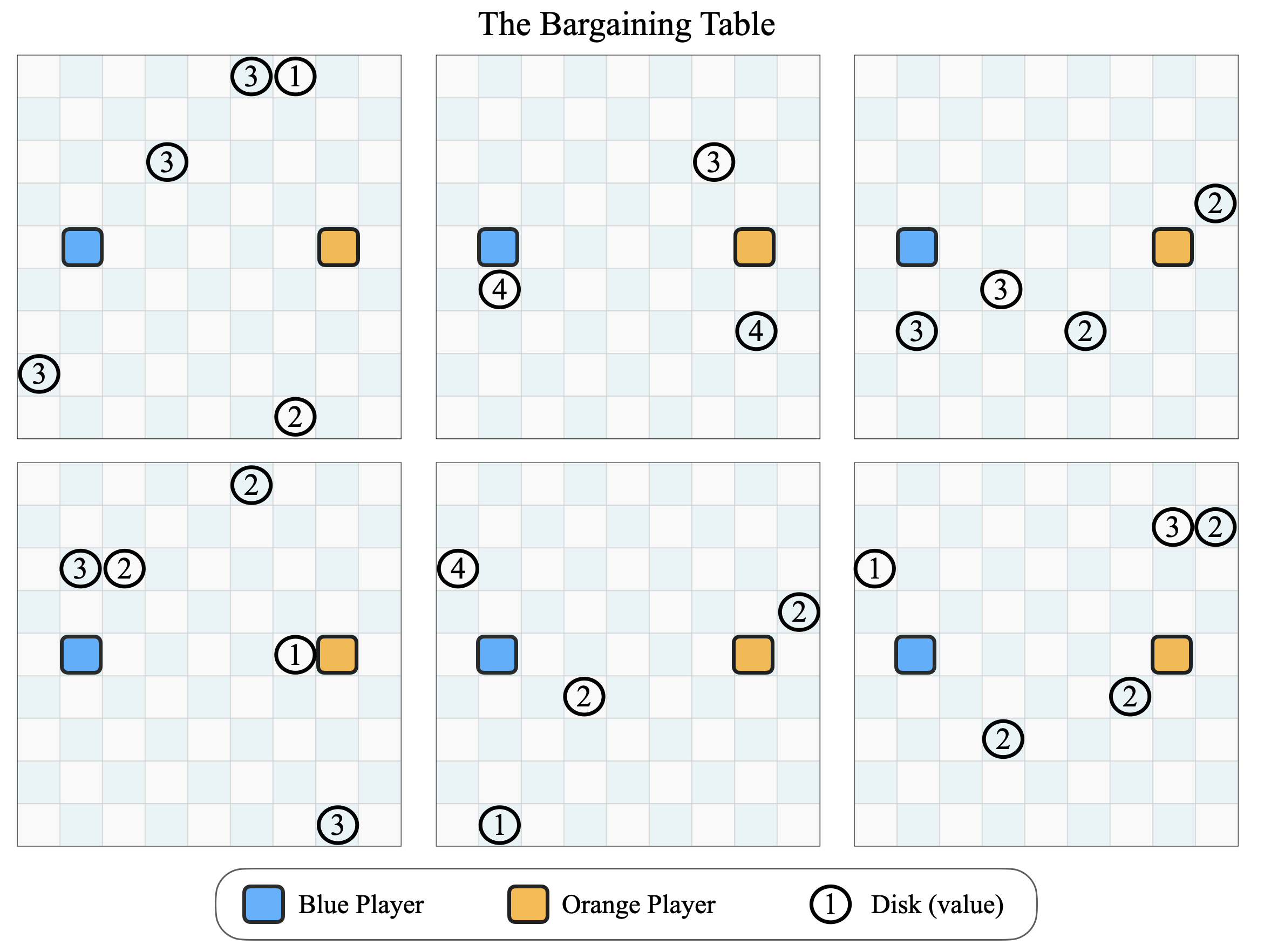}
    \caption{Examples of Bargaining Table games as per~\cite{mizrahi2020using}.}
    \label{fig:appendix-bargaining-examples}
\end{figure*}

We hereby report results for the Bargaining Table game. In Figure~\ref{fig:appendix-bargaining-examples}, we illustrate some games as per~\citet{mizrahi2020using}.
\subsubsection{Results for the Orange Player}\label{a:yellow-bargaining}
\begin{figure*}
    \centering
    \includegraphics[width=1\linewidth]{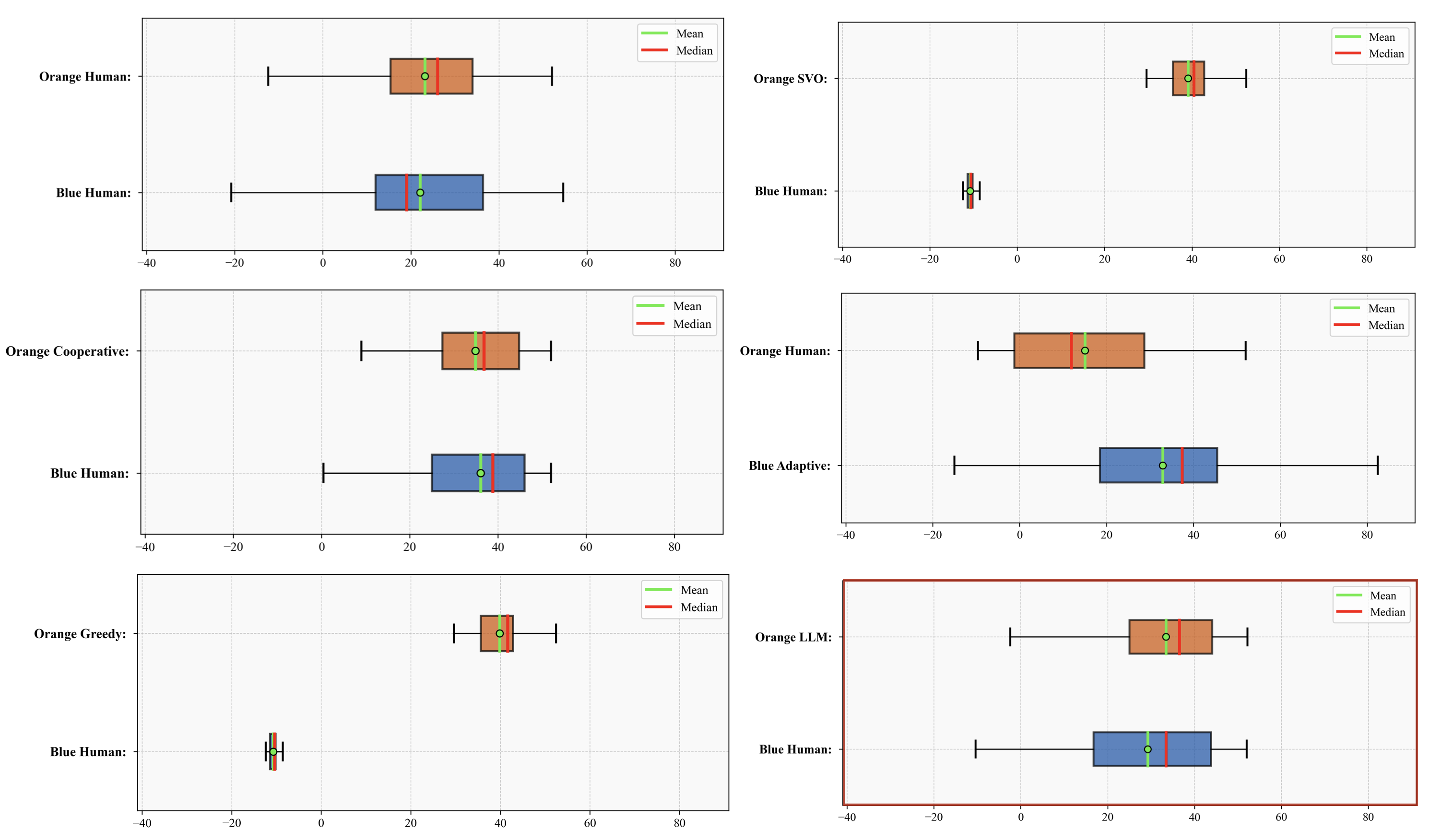}
    \caption{Each barplot reports the mean and median payoff of the \emph{blue} and \emph{orange} players in $100$ iterations, per typology of game, of the Bargaining Table. While the orange agent changes her strategy, the data for the blue player is that of humans who played the game and comes from~\cite{mizrahi2020using}.
    Bottom-right: when the \emph{orange player} is an LLM (GPT-oss-120B), the payoff of both players is comparable to that of a cooperative player.}
    \label{fig:appendix-bargaining-yellow}
\end{figure*}

Figure~\ref{fig:appendix-bargaining-yellow} reports the results for the Bargaining Table when the player that is optimised is the orange. Results gathered from human data and specialised agents in~\citet{mizrahi2020using} are compared to GPT-oss-120B. Similarly to the results in the main paper, GPT-oss behaves like a cooperative player, with high values of individual utility and welfare payoff.

\subsubsection{The Effect of Reasoning and Scale}\label{a:reasoning-scale-bargaining}
\begin{figure*}[t]
\centering
\setlength{\tabcolsep}{1pt}
\renewcommand{\arraystretch}{0}
\begin{tabular}{@{}cc@{}}
\includegraphics[width=0.49\textwidth]{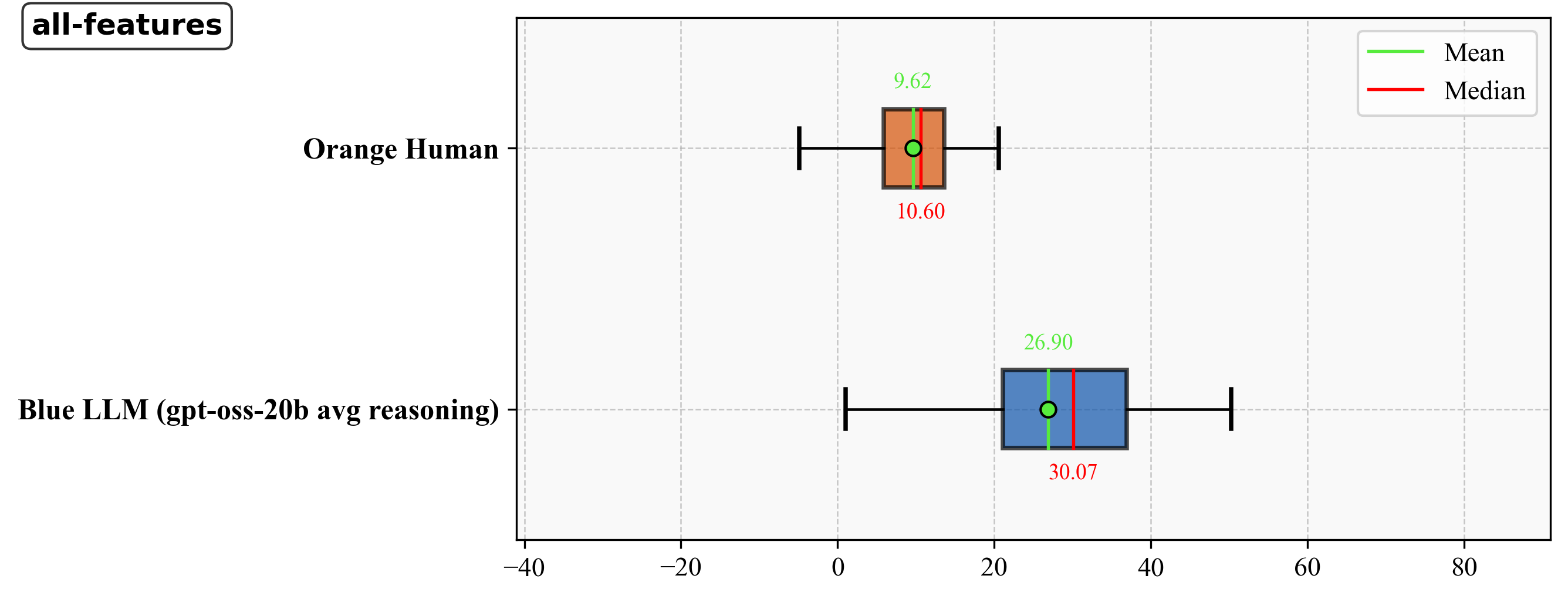} &
\includegraphics[width=0.49\textwidth]{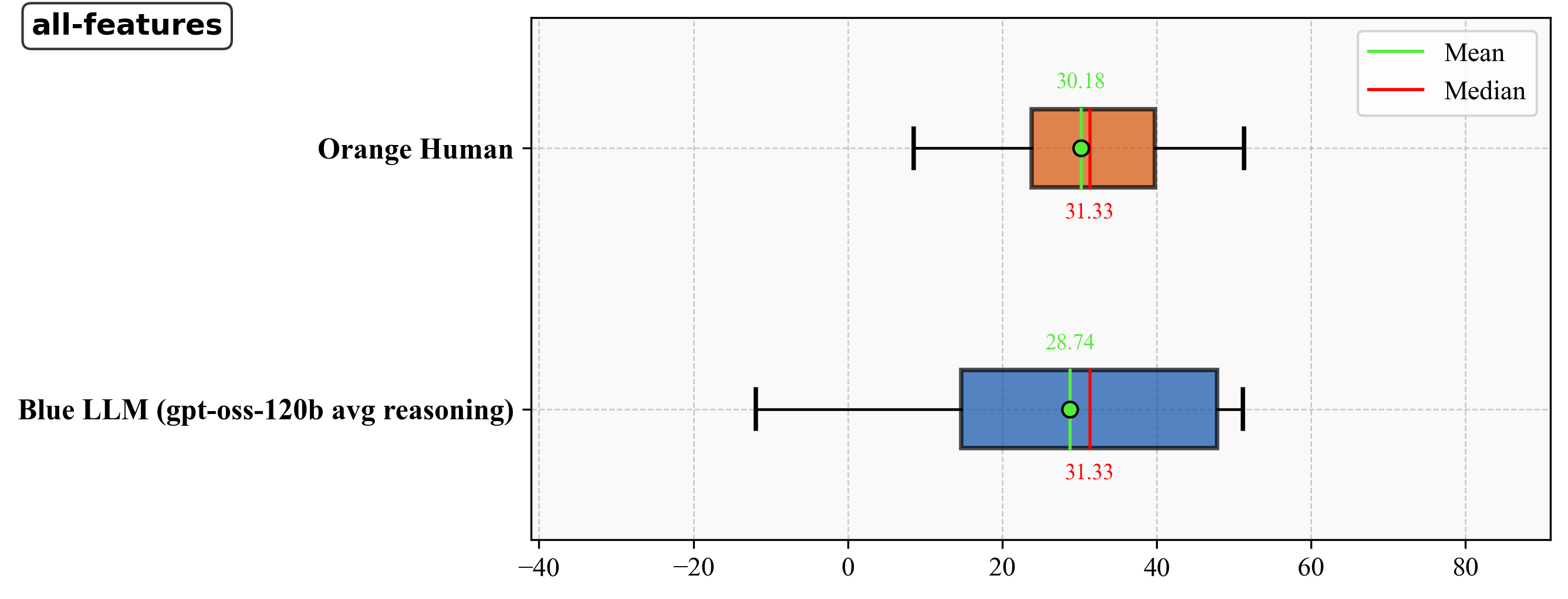} \\
\includegraphics[width=0.49\textwidth]{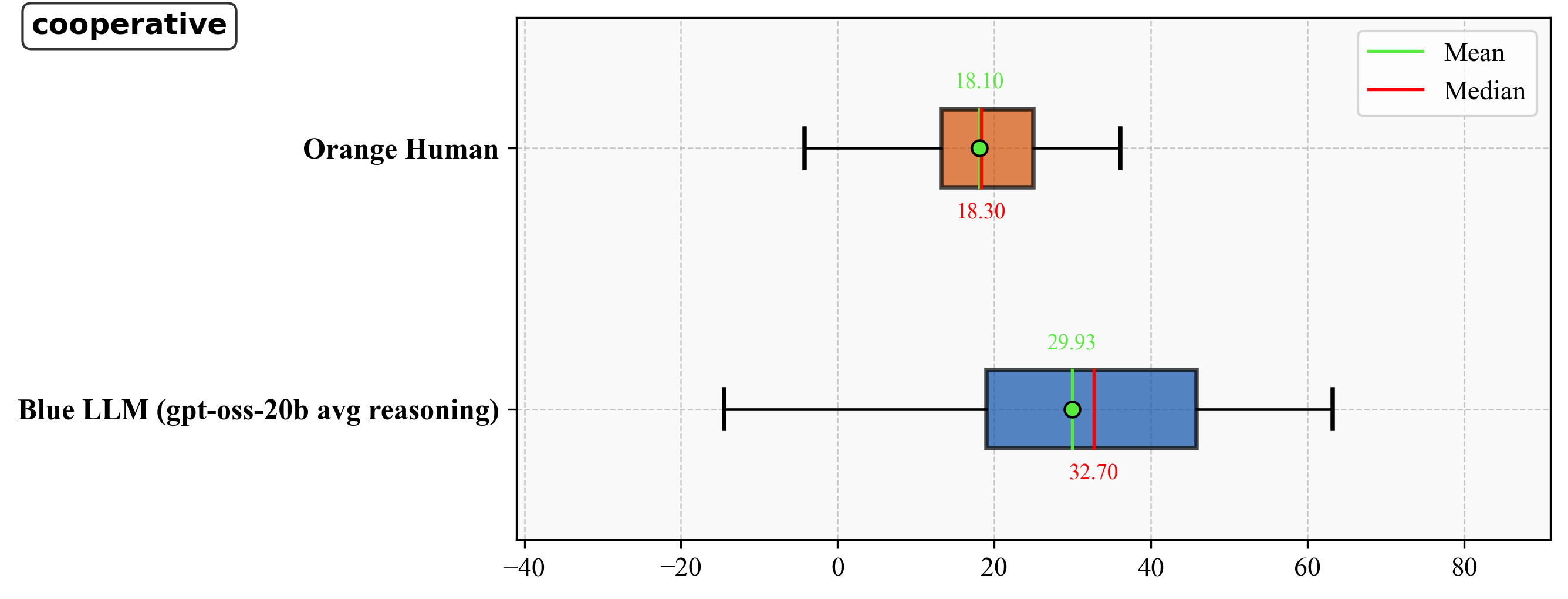} &
\includegraphics[width=0.49\textwidth]{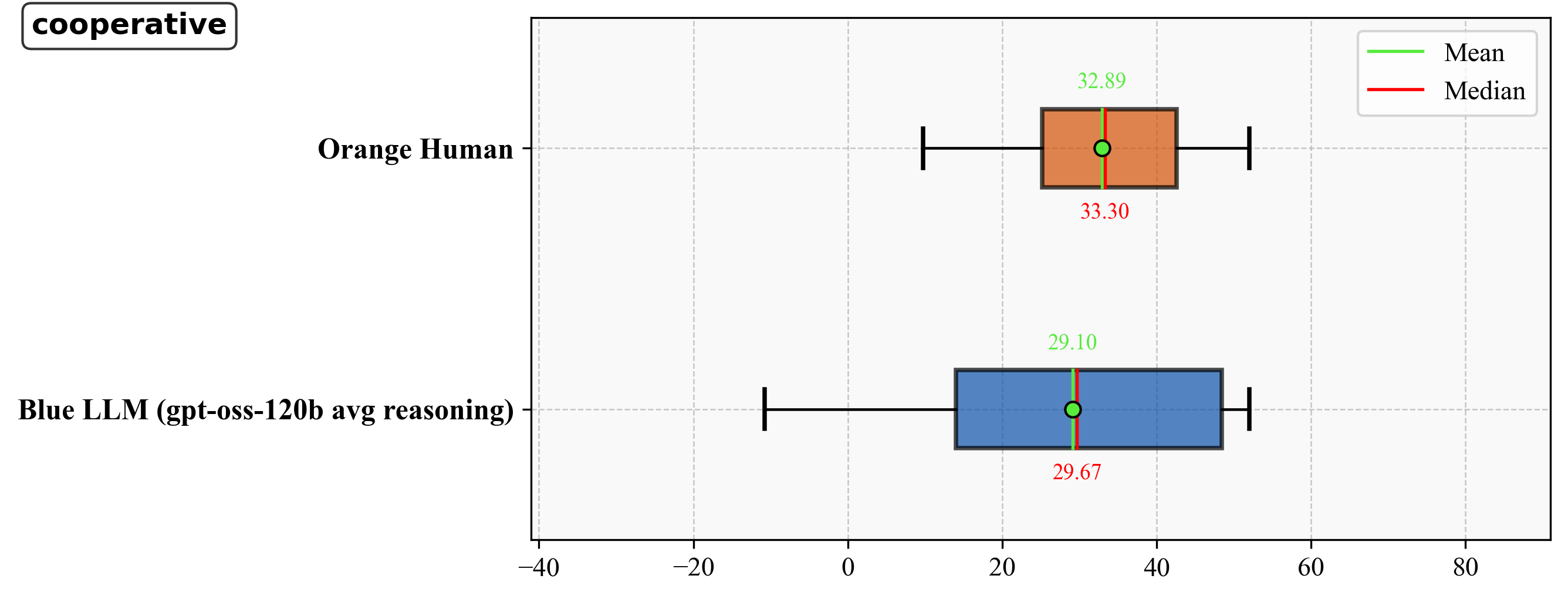} \\
\includegraphics[width=0.49\textwidth]{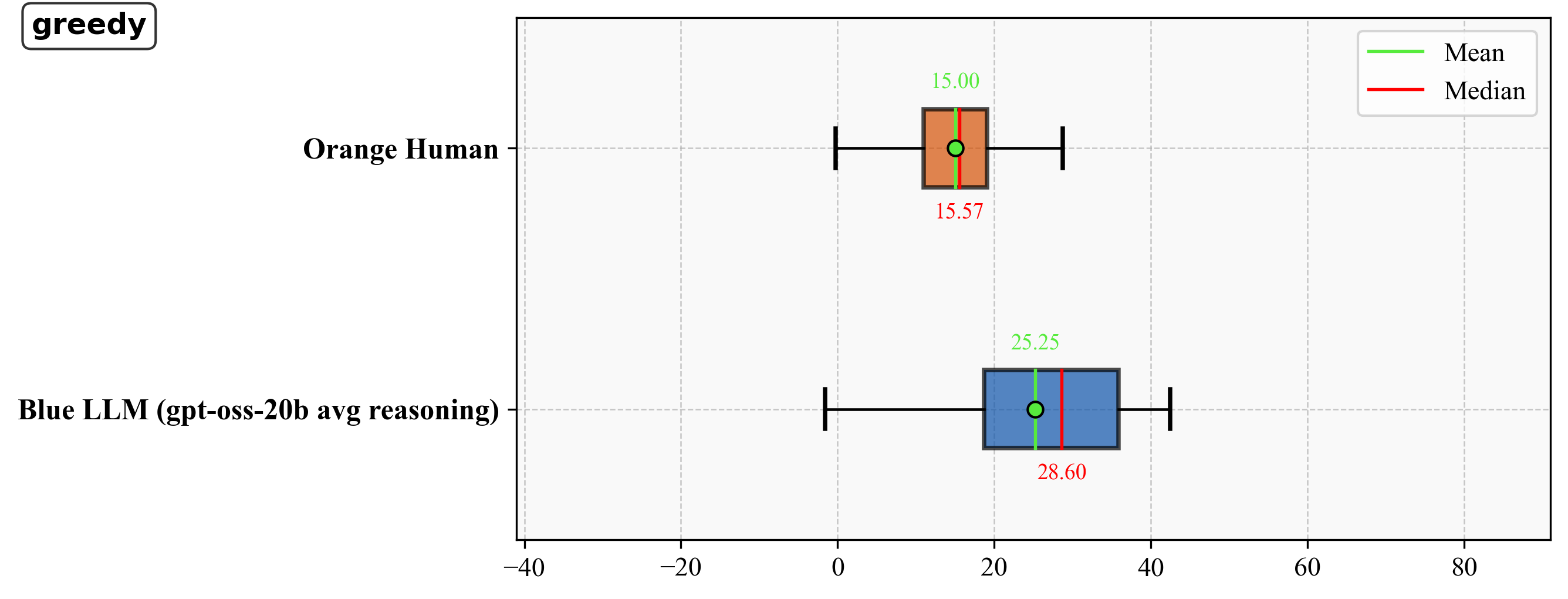} &
\includegraphics[width=0.49\textwidth]{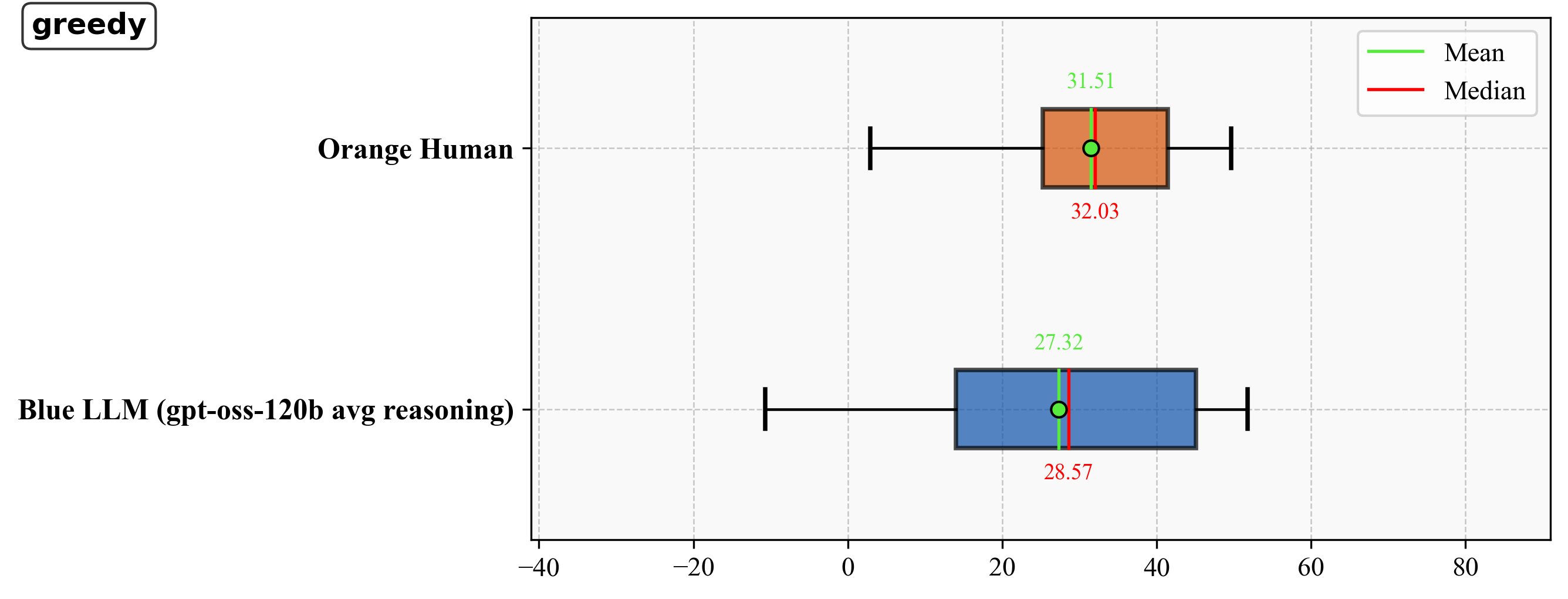} \\
\includegraphics[width=0.49\textwidth]{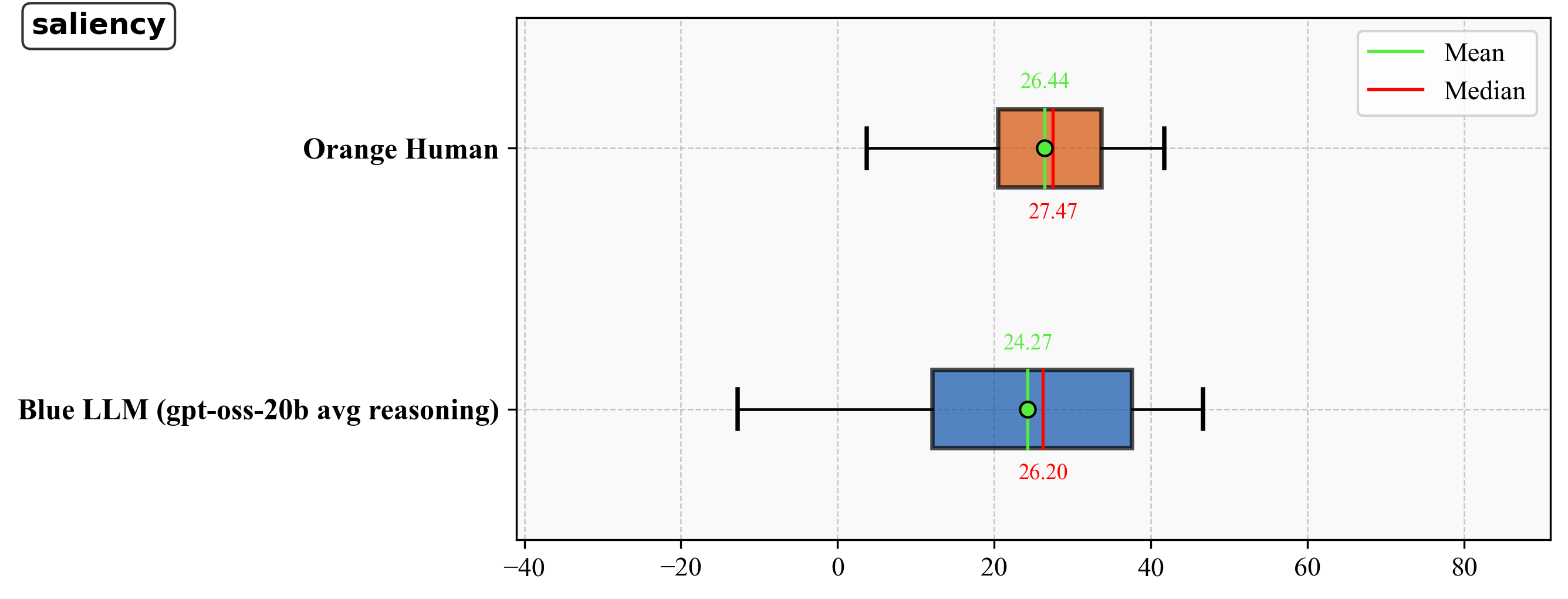} &
\includegraphics[width=0.49\textwidth]{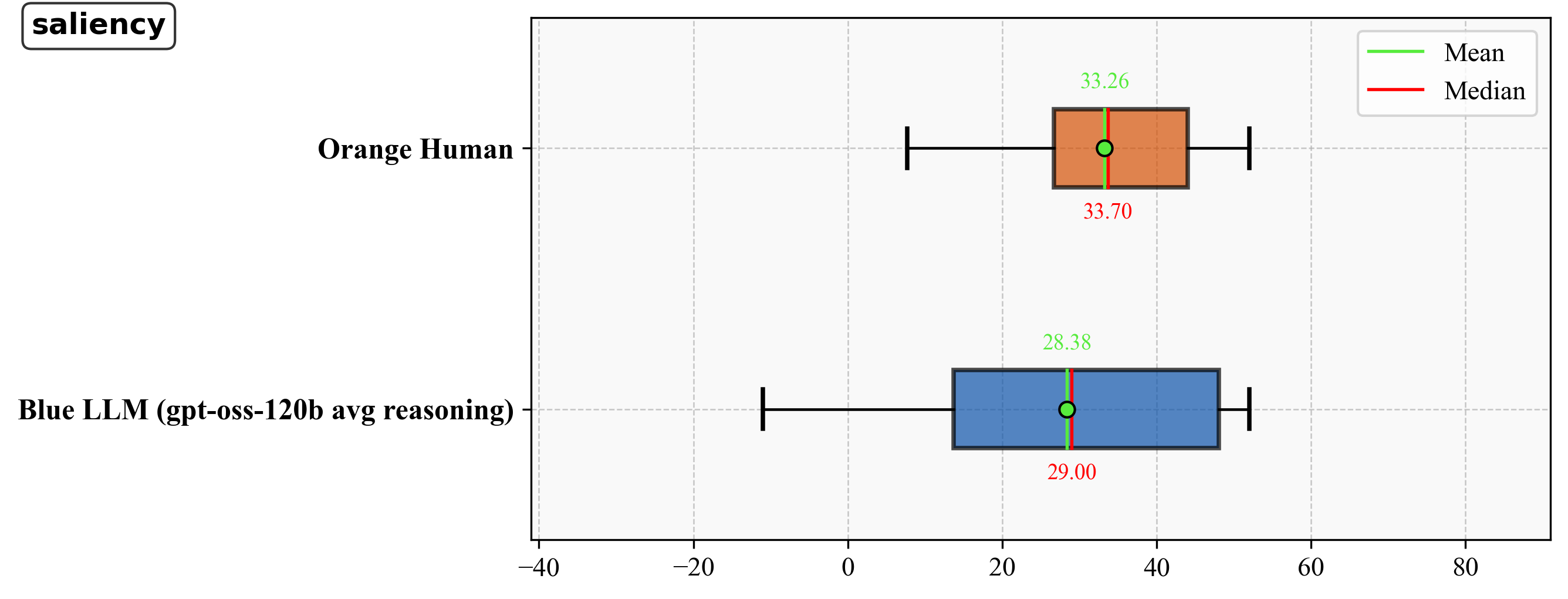} \\
\includegraphics[width=0.49\textwidth]{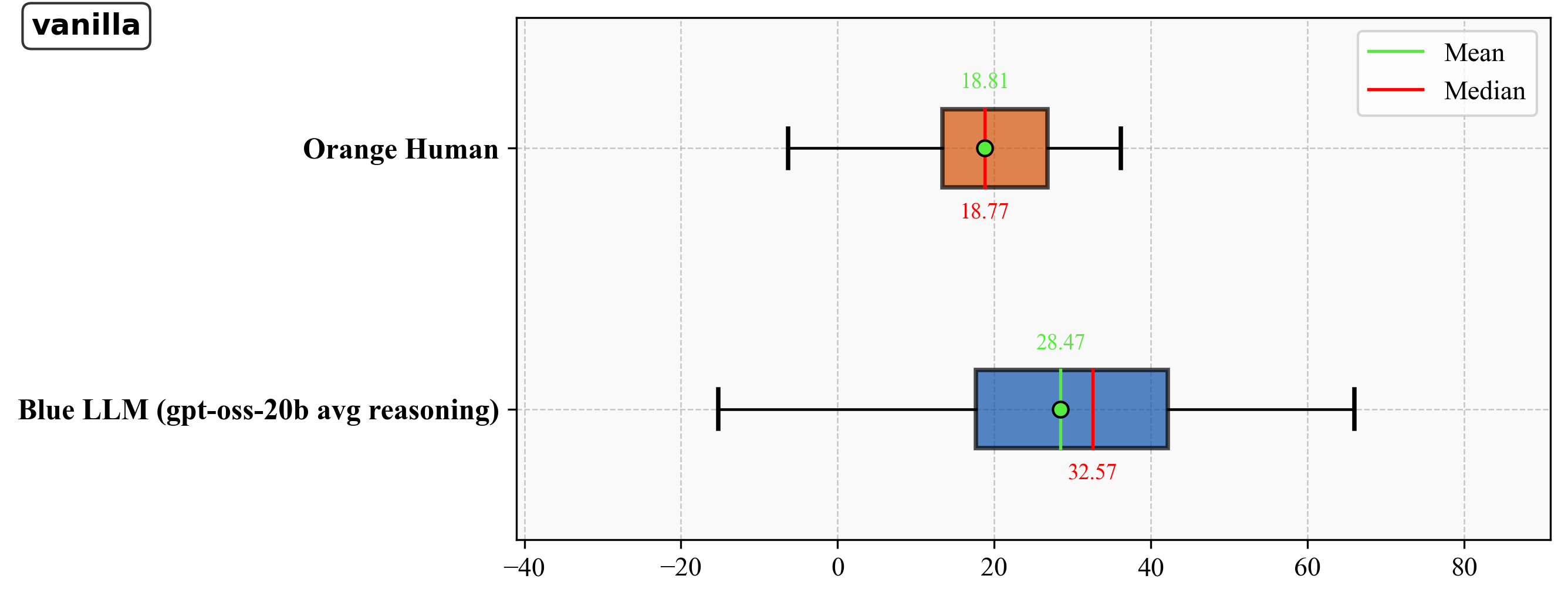} &
\includegraphics[width=0.49\textwidth]{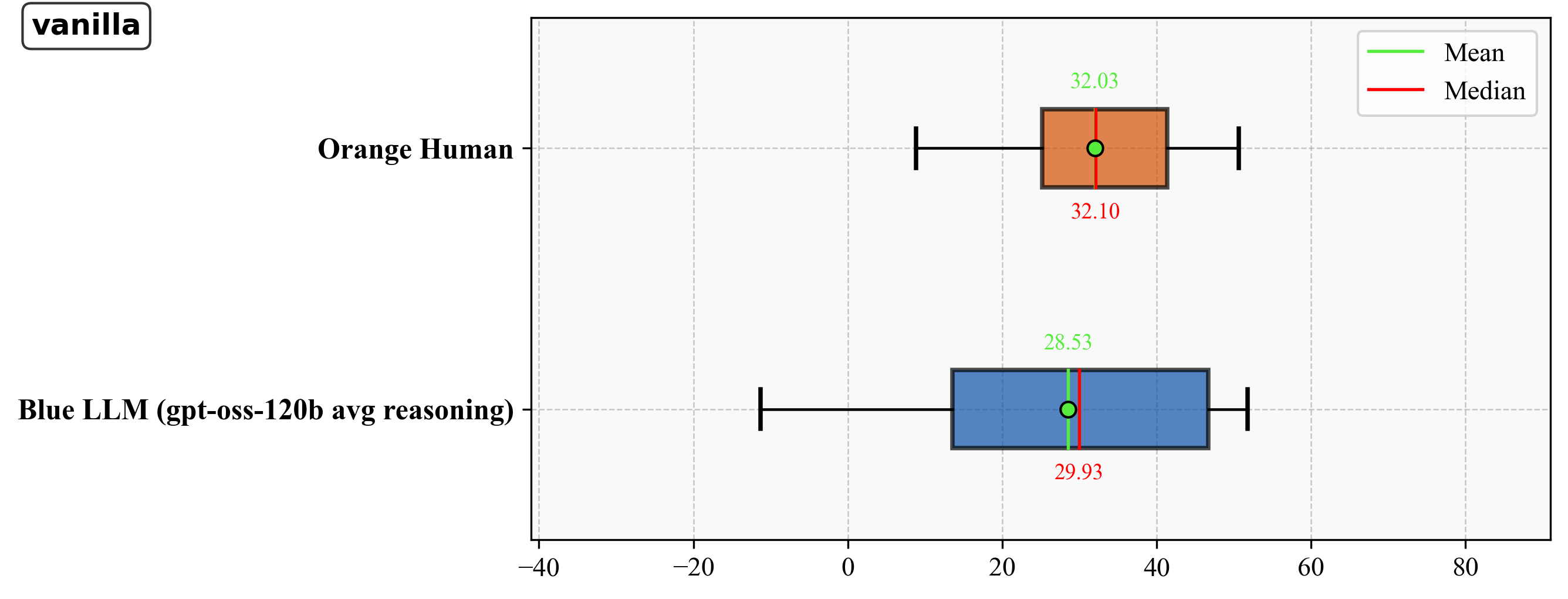} \\
\end{tabular}
\vspace{-2mm}
\caption{Bargaining Table averages by model size: gpt-oss-20b (left) vs gpt-oss-120b (right) for all-features, cooperative, greedy, saliency, and vanilla. The LLM is the blue player.}
\label{fig:bt-avg-across-reasoning-p1-llm-all10}
\end{figure*}

\begin{figure*}[t]
\centering
\setlength{\tabcolsep}{1pt}
\renewcommand{\arraystretch}{0}
\begin{tabular}{@{}cc@{}}
\includegraphics[width=0.49\textwidth]{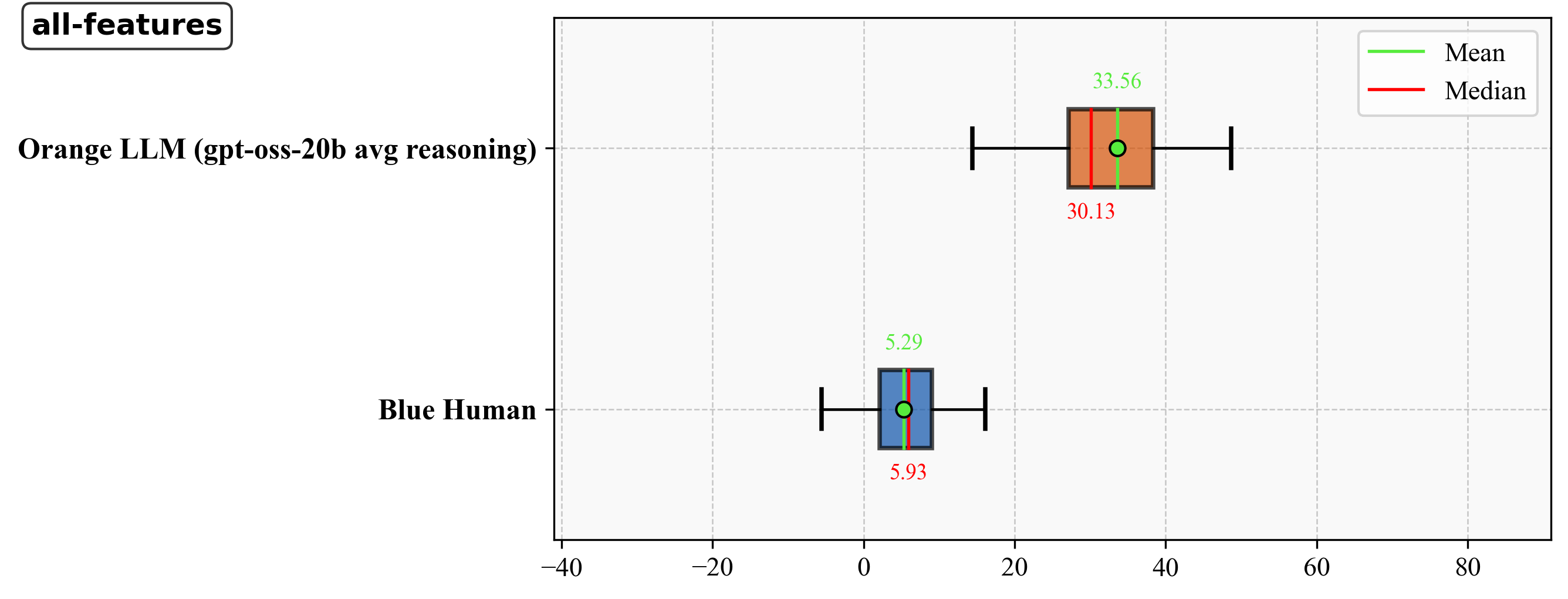} &
\includegraphics[width=0.49\textwidth]{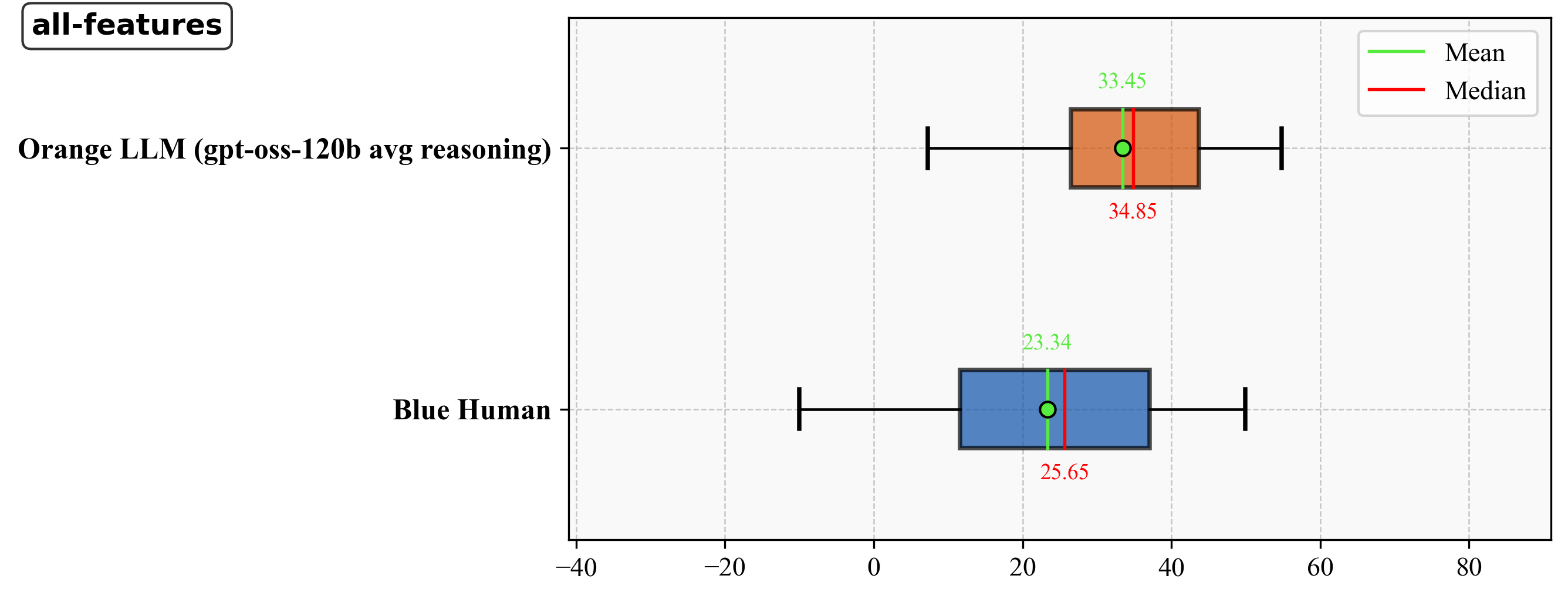} \\
\includegraphics[width=0.49\textwidth]{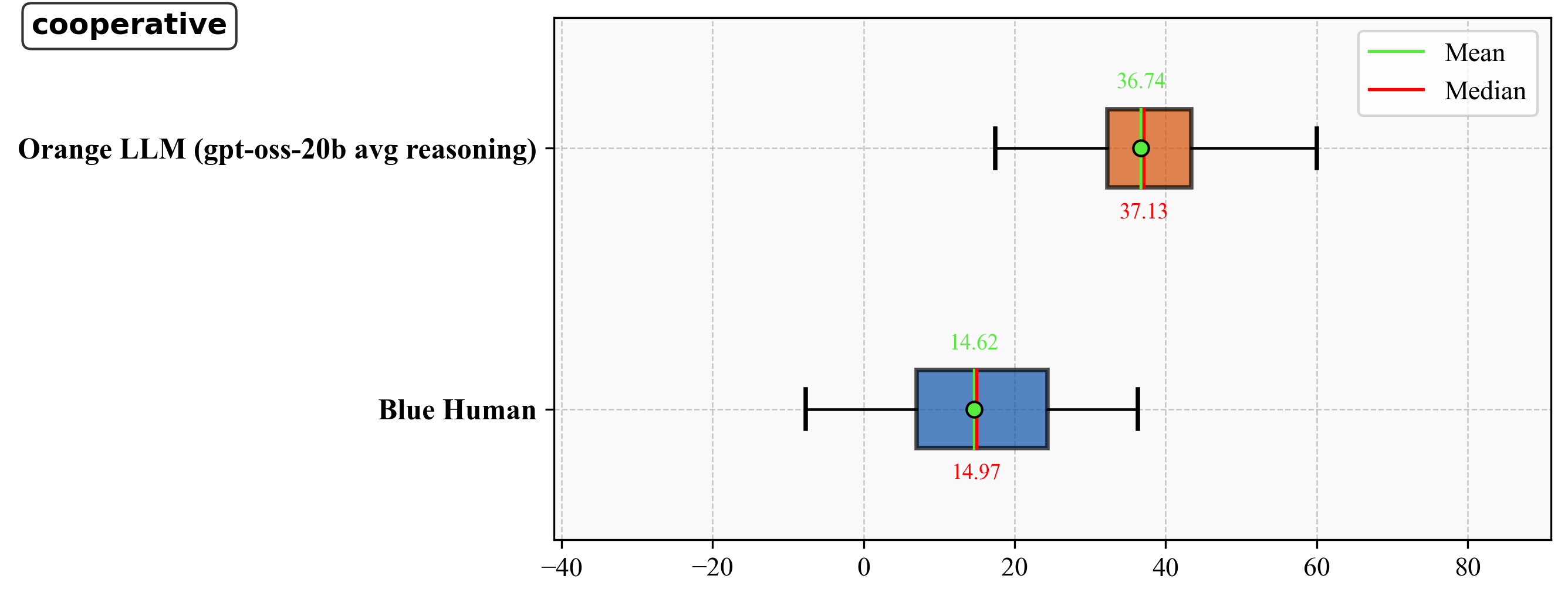} &
\includegraphics[width=0.49\textwidth]{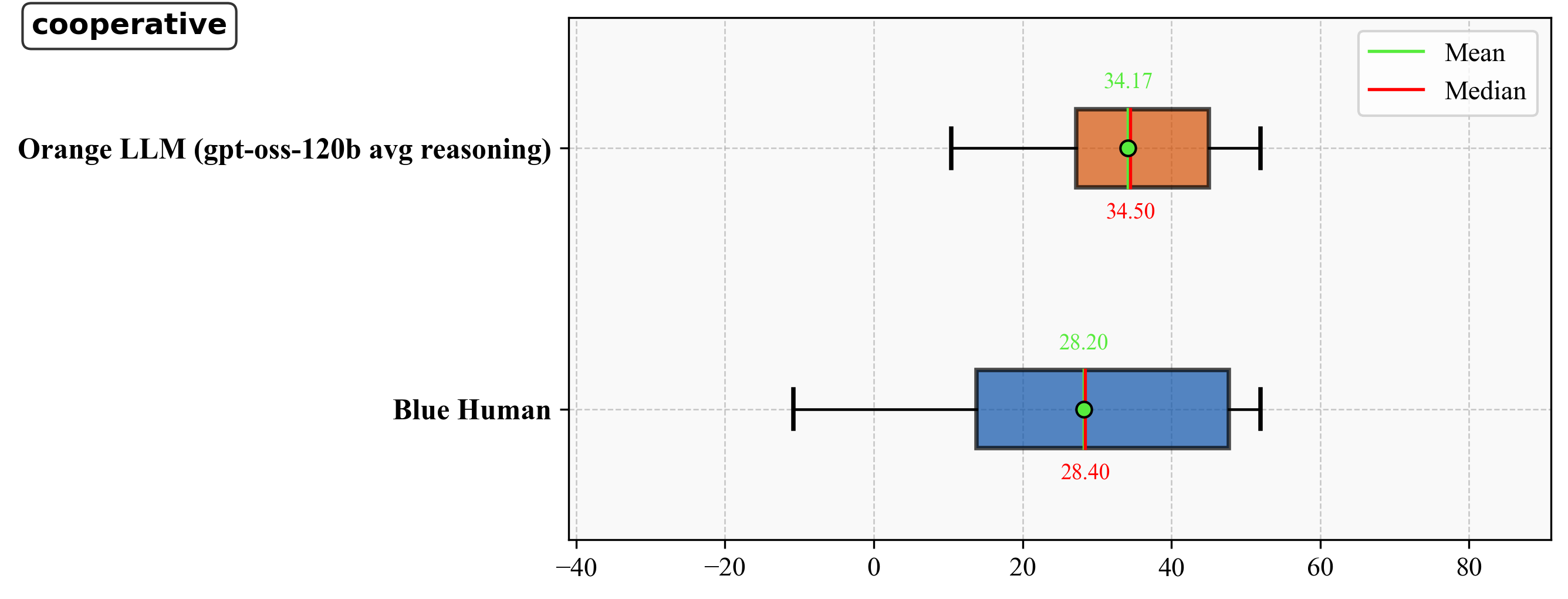} \\
\includegraphics[width=0.49\textwidth]{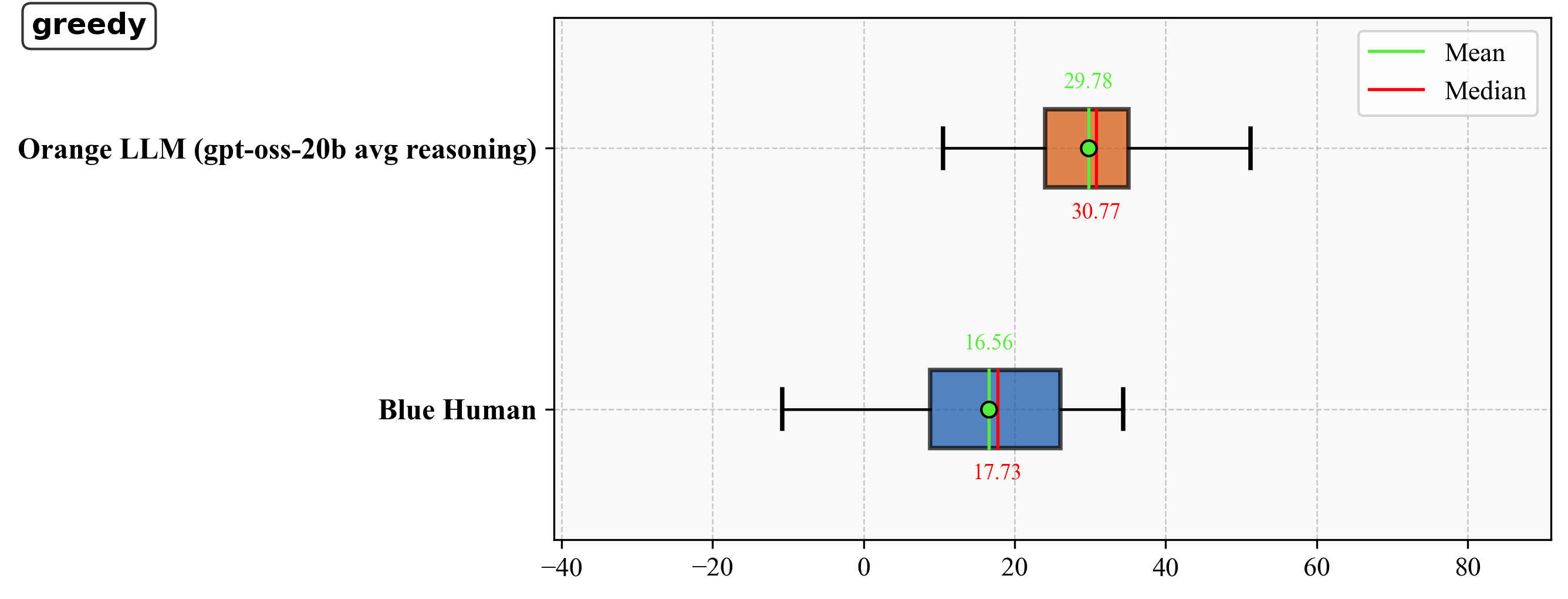} &
\includegraphics[width=0.49\textwidth]{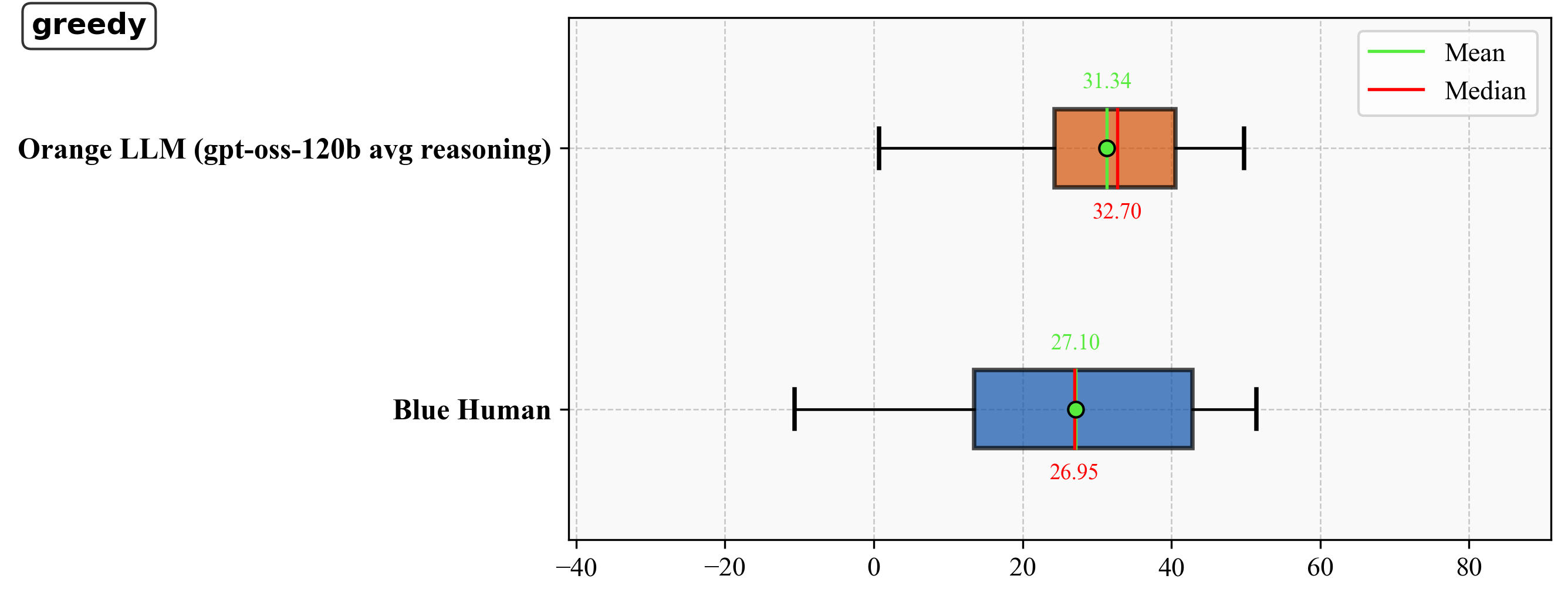} \\
\includegraphics[width=0.49\textwidth]{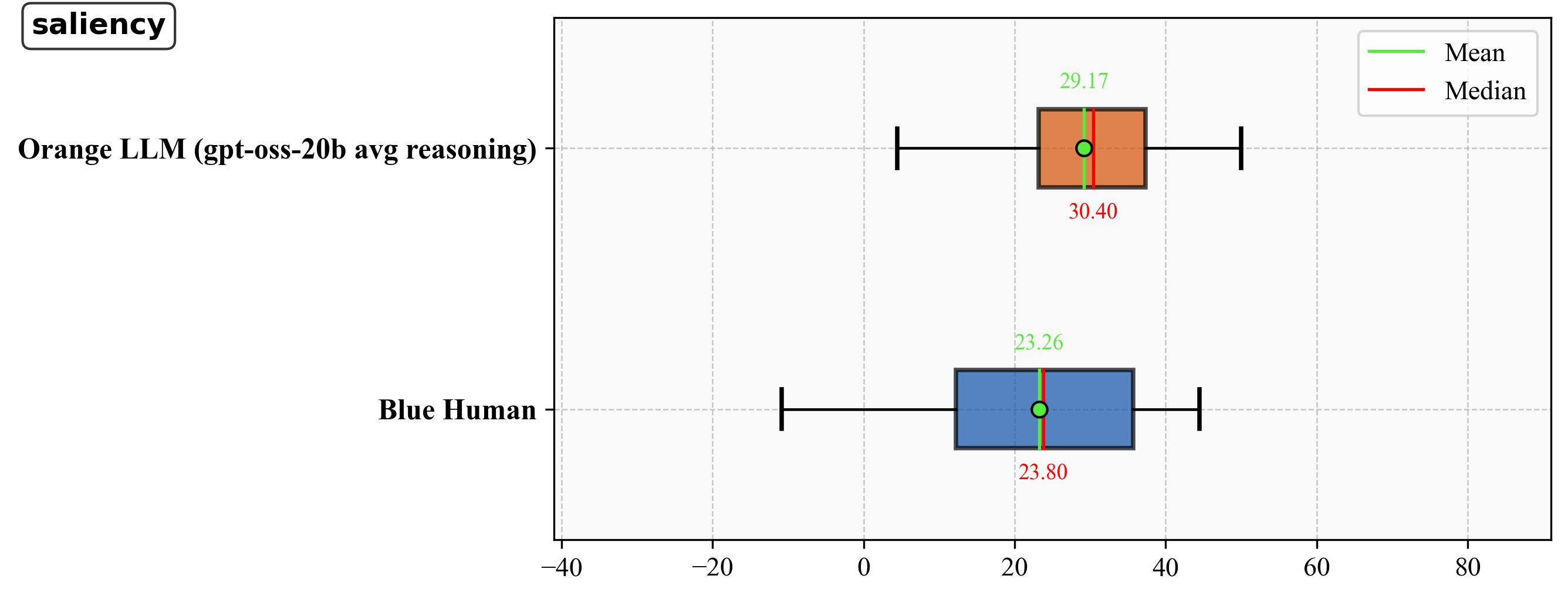} &
\includegraphics[width=0.49\textwidth]{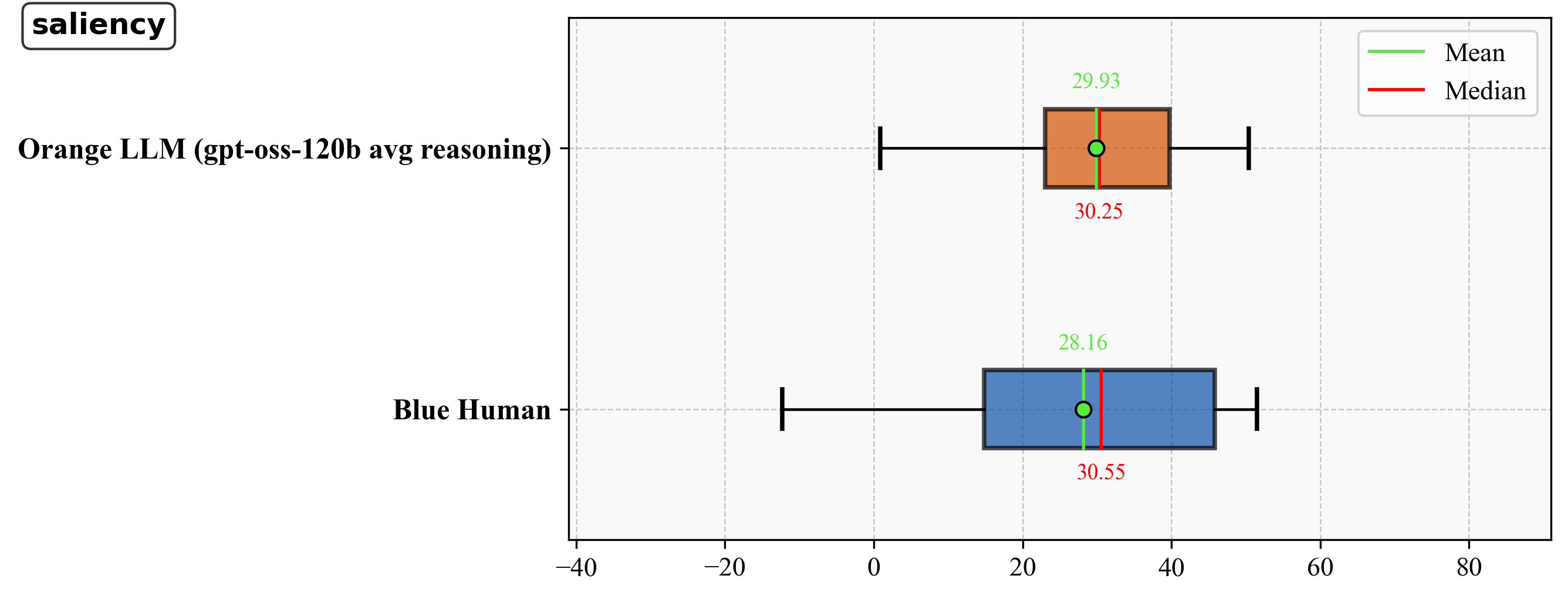} \\
\includegraphics[width=0.49\textwidth]{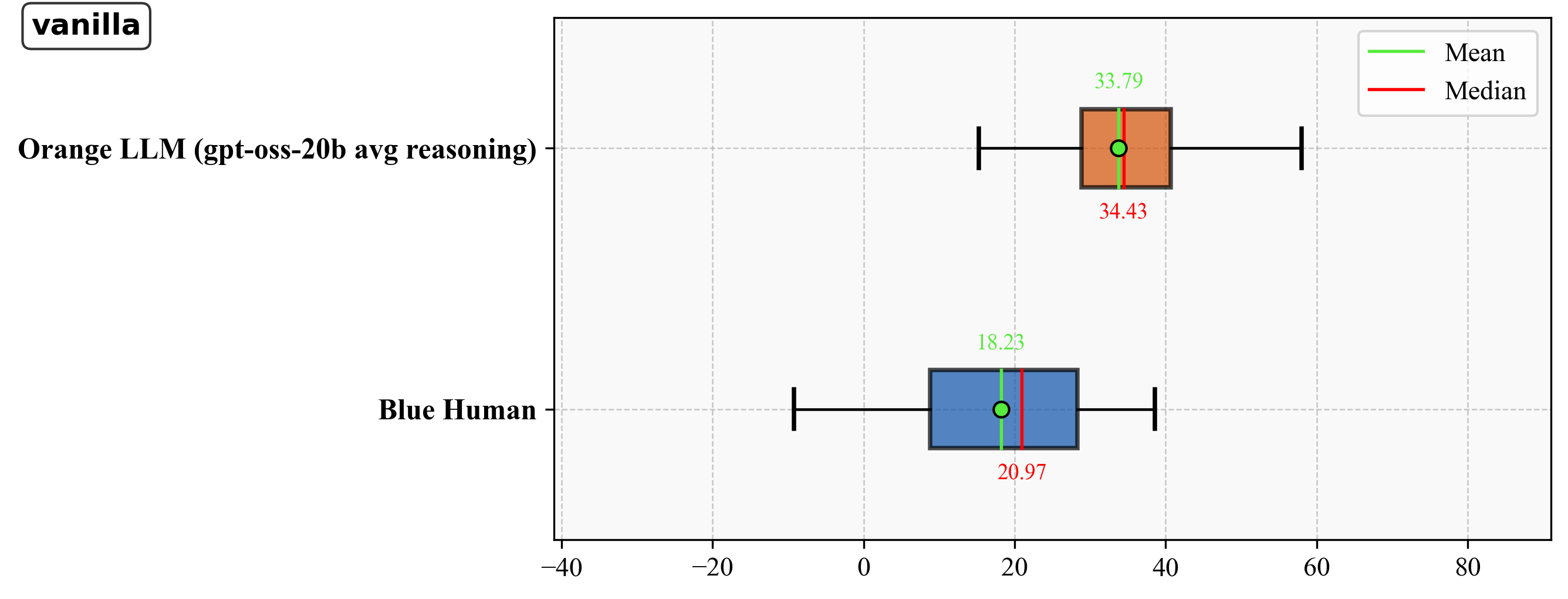} &
\includegraphics[width=0.49\textwidth]{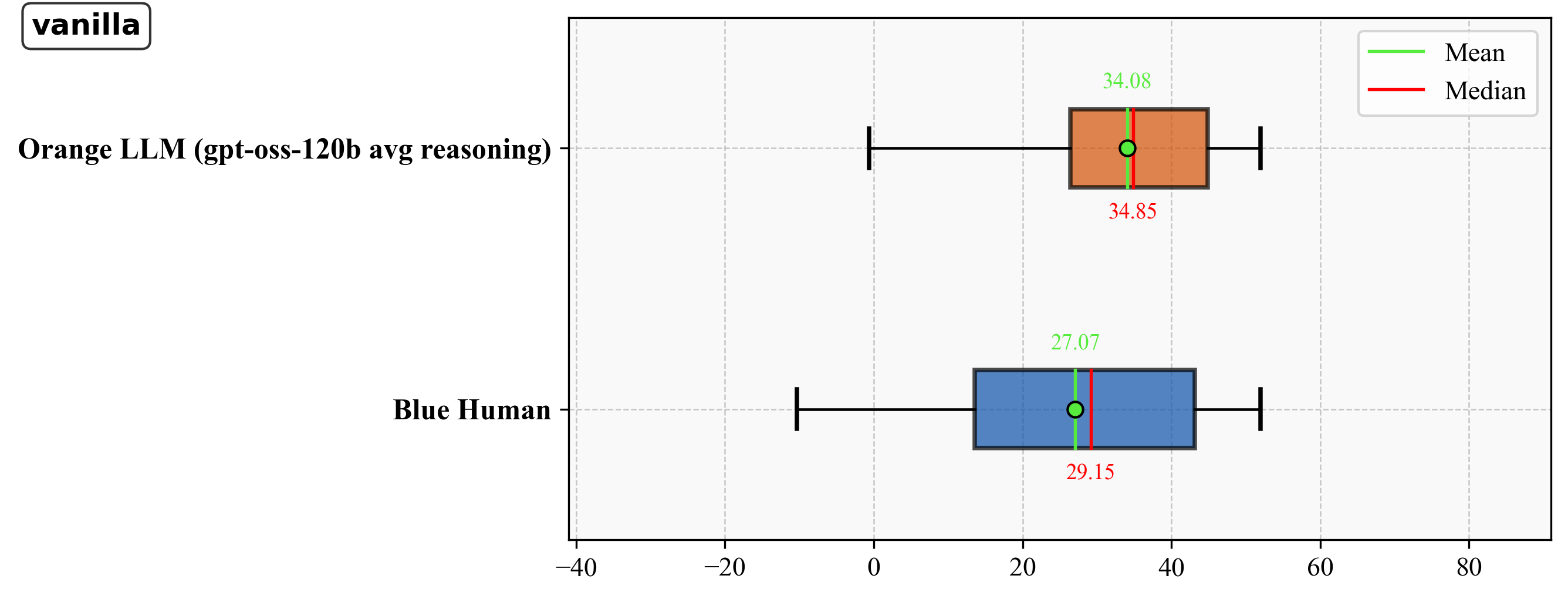} \\
\end{tabular}
\vspace{-2mm}
\caption{Bargaining Table averages by model size: gpt-oss-20b (left) vs gpt-oss-120b (right) for all-features, cooperative, greedy, saliency, and vanilla. The LLM is the orange player.}
\label{fig:bt-avg-across-reasoning-p2-llm-all10}
\end{figure*}

\paragraph{Comparison by model size}
Referencing the main paper's results, averaging over model size (oss-20B and oss-120B), Table~\ref{tab:bargaining-by-reasoning} (top) presents the mean and median scores by variant for the orange (human) and blue (LLM) players. The \textbf{Orange (Human) exhibits a strong model-size effect} across all variants, whereas \textbf{Blue (LLM) is comparatively stable across sizes}. For Orange, oss-20b means span \([9.62,\,26.44]\) while oss-120b means span \([30.18,\,33.26]\), with the largest gain for \texttt{all-features} (mean \(9.62 \rightarrow 30.18\), \(+20.56\)). In contrast, Blue means remain tightly clustered (oss-20b: \([24.27,\,29.93]\); oss-120b: \([27.32,\,29.10]\)), indicating limited sensitivity to scaling. Variant effects differ by player: under oss-20b, \texttt{saliency} maximizes Orange (mean \(26.44\)) while minimizing Blue (mean \(24.27\)), consistent with a player-specific trade-off. Notably, the \texttt{saliency} prompt explicitly instructs the agent to anticipate the other player's moves and prefer discs the other player is unlikely to select; under this interpretation, the higher Blue performance at oss-120b (mean \(28.38\) vs.\ \(24.27\) at oss-20b) suggests improved opponent-modeling accuracy at a larger scale. Relative to the \texttt{vanilla} baseline, the strongest improvement for Orange is obtained by \texttt{saliency} at oss-20b (mean \(18.81 \rightarrow 26.44\), \(+7.63\)) and by \texttt{saliency} at oss-120b (mean \(32.03 \rightarrow 33.26\), \(+1.23\)), while for Blue the largest gains over \texttt{vanilla} arise from \texttt{cooperative} at oss-20b (mean \(28.47 \rightarrow 29.93\), \(+1.46\)) and \texttt{cooperative} at oss-120b (mean \(28.53 \rightarrow 29.10\), \(+0.57\)). All graphs are shown in Figures~\ref{fig:bt-avg-across-reasoning-p1-llm-all10} and~\ref{fig:bt-avg-across-reasoning-p2-llm-all10}.

\begin{figure*}[t]
\centering
\setlength{\tabcolsep}{1pt}
\renewcommand{\arraystretch}{0}
\begin{tabular}{@{}ccc@{}}
\includegraphics[width=0.325\textwidth]{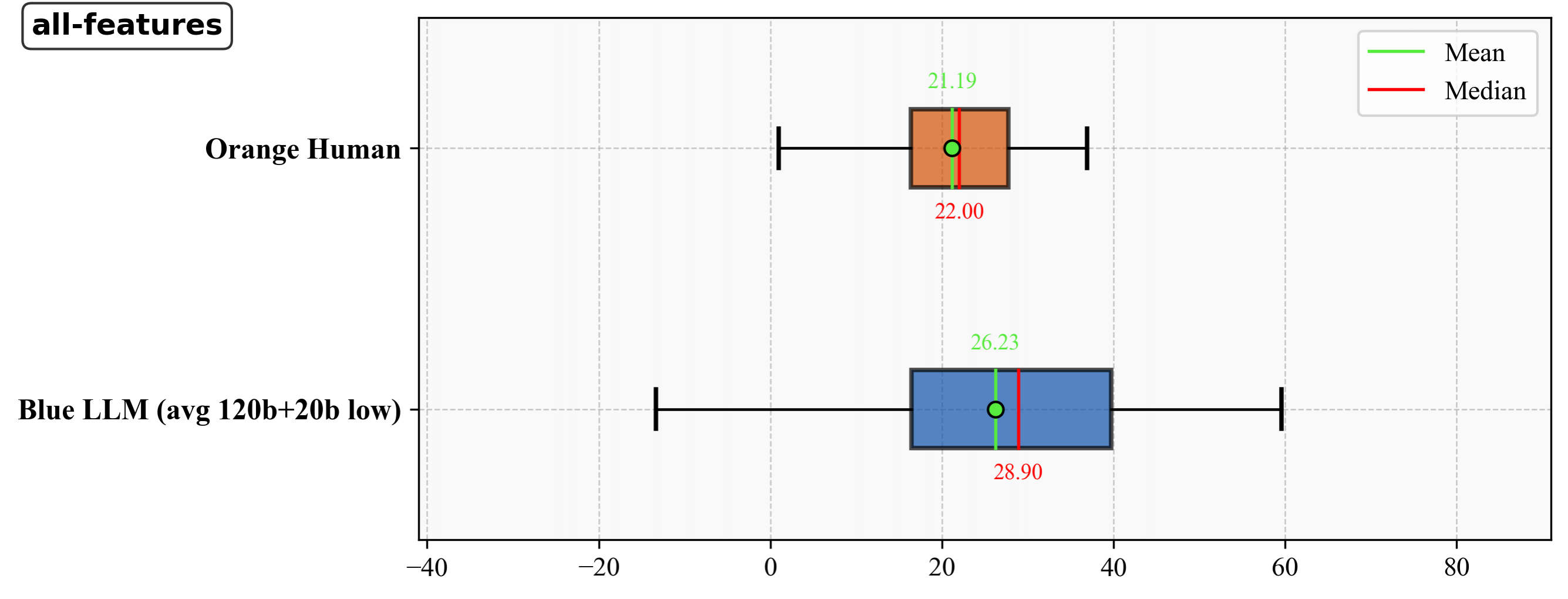} &
\includegraphics[width=0.325\textwidth]{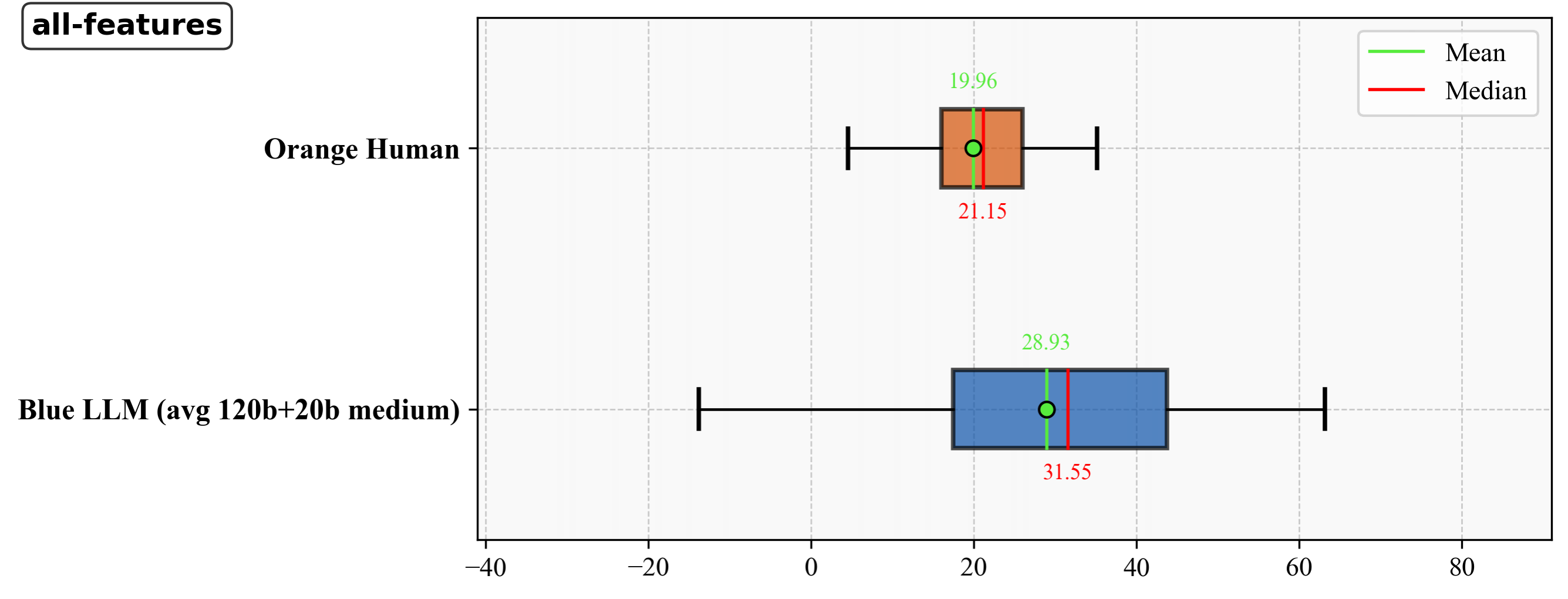} &
\includegraphics[width=0.325\textwidth]{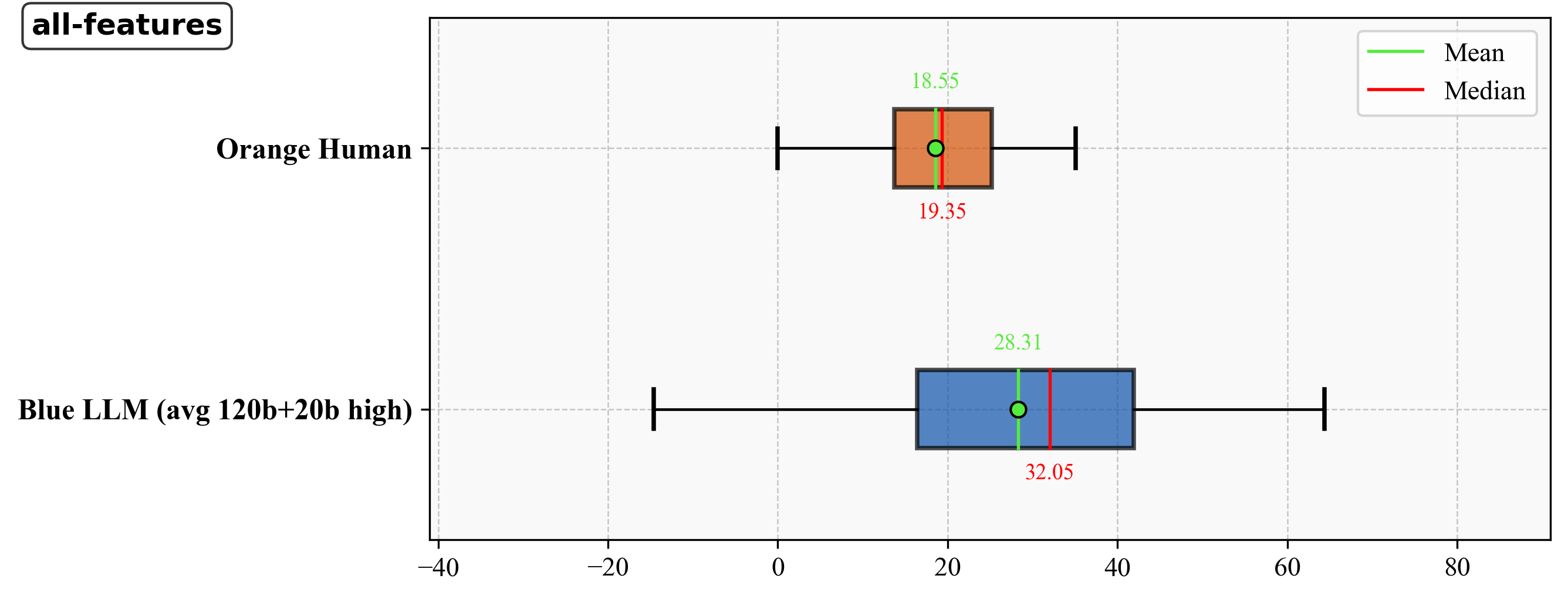} \\
\includegraphics[width=0.325\textwidth]{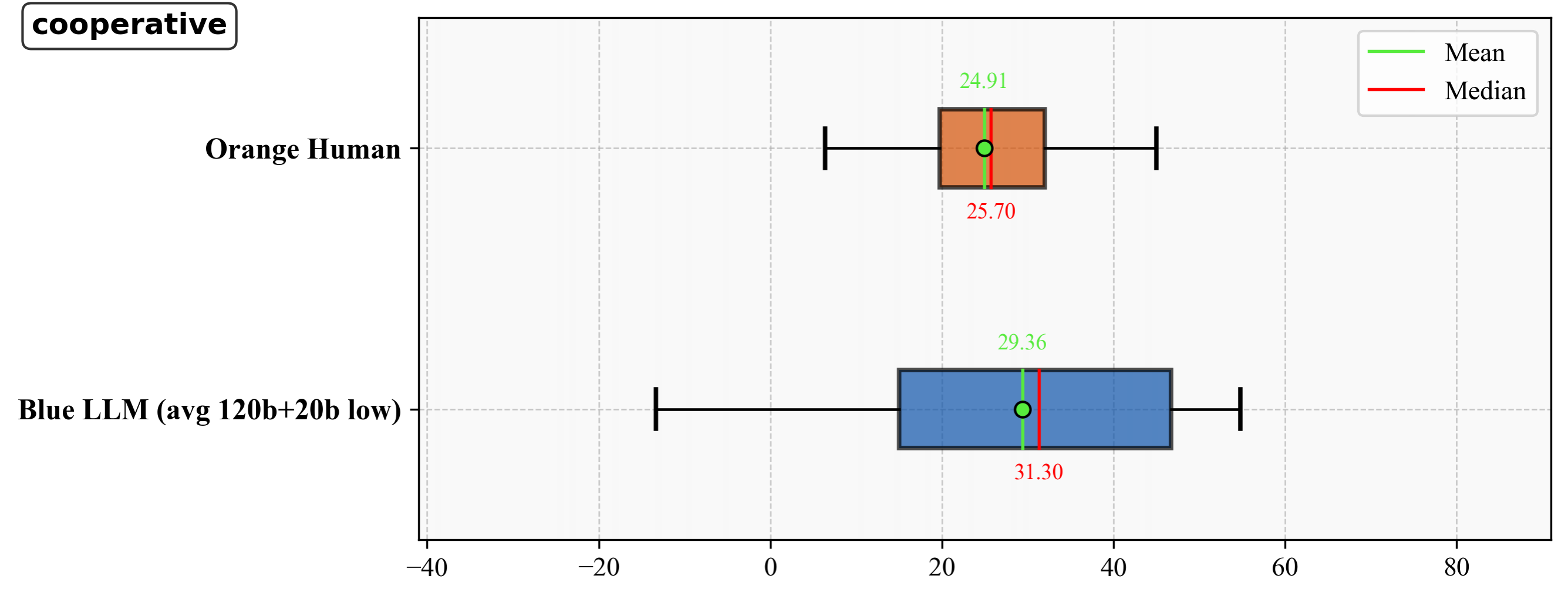} &
\includegraphics[width=0.325\textwidth]{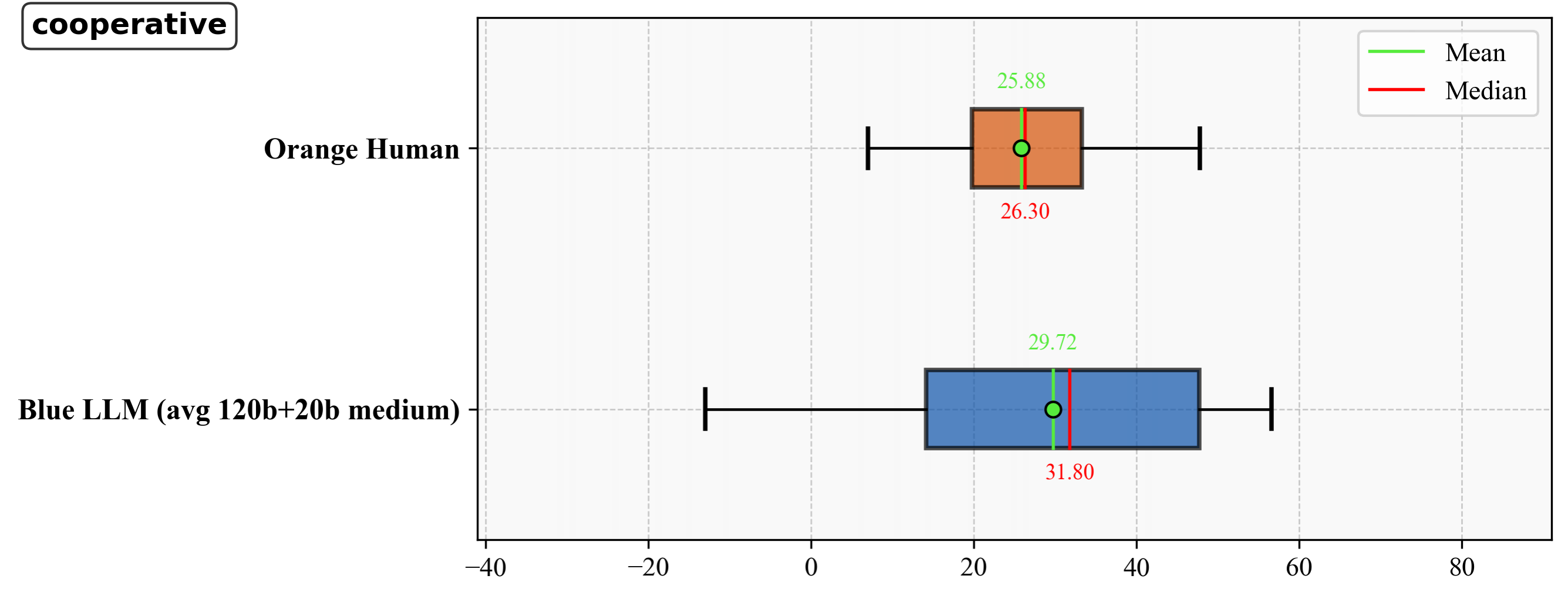} &
\includegraphics[width=0.325\textwidth]{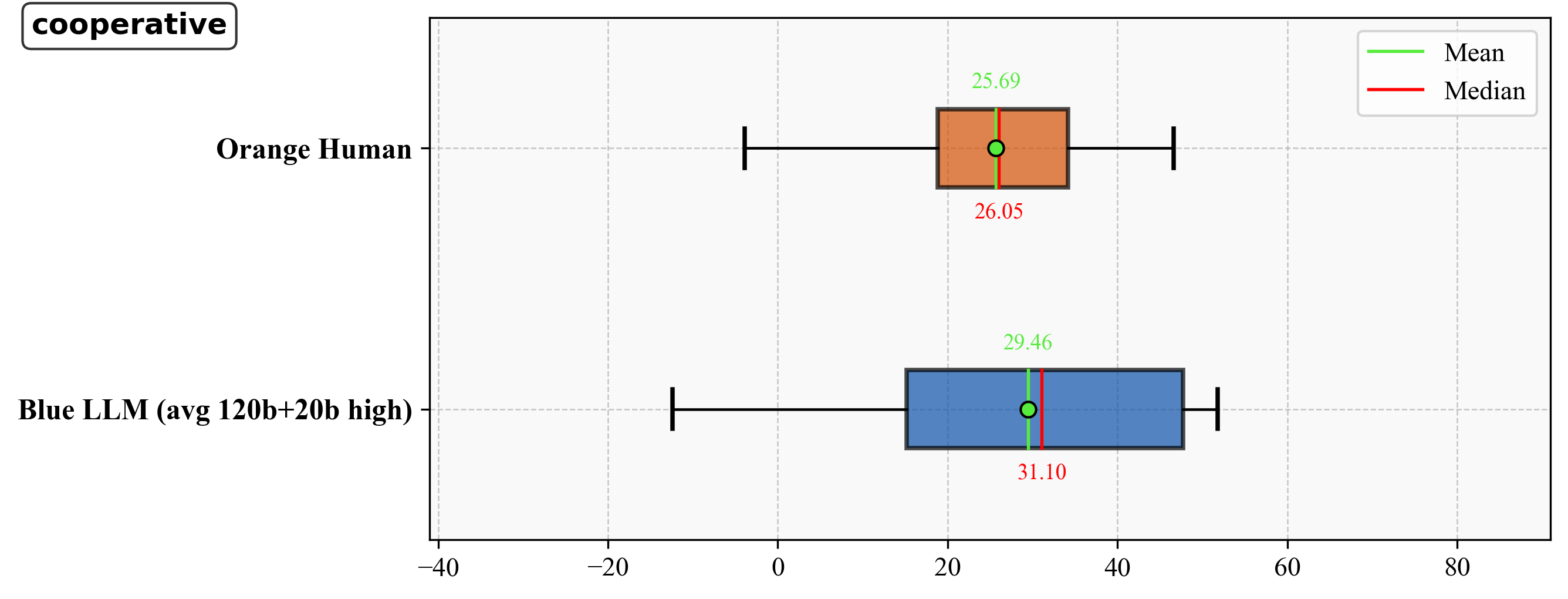} \\
\includegraphics[width=0.325\textwidth]{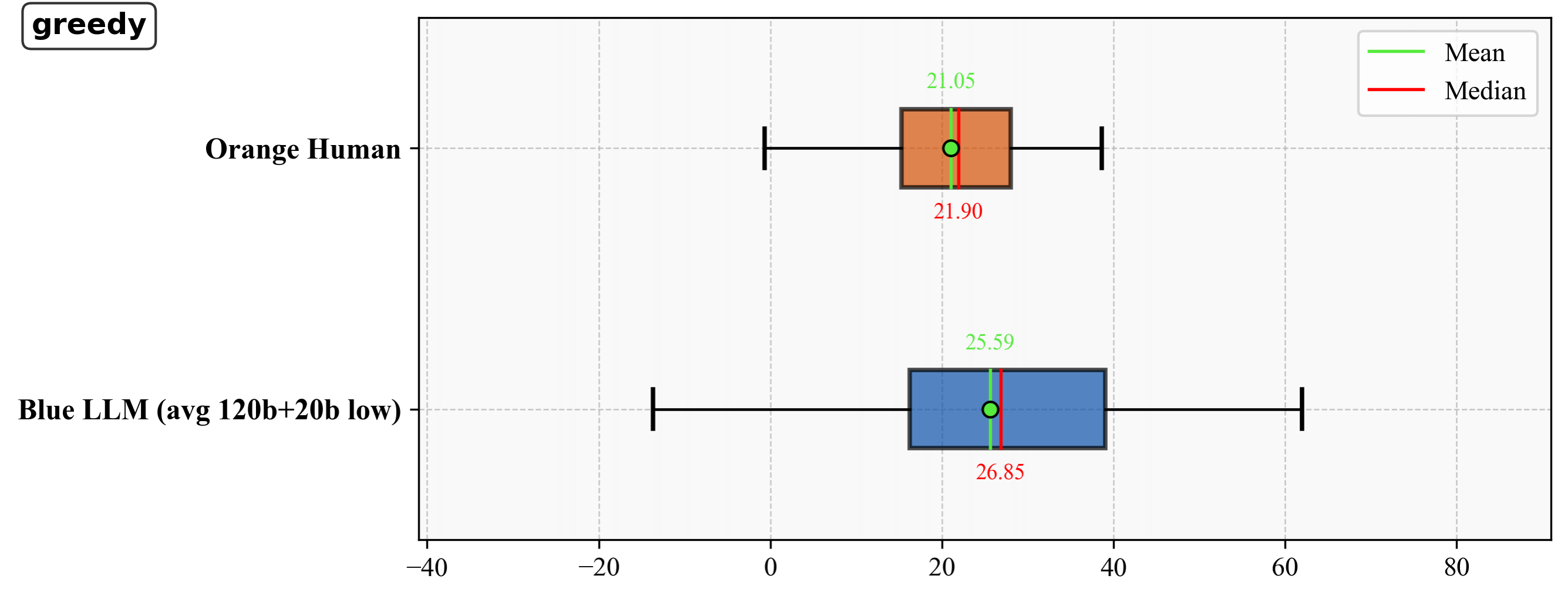} &
\includegraphics[width=0.325\textwidth]{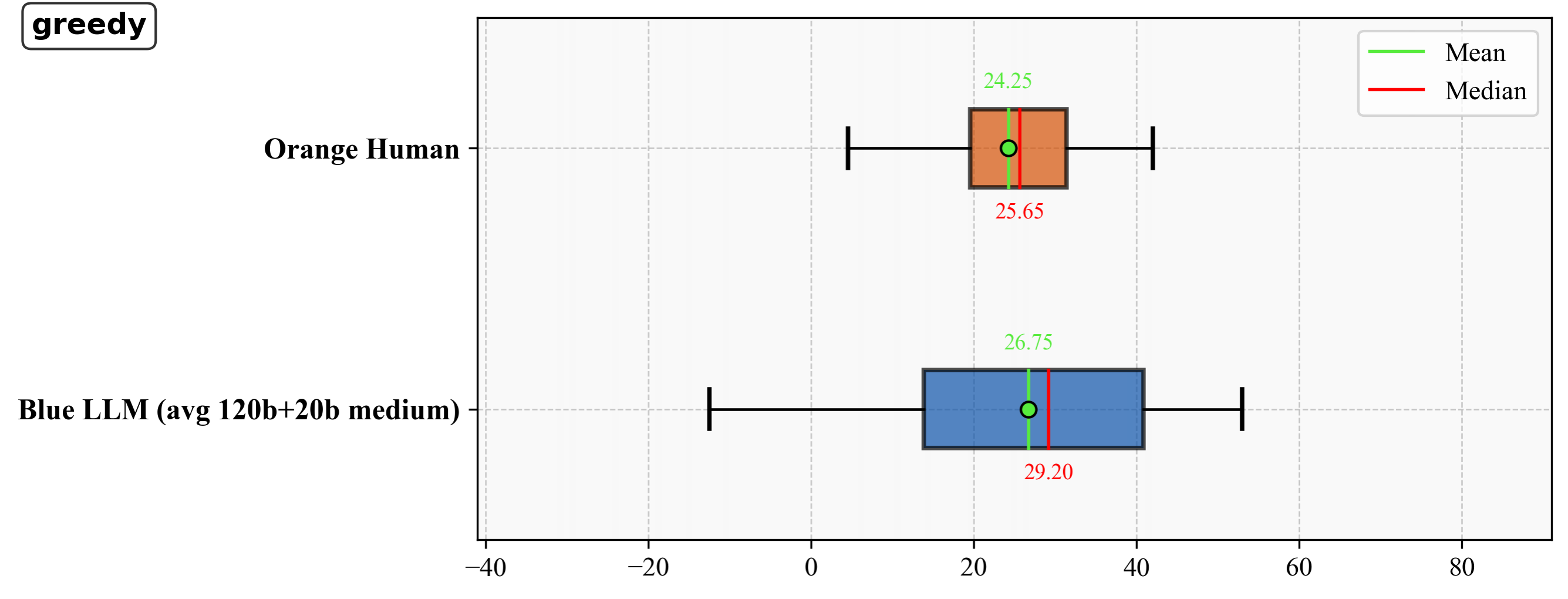} &
\includegraphics[width=0.325\textwidth]{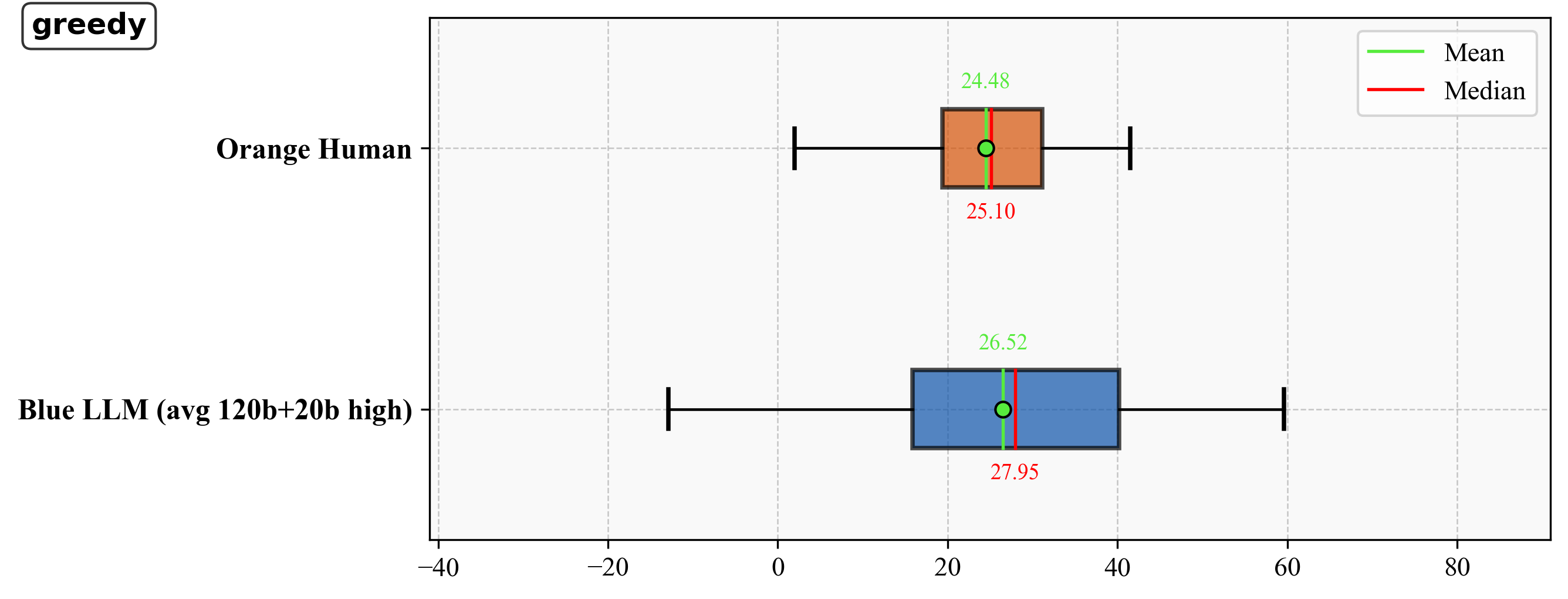} \\
\includegraphics[width=0.325\textwidth]{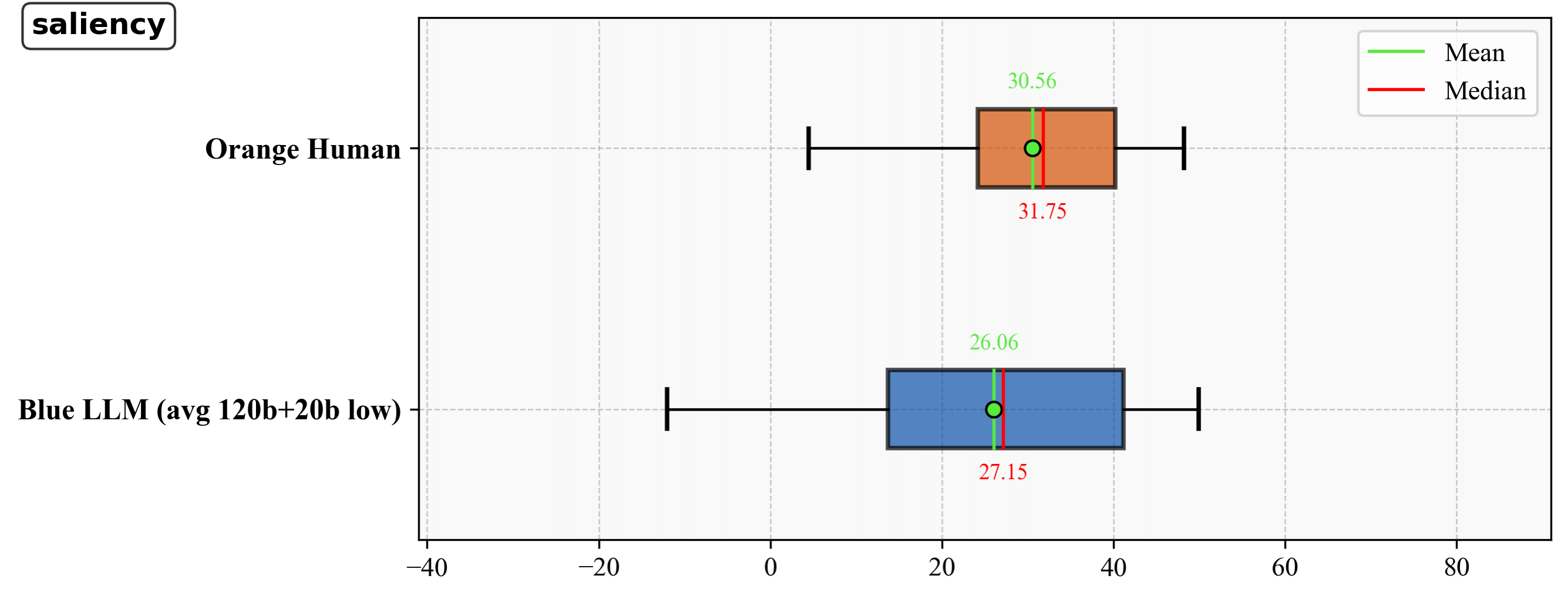} &
\includegraphics[width=0.325\textwidth]{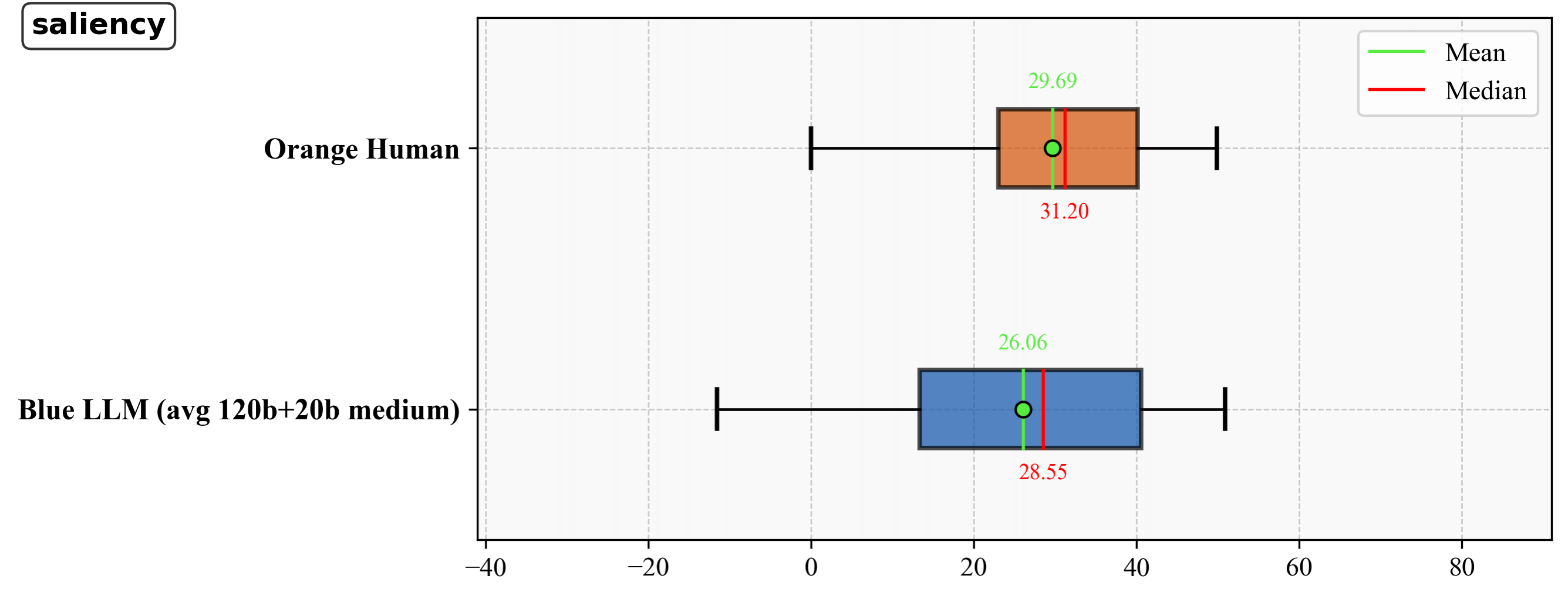} &
\includegraphics[width=0.325\textwidth]{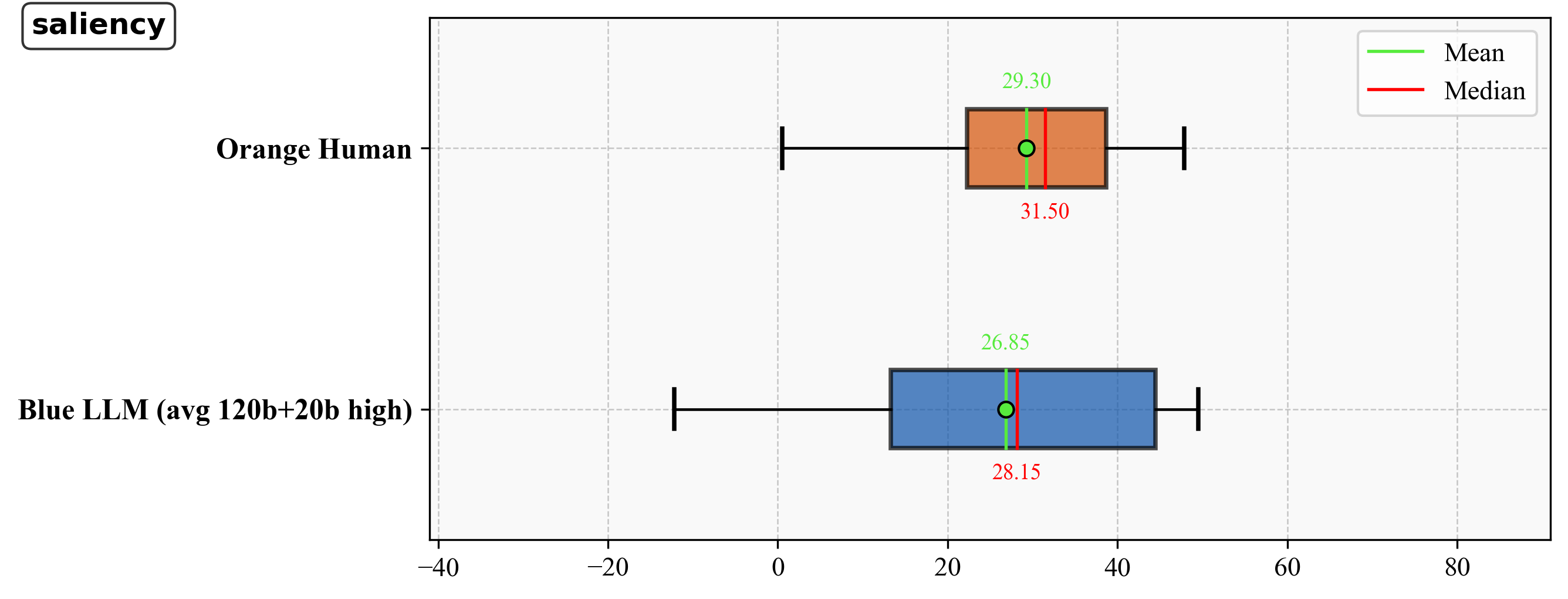} \\
\includegraphics[width=0.325\textwidth]{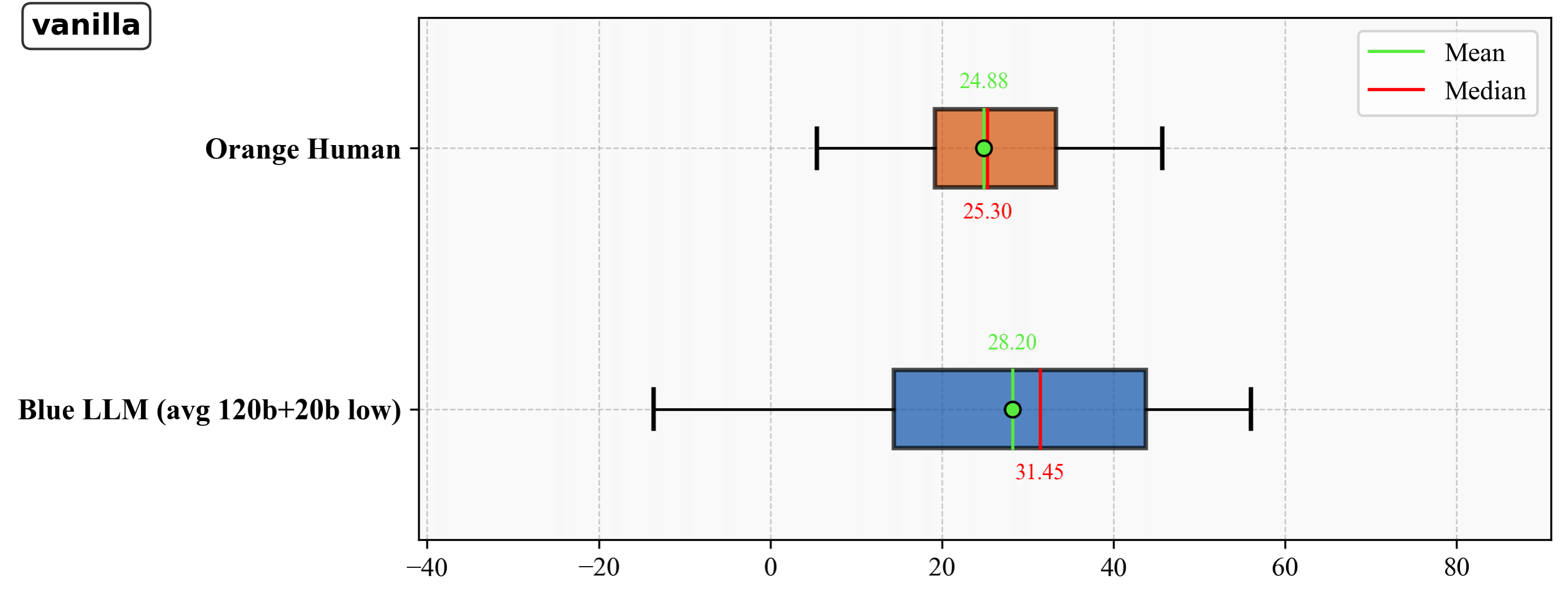} &
\includegraphics[width=0.325\textwidth]{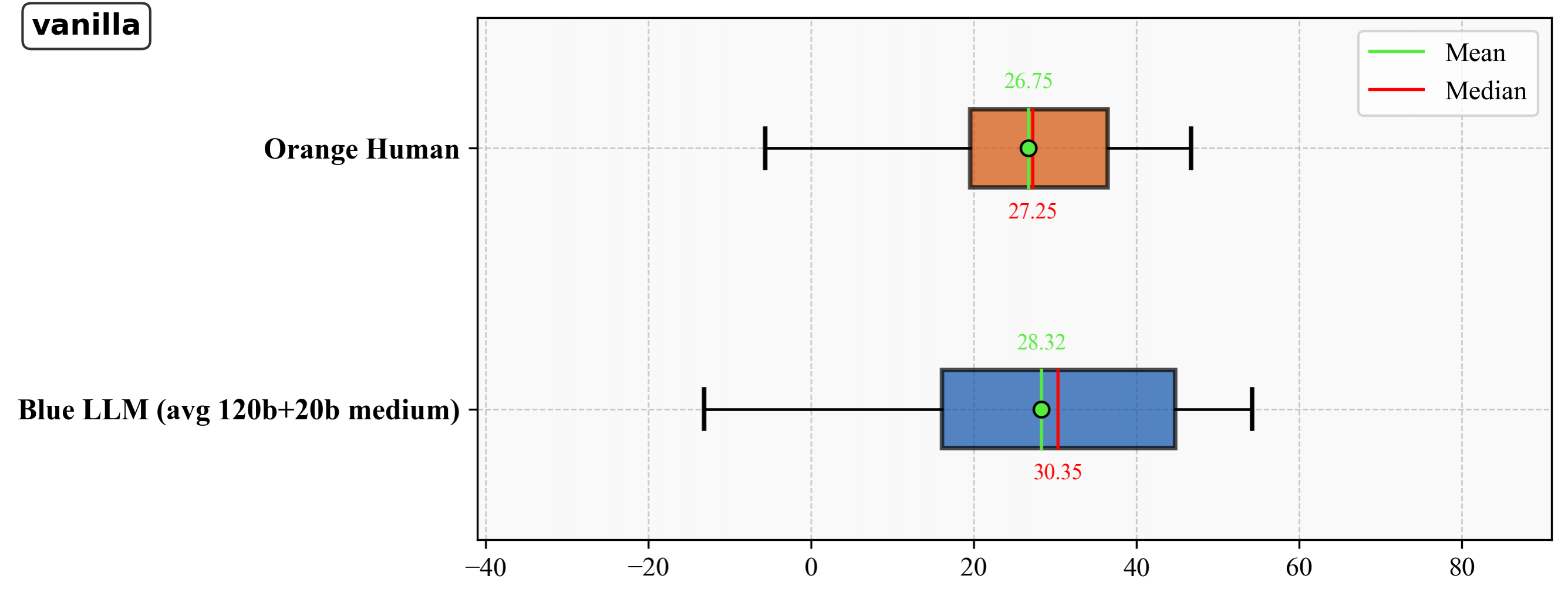} &
\includegraphics[width=0.325\textwidth]{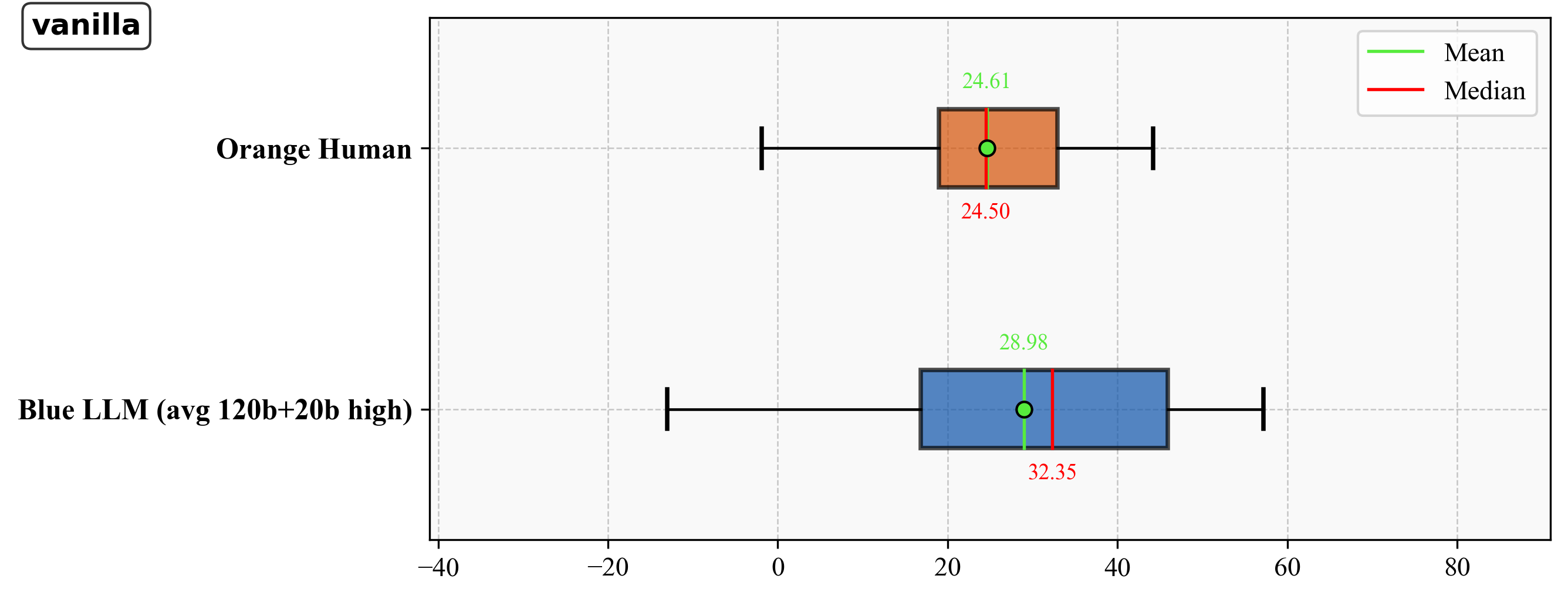} \\
\end{tabular}
\vspace{-2mm}
\caption{Bargaining Table averages, grouped by reasoning level. Columns correspond to low, medium, and high reasoning; rows correspond to all-features, cooperative, greedy, saliency, and vanilla (averaged over gpt-oss-20b and gpt-oss-120b). The LLM is the blue player.}
\label{fig:bt-avg-by-reasoning-p1-llm-low-med-high}
\end{figure*}

\begin{figure*}[t]
\centering
\setlength{\tabcolsep}{1pt}
\renewcommand{\arraystretch}{0}
\begin{tabular}{@{}ccc@{}}
\includegraphics[width=0.325\textwidth]{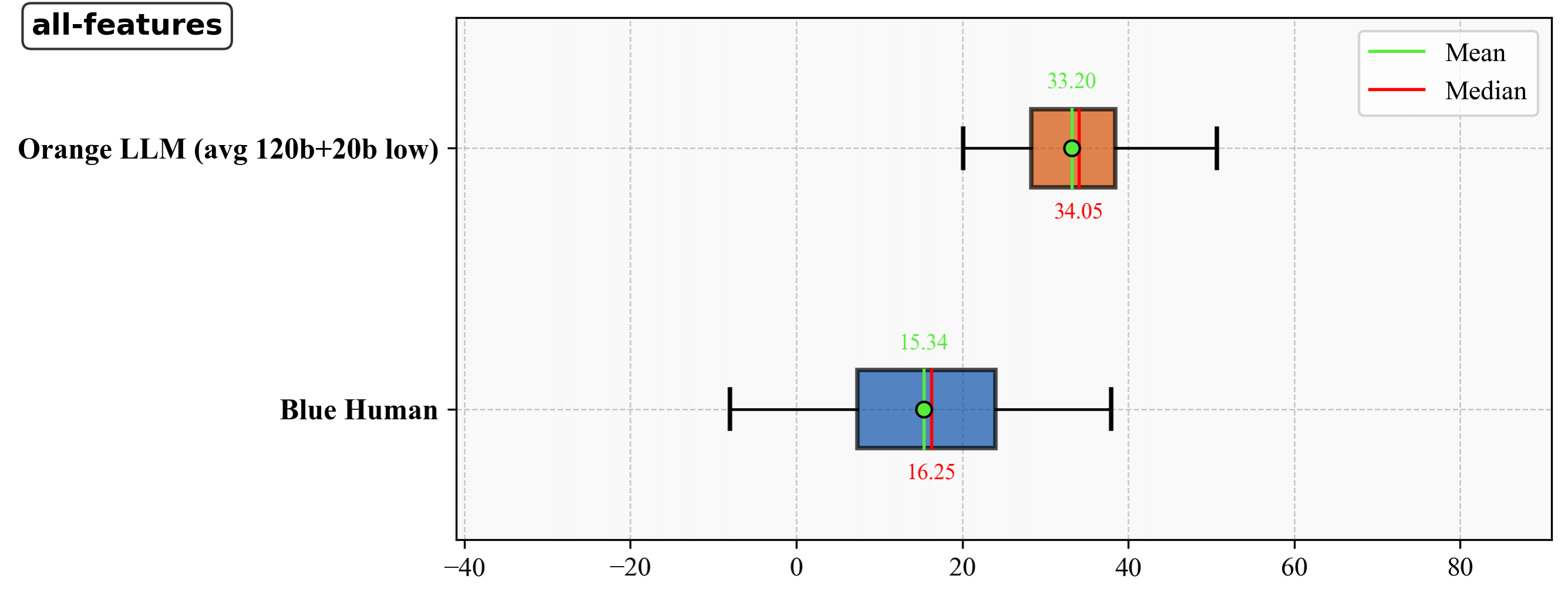} &
\includegraphics[width=0.325\textwidth]{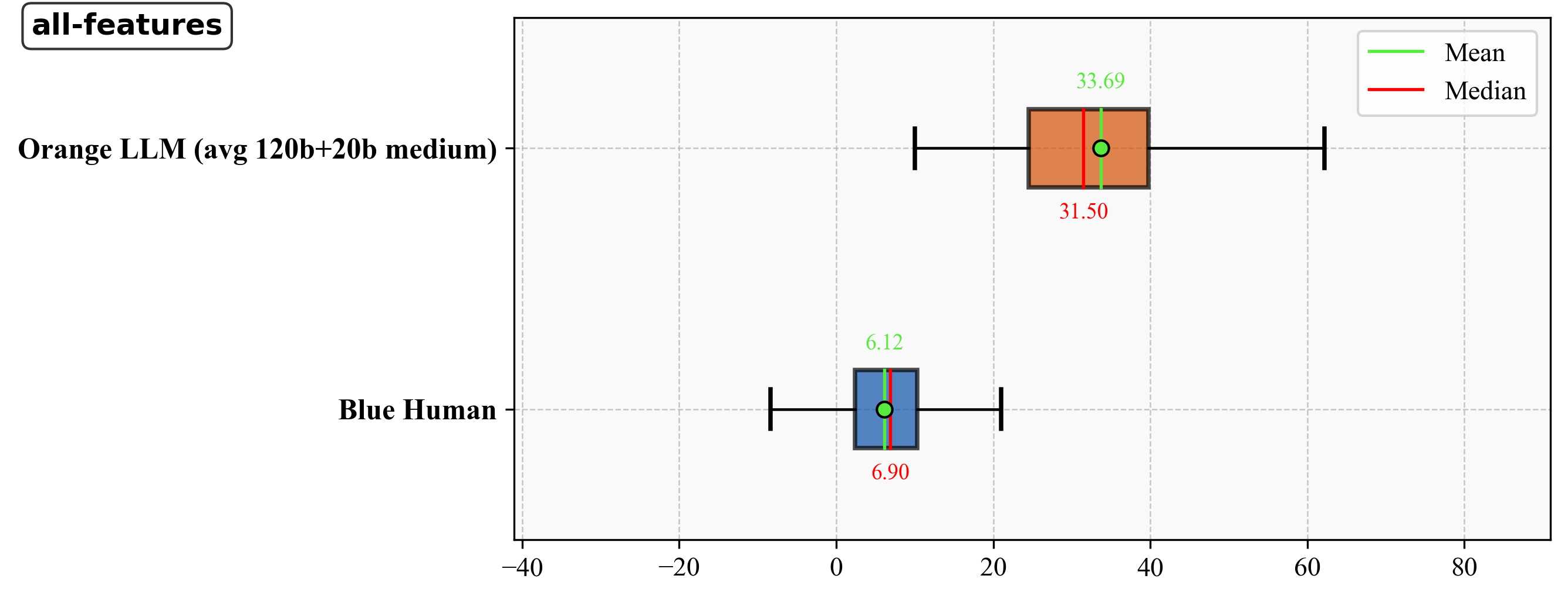} &
\includegraphics[width=0.325\textwidth]{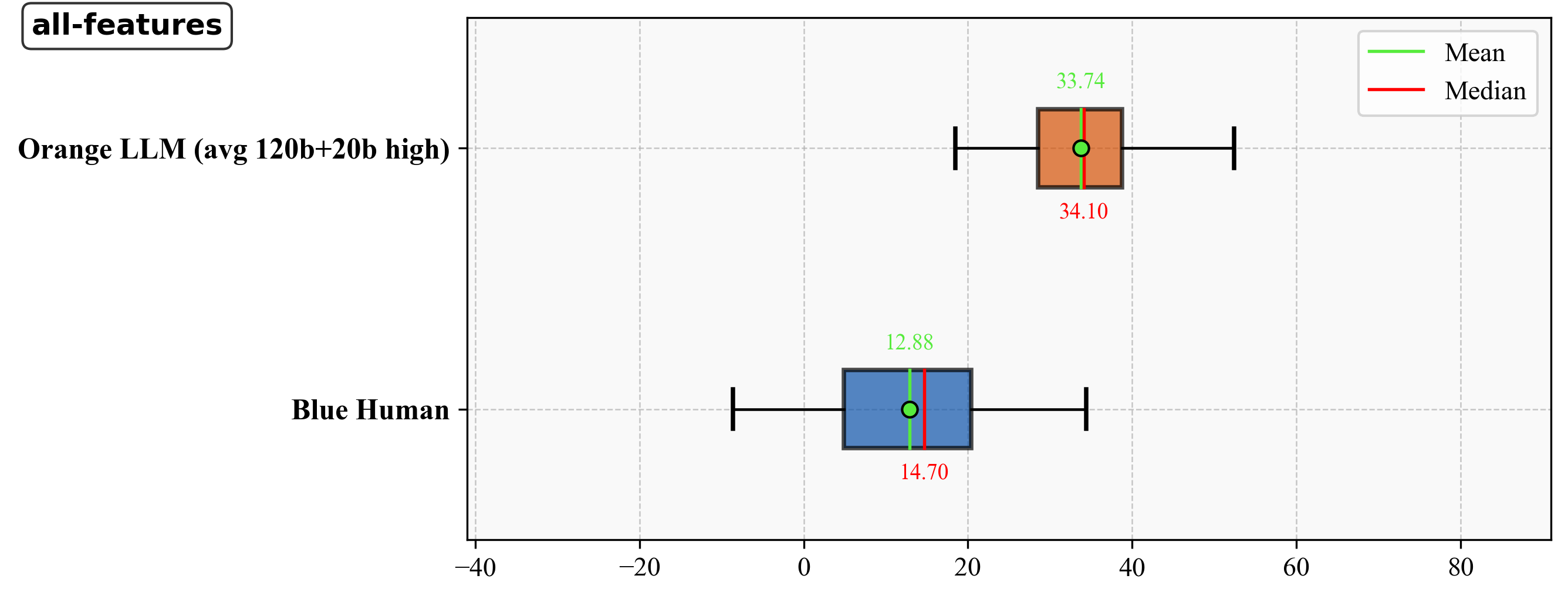} \\
\includegraphics[width=0.325\textwidth]{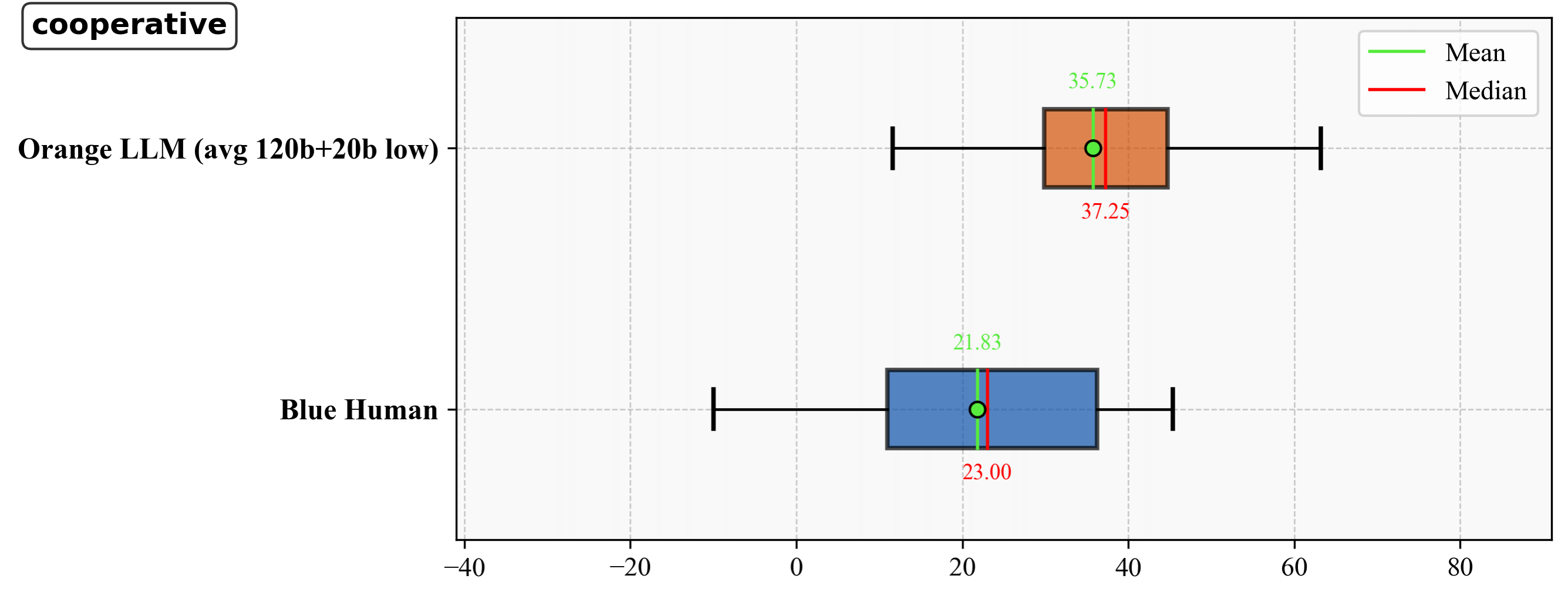} &
\includegraphics[width=0.325\textwidth]{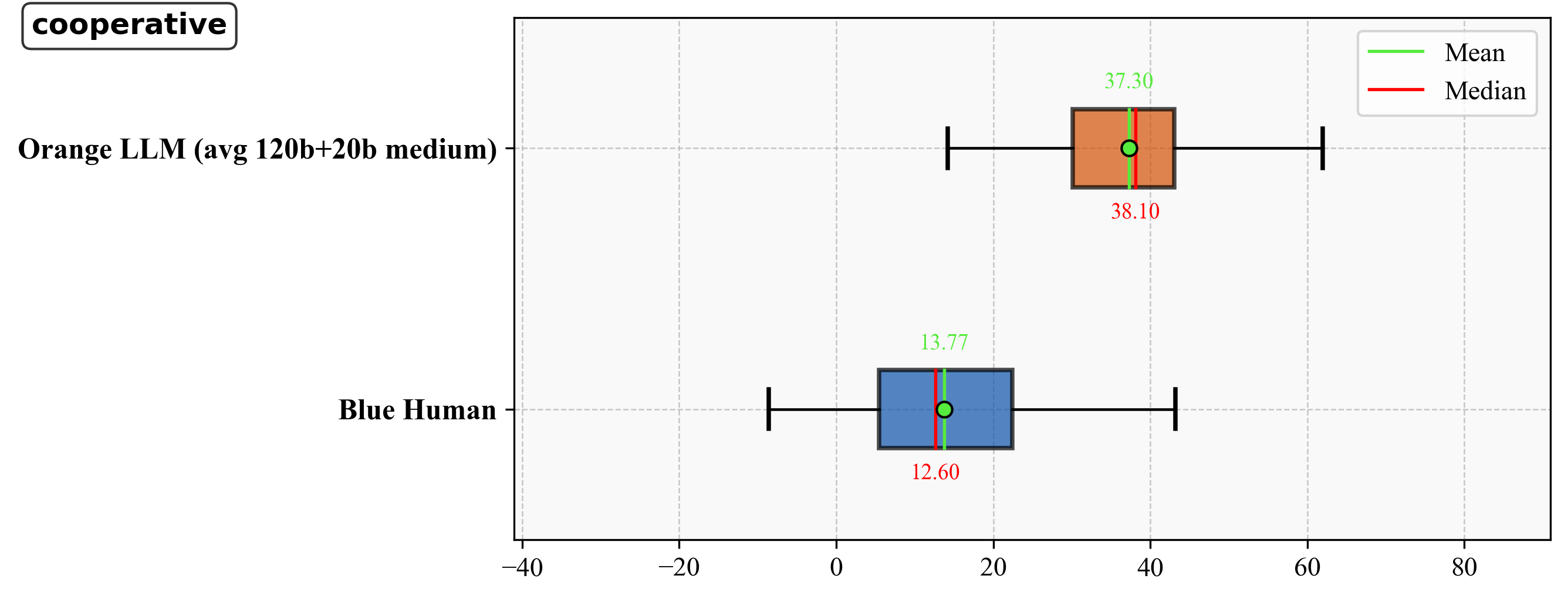} &
\includegraphics[width=0.325\textwidth]{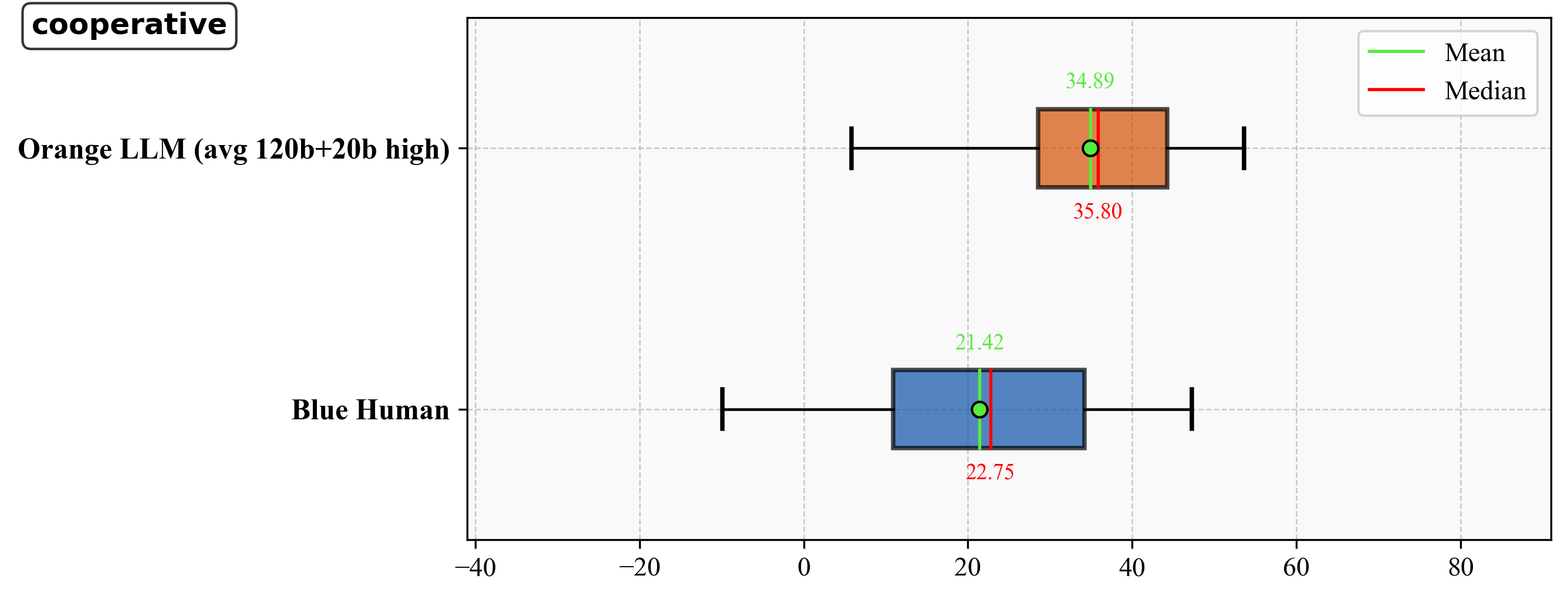} \\
\includegraphics[width=0.325\textwidth]{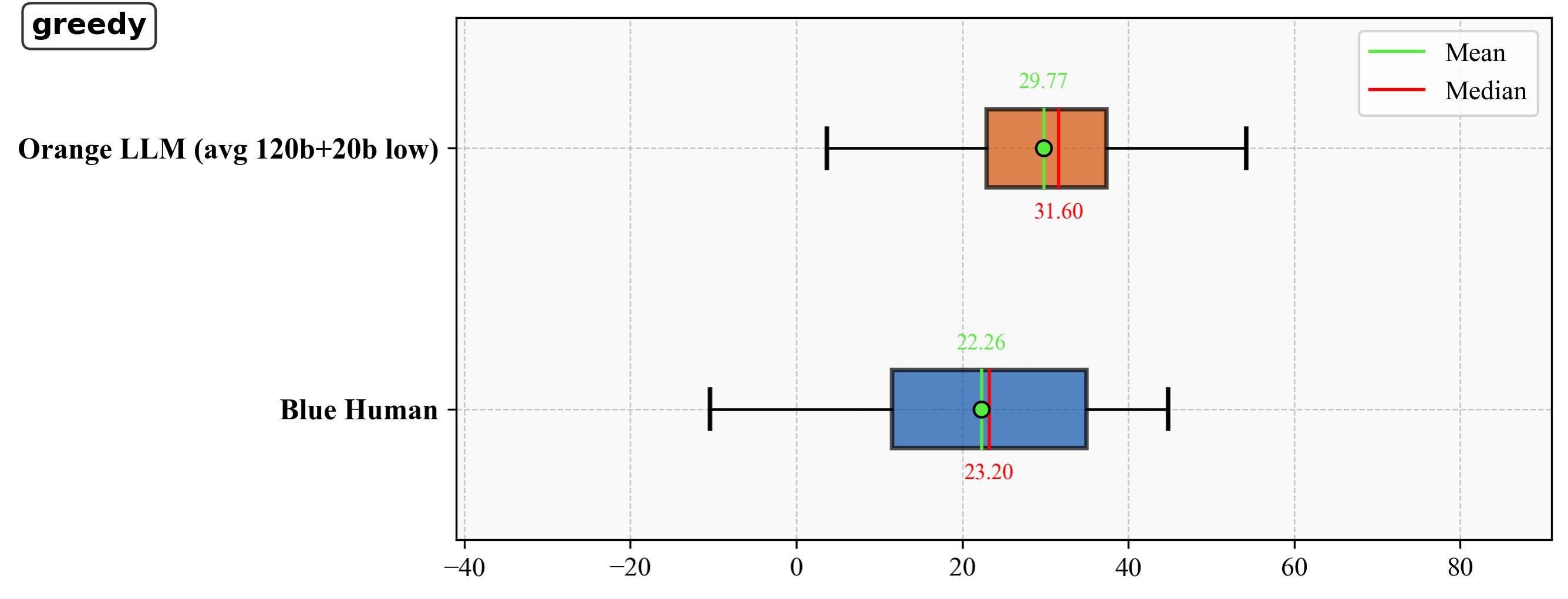} &
\includegraphics[width=0.325\textwidth]{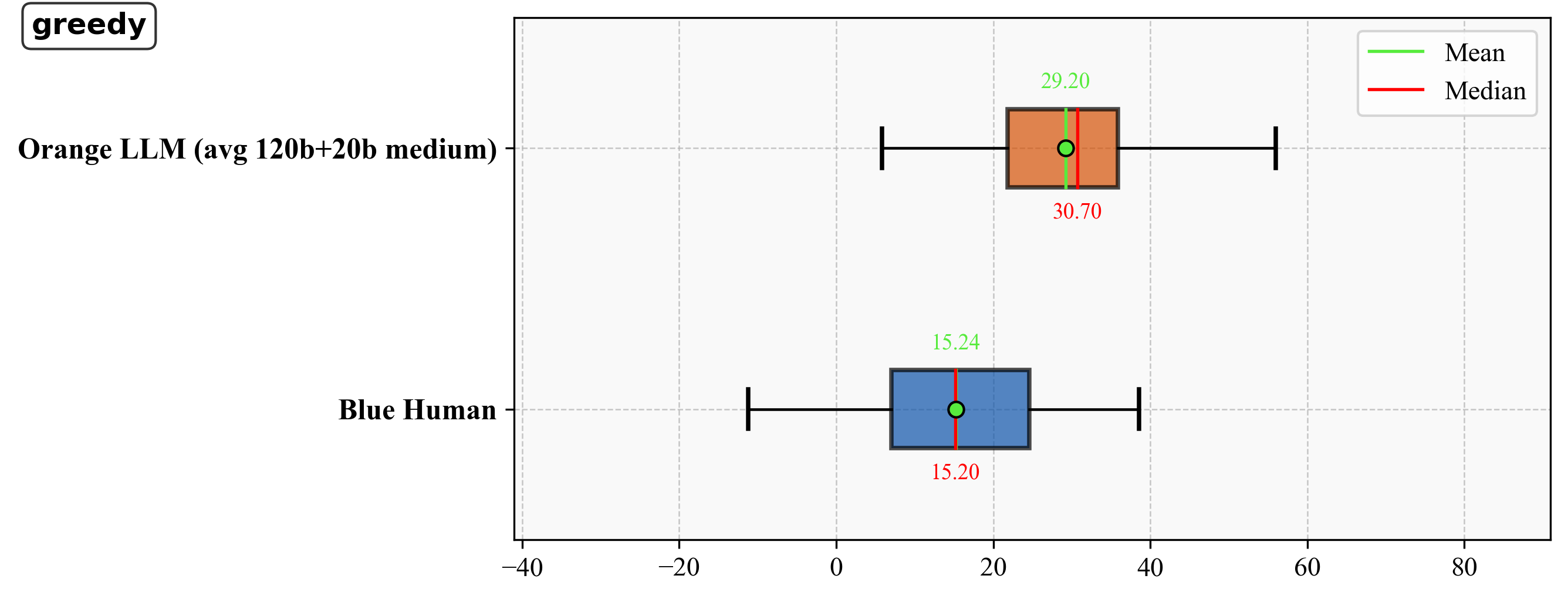} &
\includegraphics[width=0.325\textwidth]{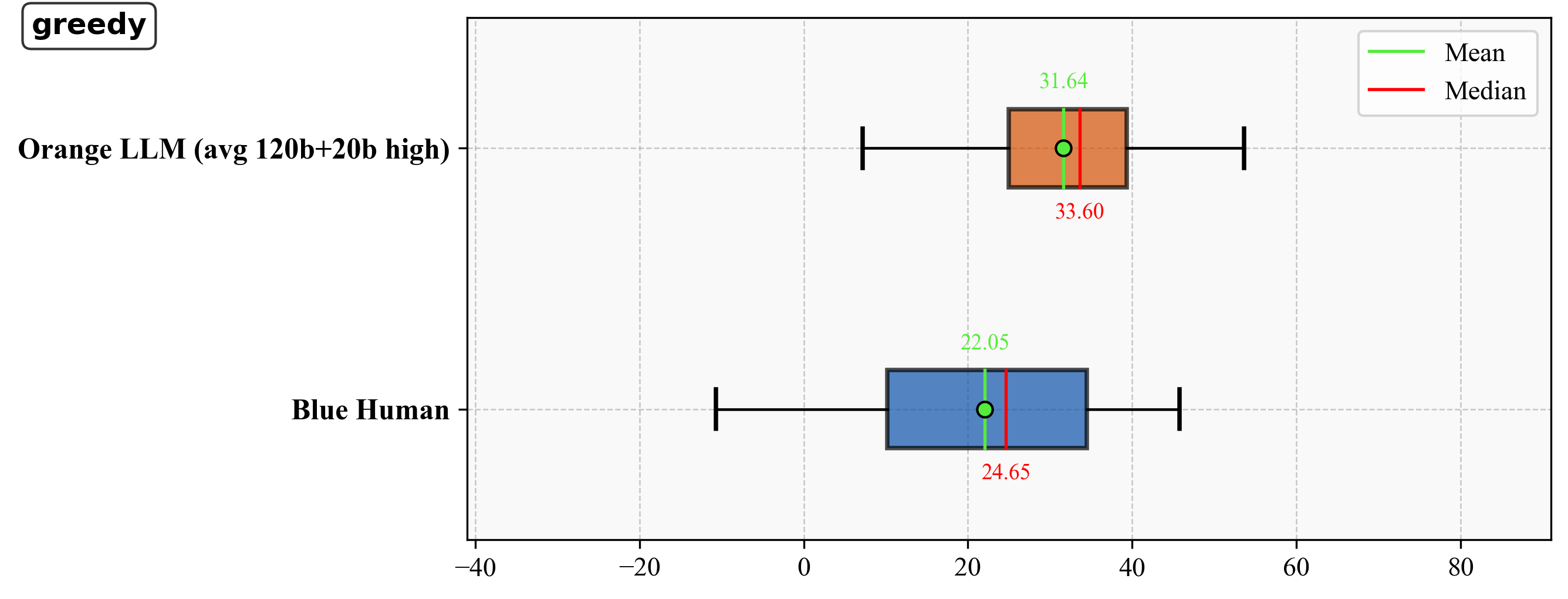} \\
\includegraphics[width=0.325\textwidth]{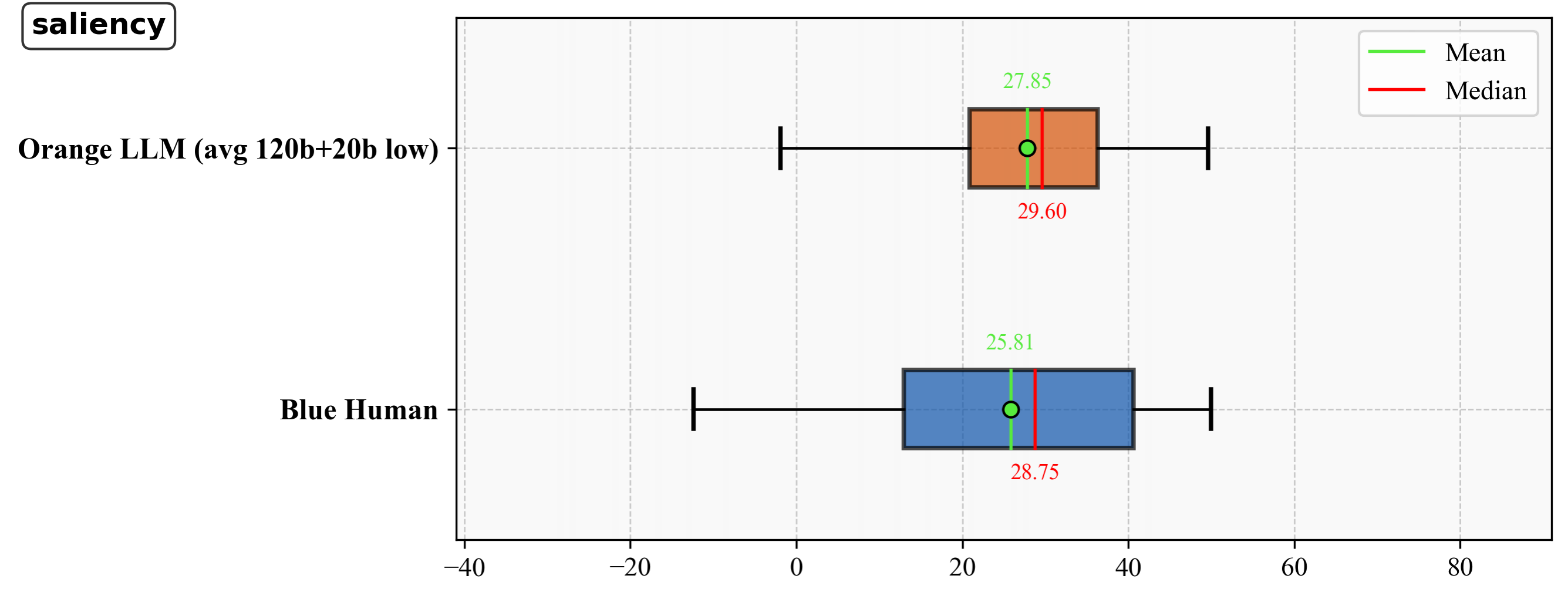} &
\includegraphics[width=0.325\textwidth]{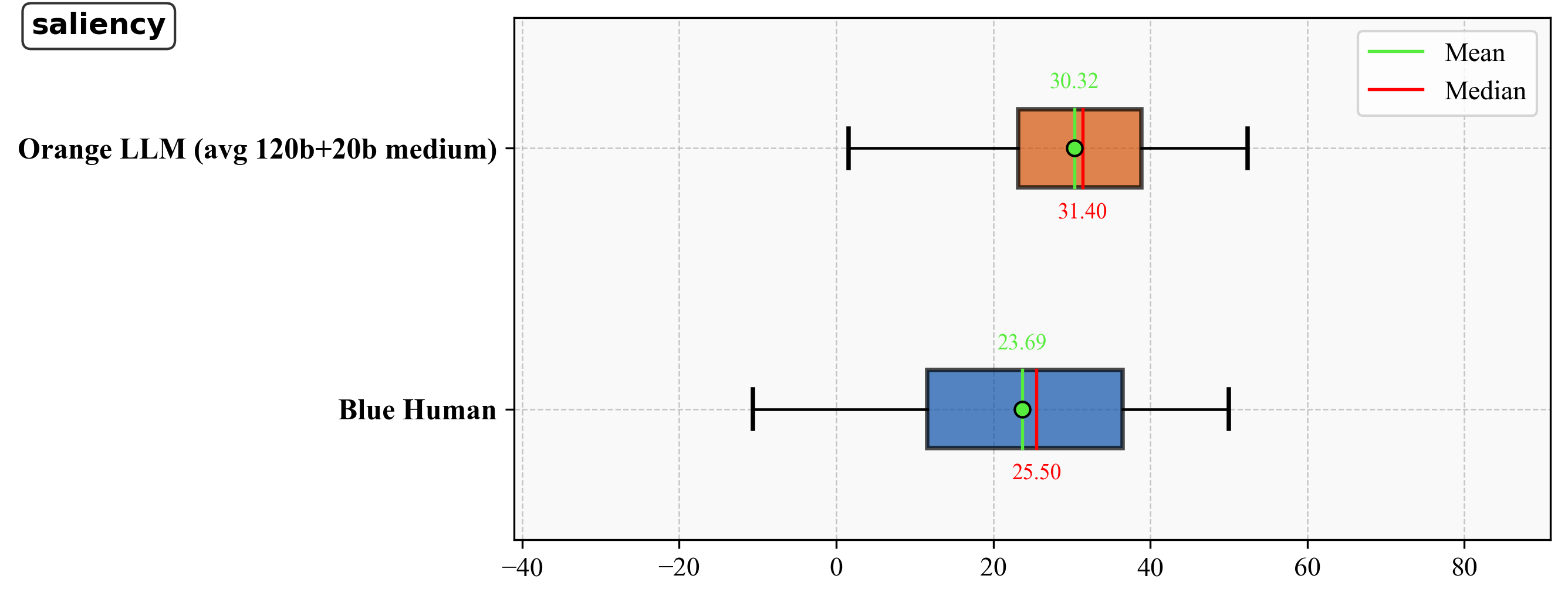} &
\includegraphics[width=0.325\textwidth]{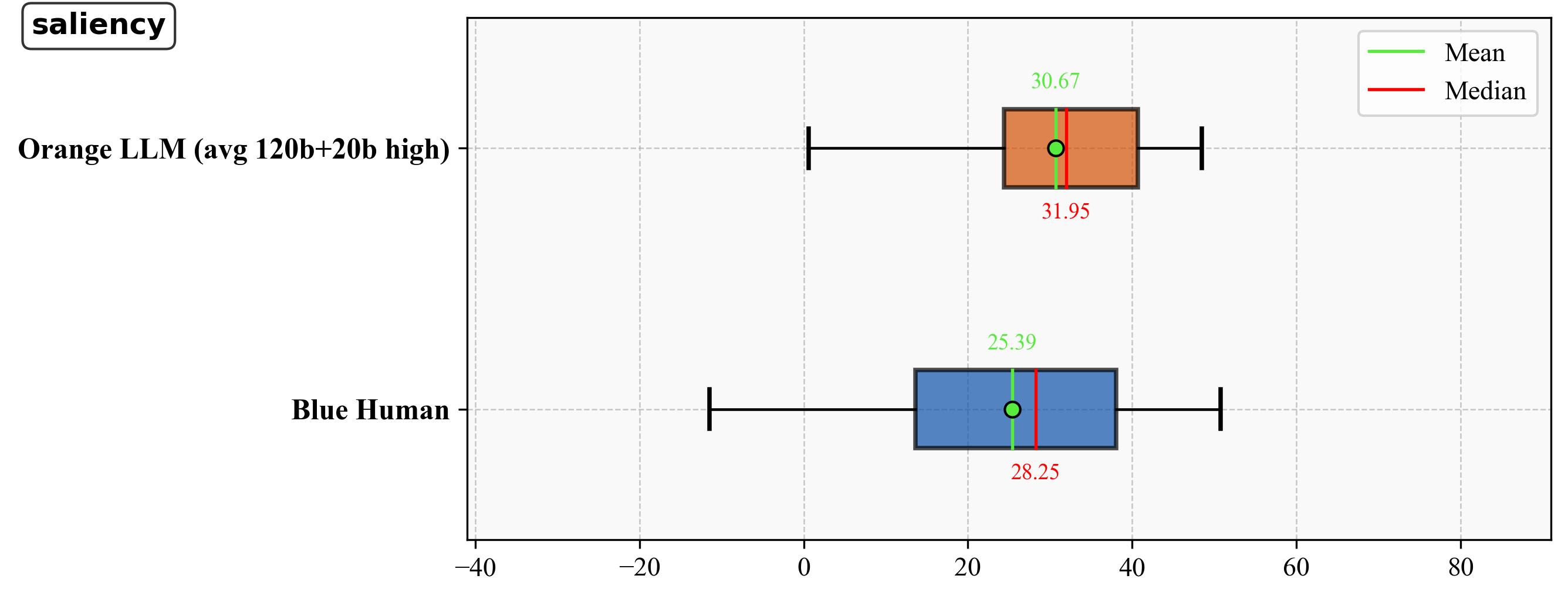} \\
\includegraphics[width=0.325\textwidth]{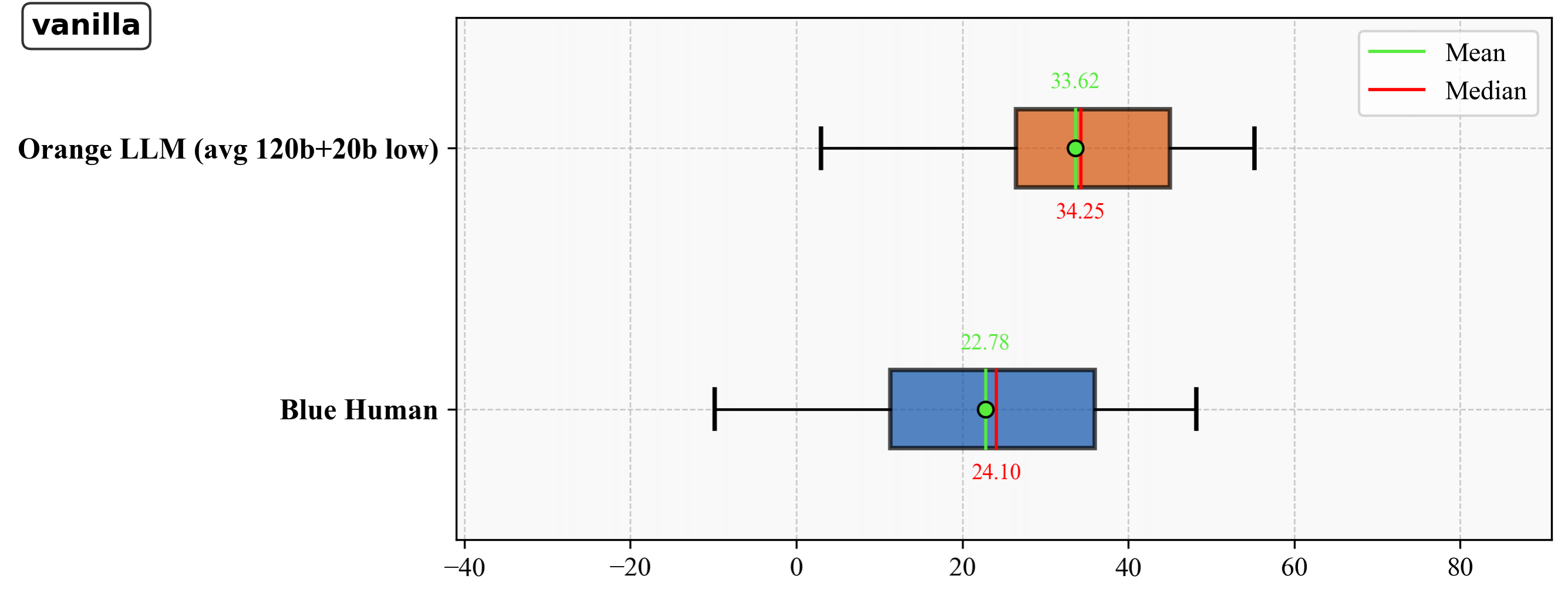} &
\includegraphics[width=0.325\textwidth]{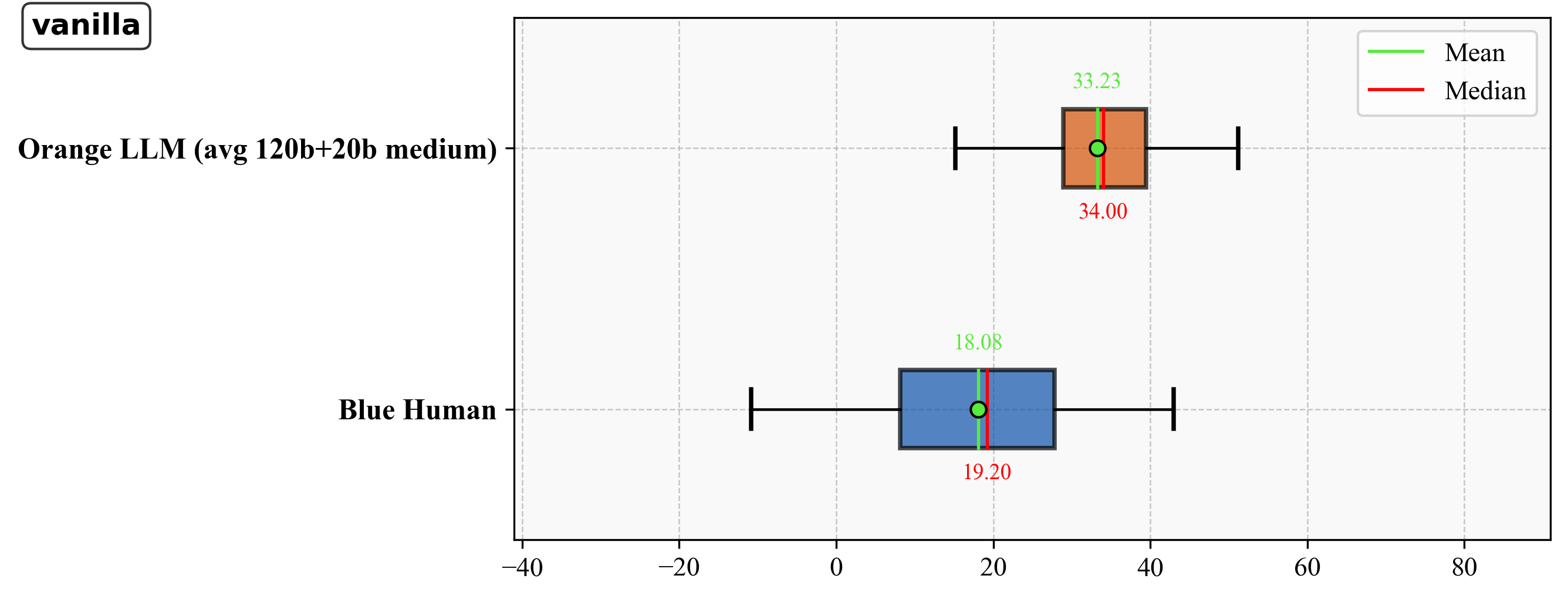} &
\includegraphics[width=0.325\textwidth]{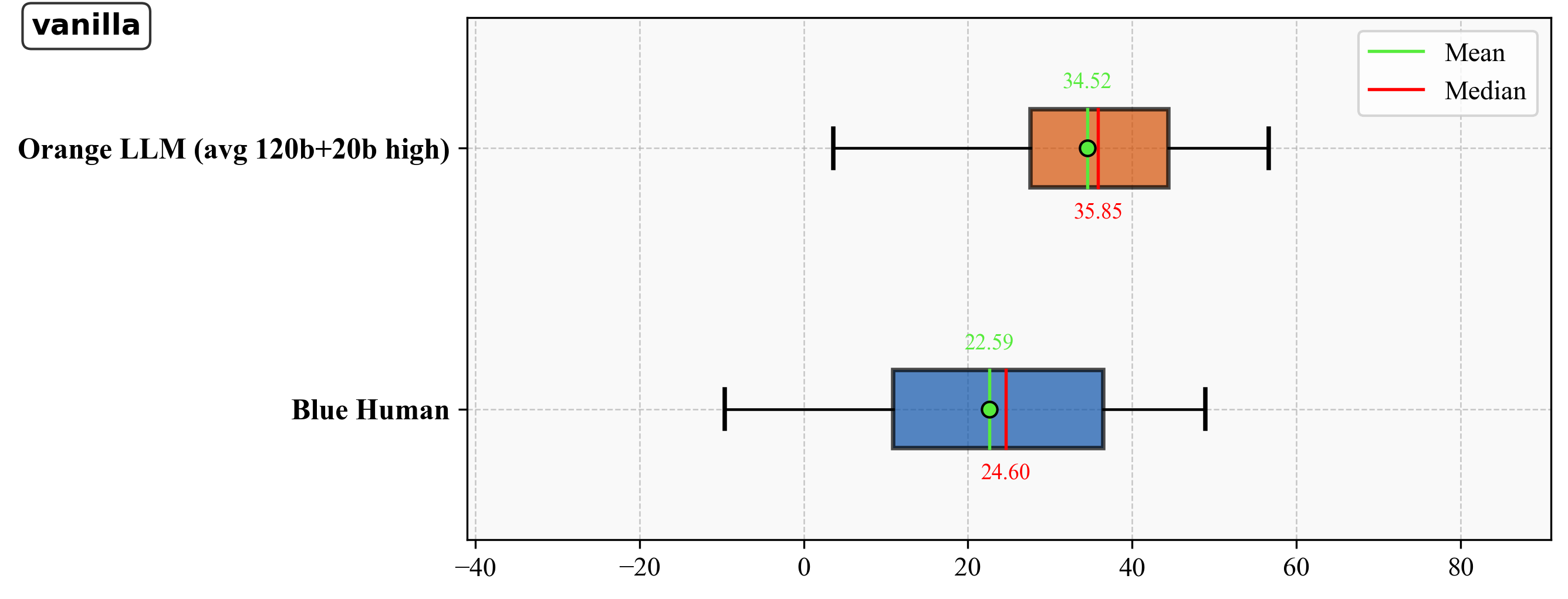} \\
\end{tabular}
\vspace{-2mm}
\caption{Bargaining Table averages, grouped by reasoning level. Columns correspond to low, medium, and high reasoning; rows correspond to all-features, cooperative, greedy, saliency, and vanilla (averaged over gpt-oss-20b and gpt-oss-120b). The LLM is the orange player.}
\label{fig:bt-avg-by-reasoning-p2-llm-low-med-high}
\end{figure*}

\paragraph{Comparison by reasoning level}
Referencing the main paper's results, averaging over reasoning levels (low, medium, high), Table~\ref{tab:bargaining-by-reasoning} (bottom) presents the mean and median scores by variant for the orange (human) and blue (LLM) players. \textbf{Reasoning level primarily modulates Orange}, with \textbf{weaker and more variant-dependent changes for Blue}. For Orange, \texttt{greedy} increases substantially from low to medium/high (means \(21.05 \rightarrow 24.25/24.48\)), while \texttt{all-features} decreases with higher reasoning (means \(21.19 \rightarrow 19.96 \rightarrow 18.55\)). \texttt{saliency} is robust for Orange, remaining consistently high across reasoning levels (means \(30.56,\,29.69,\,29.30\)). For Blue, shifts are comparatively modest: \texttt{vanilla} exhibits a mild improvement with higher reasoning (mean \(28.20 \rightarrow 28.32 \rightarrow 28.98\), median peaking at high: \(32.35\)), whereas under \texttt{saliency} Blue is near-flat in the mean with a small gain at high reasoning (mean \(26.06 \rightarrow 26.85\); median peaking at medium: \(28.55\)). All graphs are shown in Figures~\ref{fig:bt-avg-by-reasoning-p1-llm-low-med-high} and~\ref{fig:bt-avg-by-reasoning-p2-llm-low-med-high}.

\smallskip
\noindent\textbf{Improvements relative to \texttt{vanilla}.}
For Orange, \texttt{saliency} yields the largest gains over \texttt{vanilla} at every reasoning level (low: \(24.88 \rightarrow 30.56\), \(+5.68\); medium: \(26.75 \rightarrow 29.69\), \(+2.94\); high: \(24.61 \rightarrow 29.30\), \(+4.69\)). For Blue, \texttt{cooperative} provides the strongest improvement over \texttt{vanilla} across reasoning levels (low: \(28.20 \rightarrow 29.36\), \(+1.16\); medium: \(28.32 \rightarrow 29.72\), \(+1.40\); high: \(28.98 \rightarrow 29.46\), \(+0.48\)), while \texttt{saliency} remains below \texttt{vanilla} in mean Blue score at all levels despite the slight increase from low to high.

\subsubsection{Results with Other Models}\label{a:other-models-bargaining}
\begin{figure*}[t]
\centering
\setlength{\tabcolsep}{1pt}
\renewcommand{\arraystretch}{0}
\begin{tabular}{@{}cc@{}}
\includegraphics[width=0.49\textwidth]{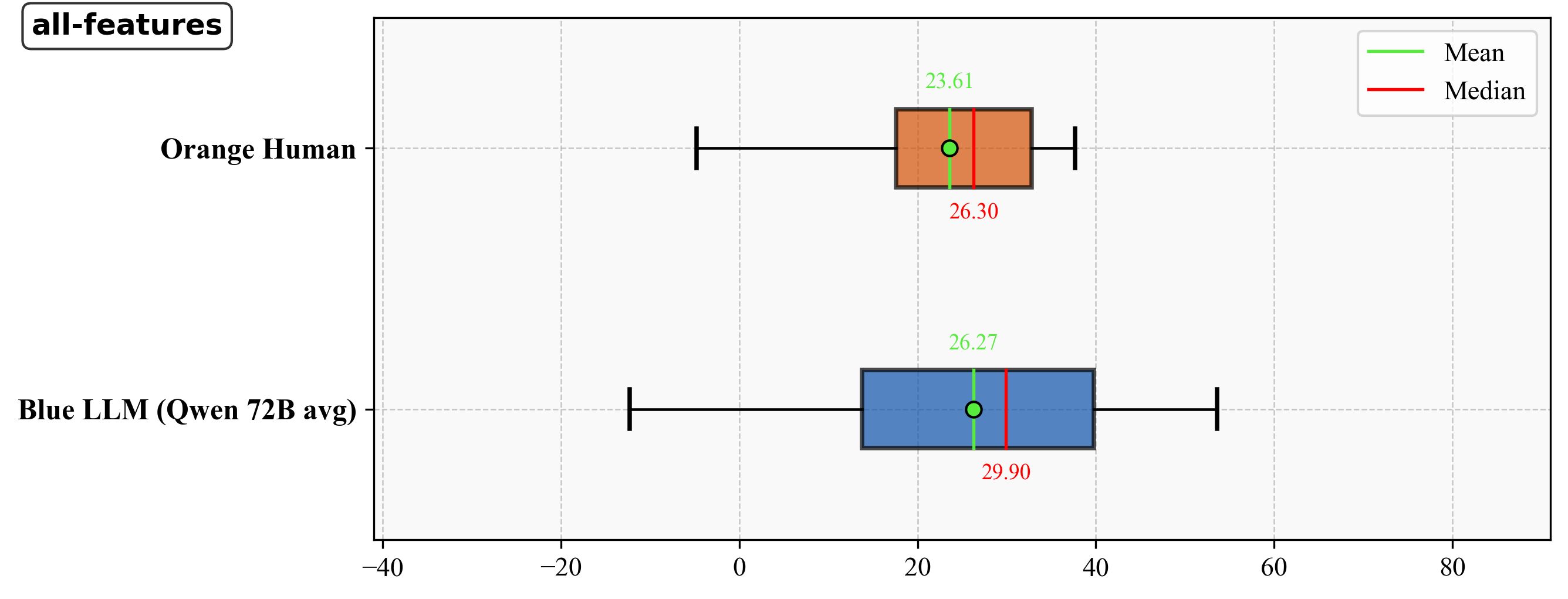} &
\includegraphics[width=0.49\textwidth]{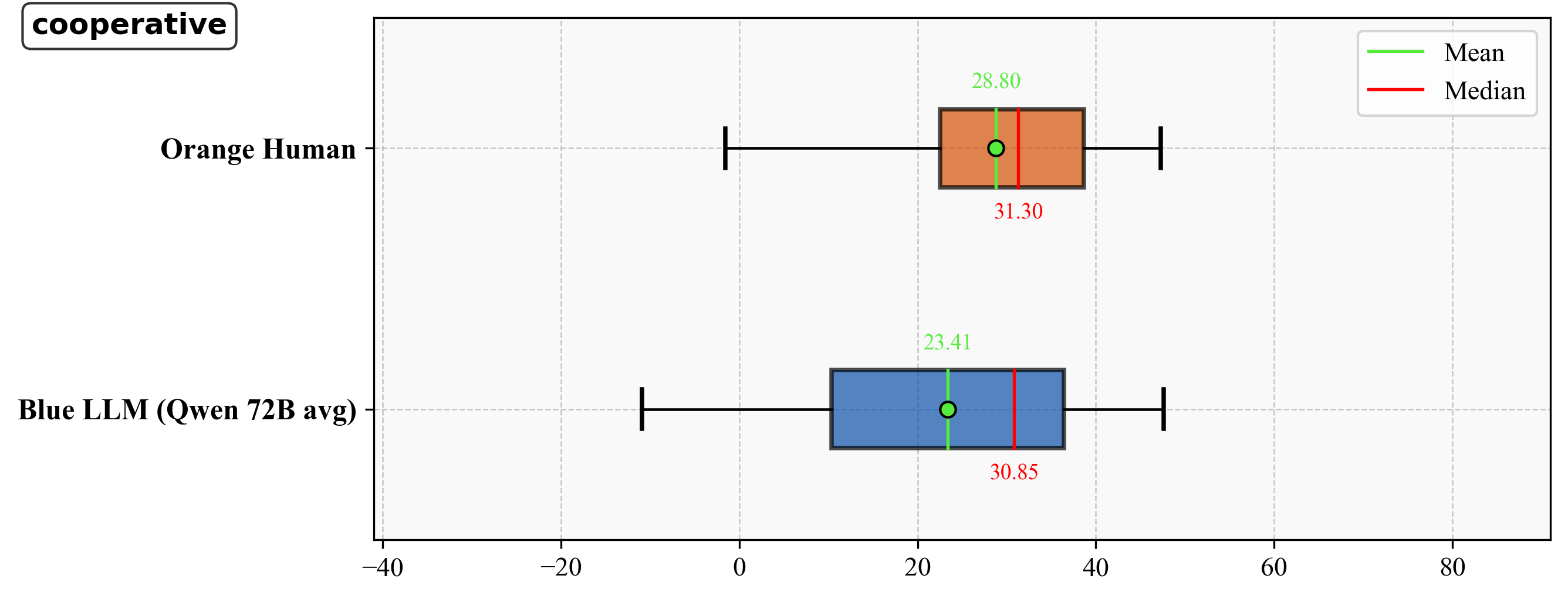} \\
\includegraphics[width=0.49\textwidth]{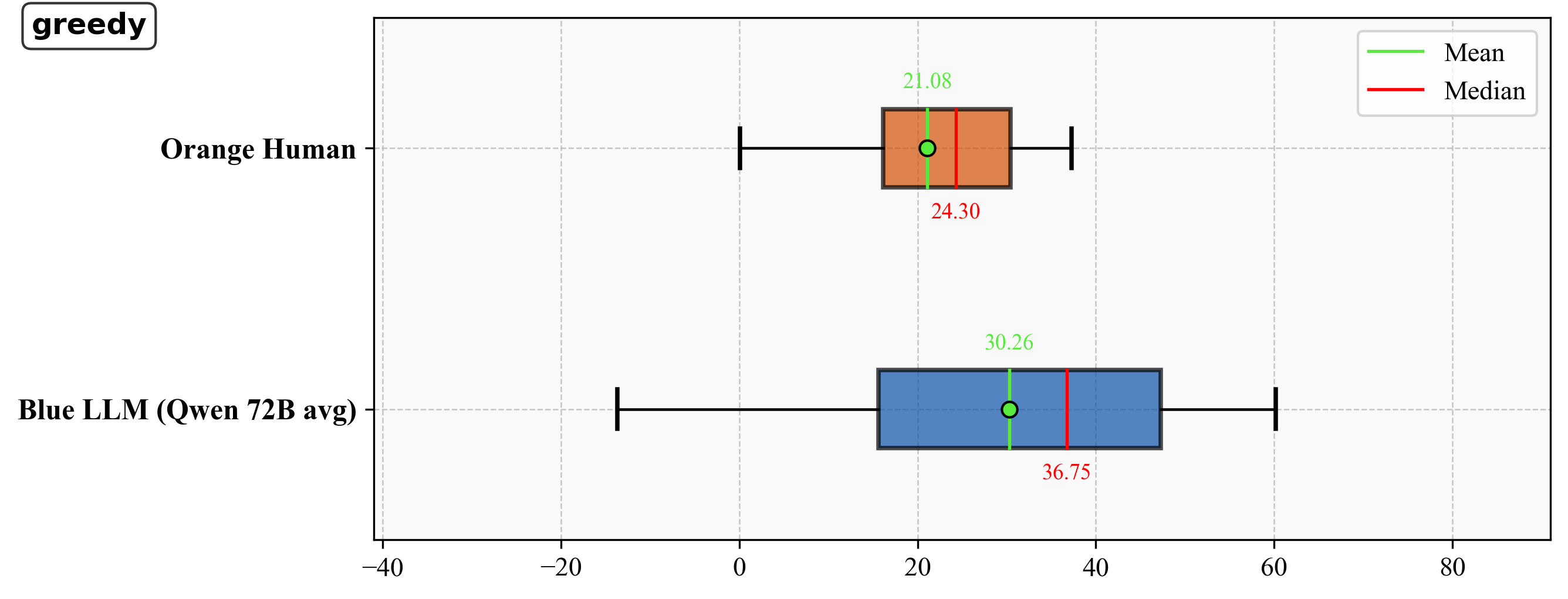} &
\includegraphics[width=0.49\textwidth]{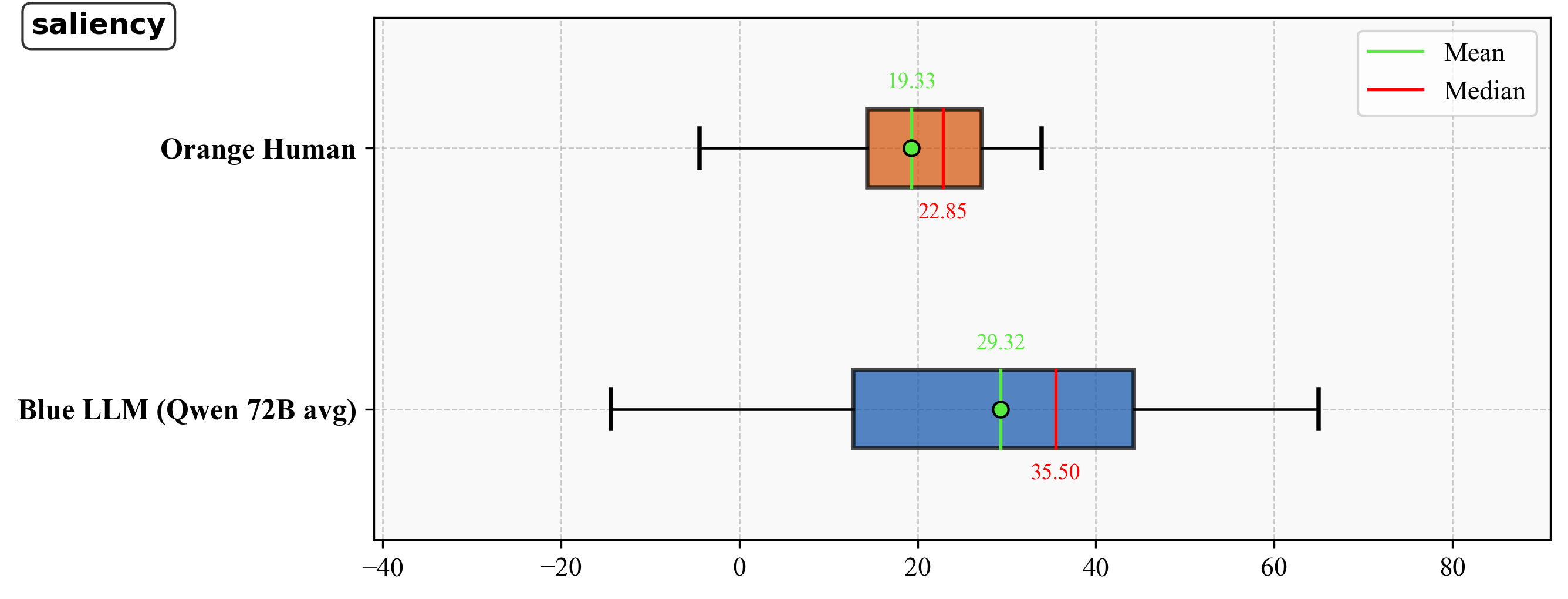} \\
\multicolumn{2}{c}{\includegraphics[width=0.49\textwidth]{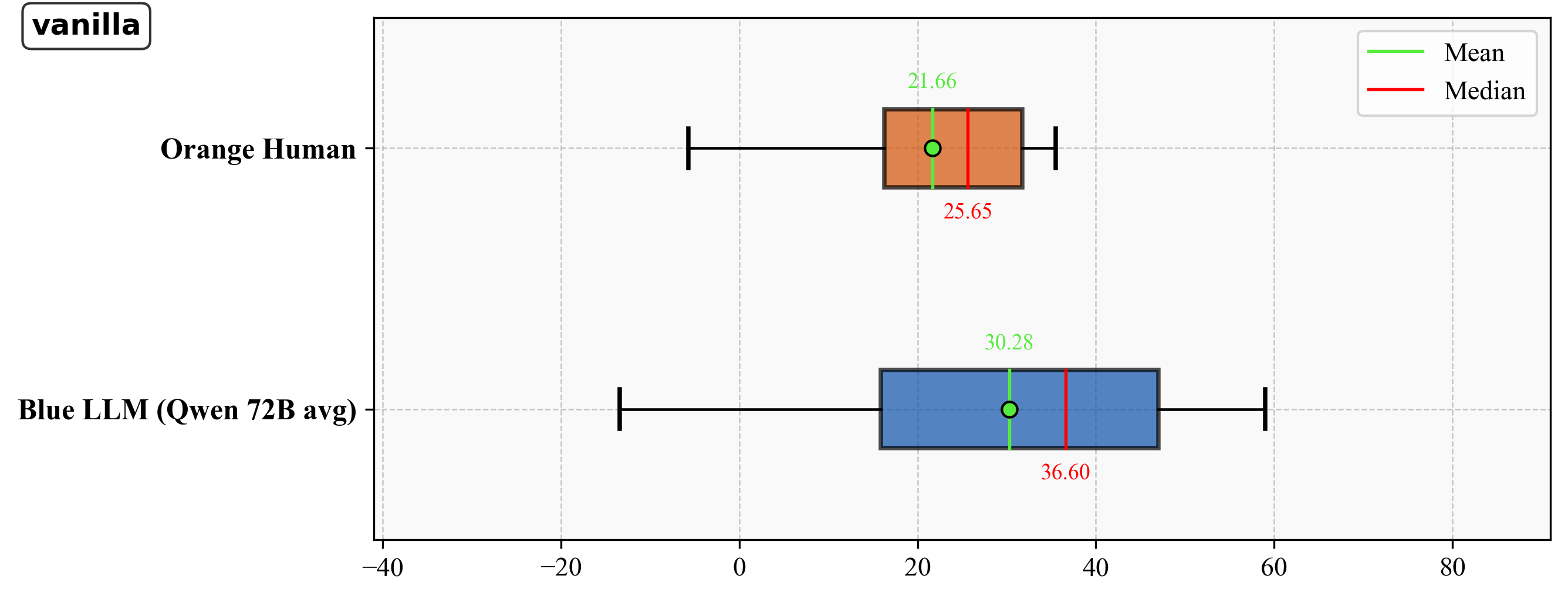}} \\
\end{tabular}
\vspace{-2mm}
\caption{Bargaining Table averages, averaged over the Qwen-72B family: Qwen2-72B-Instruct and Qwen2.5-72B-Instruct. Variants shown are all-features, cooperative, greedy, saliency, and vanilla. The LLM is the blue player.}
\label{fig:bt-avg-qwen72b-p1-llm}
\end{figure*}

\begin{figure*}[t]
\centering
\setlength{\tabcolsep}{1pt}
\renewcommand{\arraystretch}{0}
\begin{tabular}{@{}cc@{}}
\includegraphics[width=0.49\textwidth]{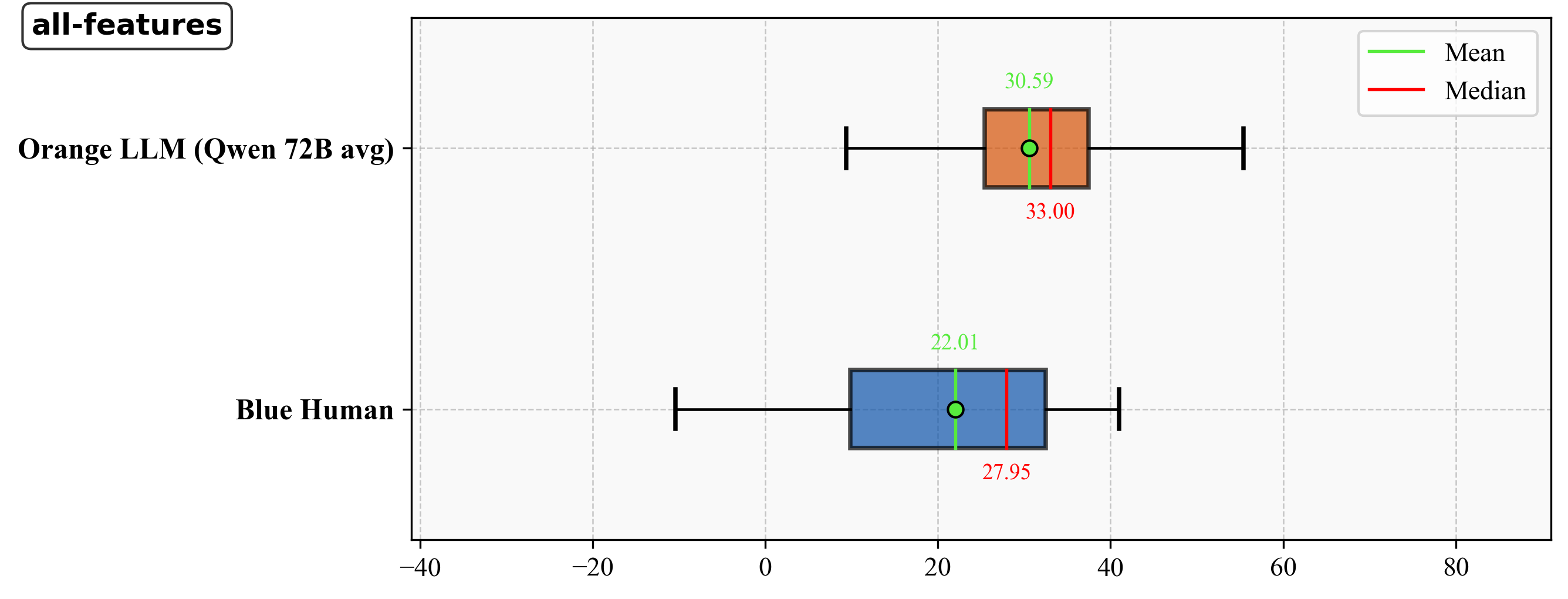} &
\includegraphics[width=0.49\textwidth]{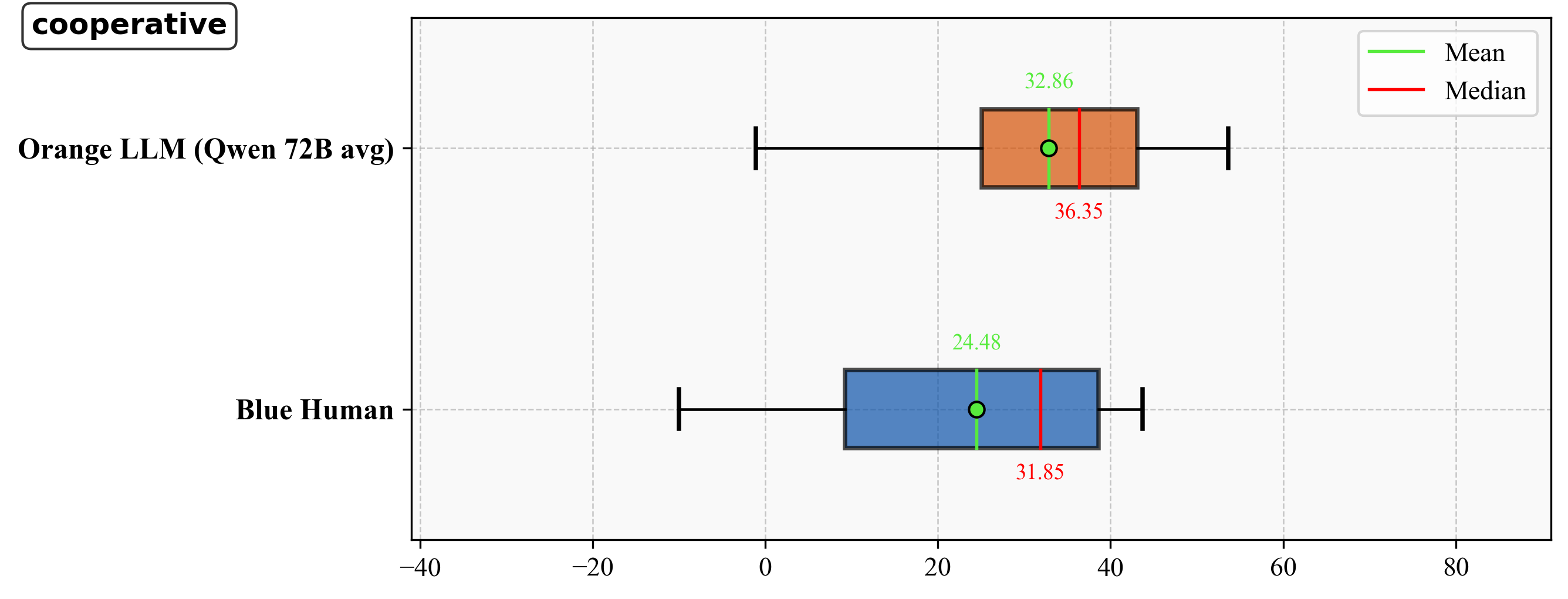} \\
\includegraphics[width=0.49\textwidth]{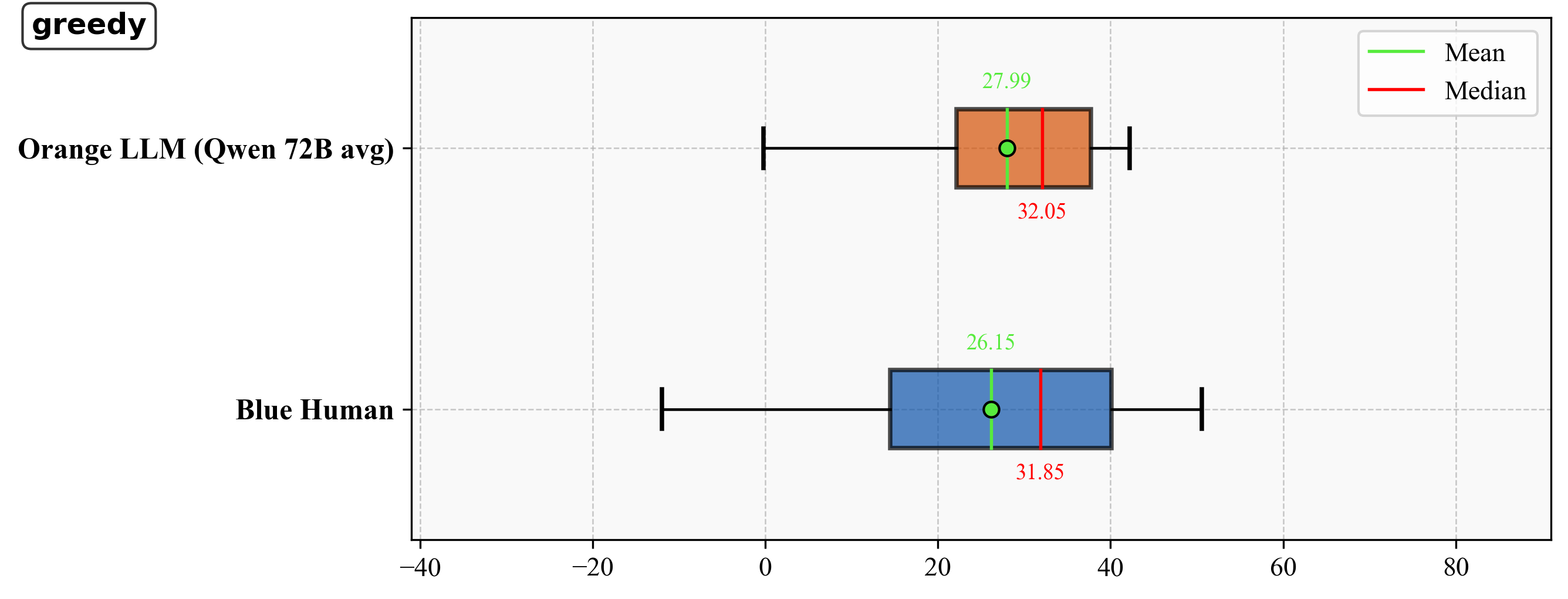} &
\includegraphics[width=0.49\textwidth]{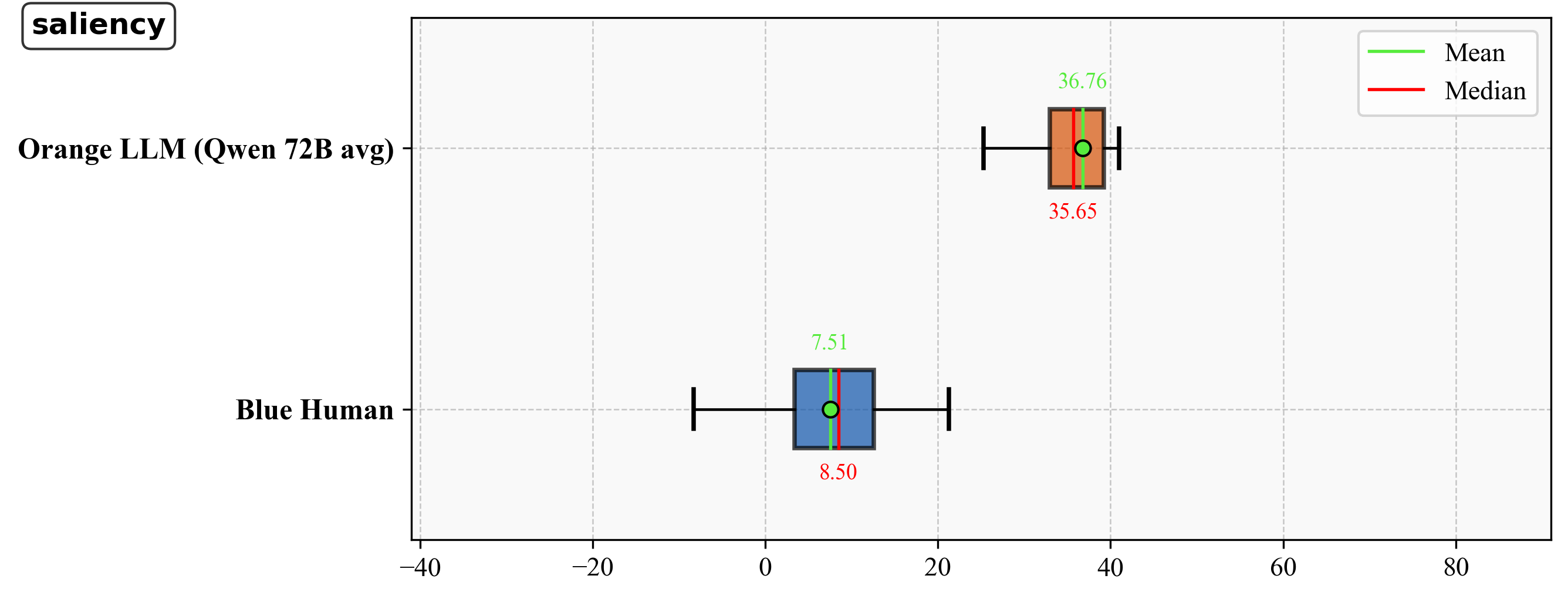} \\
\multicolumn{2}{c}{\includegraphics[width=0.49\textwidth]{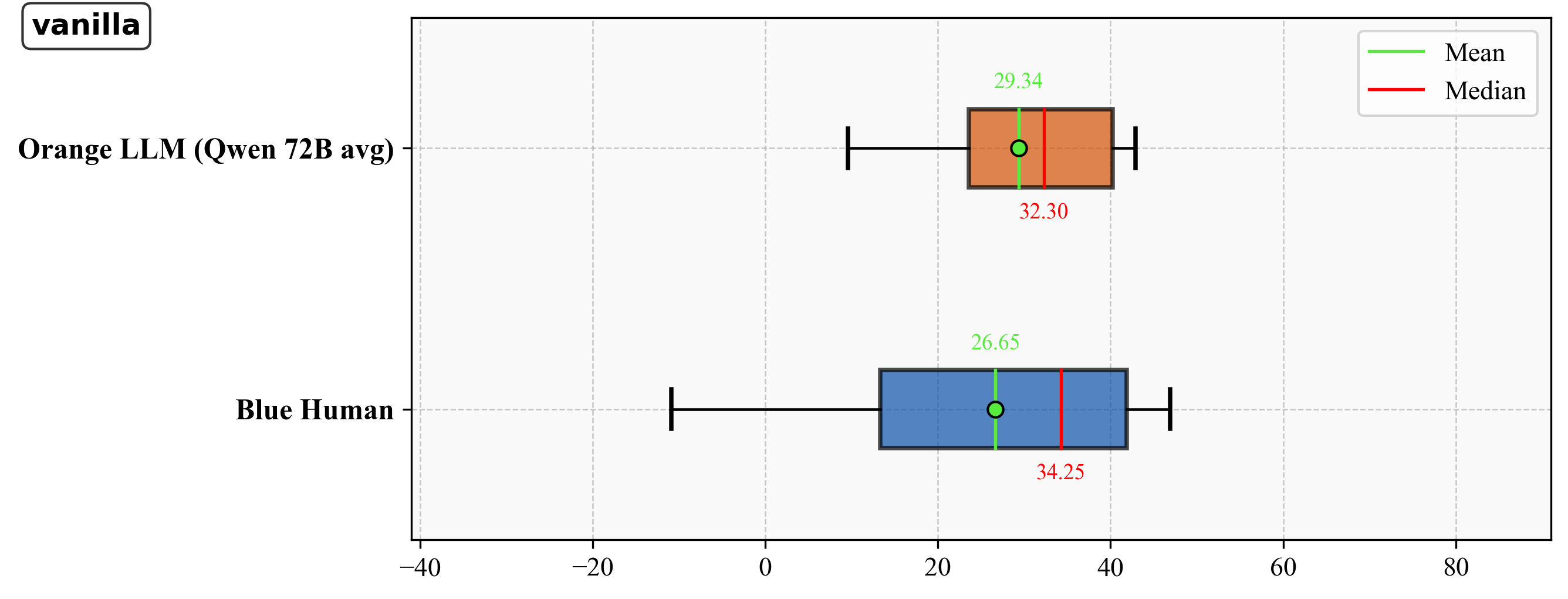}} \\
\end{tabular}
\vspace{-2mm}
\caption{Bargaining Table averages, averaged over the Qwen-72B family: Qwen2-72B-Instruct and Qwen2.5-72B-Instruct. Variants shown are all-features, cooperative, greedy, saliency, and vanilla. The LLM is the orange player.}
\label{fig:bt-avg-qwen72b-p2-llm}
\end{figure*}

The Bargaining Table averaged results for the Qwen-72B family (Qwen2-72B-Instruct and Qwen2.5-72B-Instruct), where the LLM is assigned either the blue player (P1) or the orange player (P2), are shown in Figures~\ref{fig:bt-avg-qwen72b-p1-llm} and~\ref{fig:bt-avg-qwen72b-p2-llm}.

\begin{figure*}[t]
\centering
\setlength{\tabcolsep}{1pt}
\renewcommand{\arraystretch}{0}
\begin{tabular}{@{}cc@{}}
\includegraphics[width=0.49\textwidth]{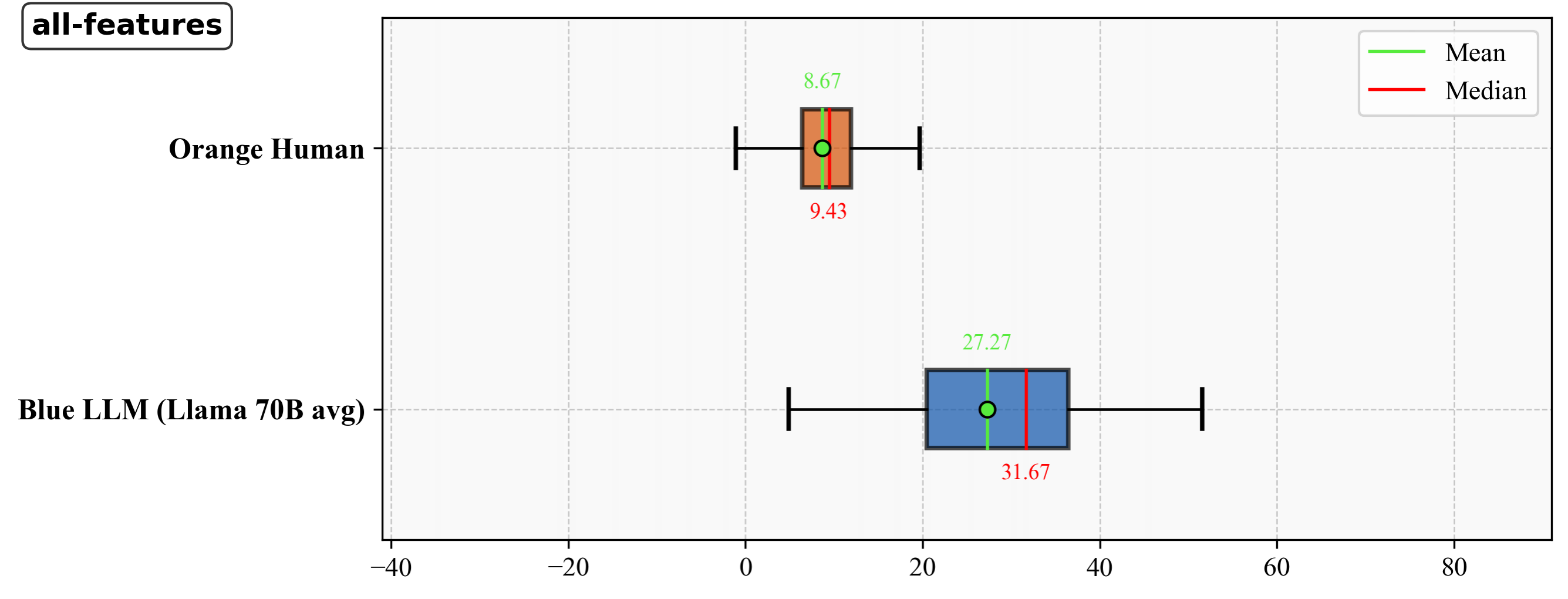} &
\includegraphics[width=0.49\textwidth]{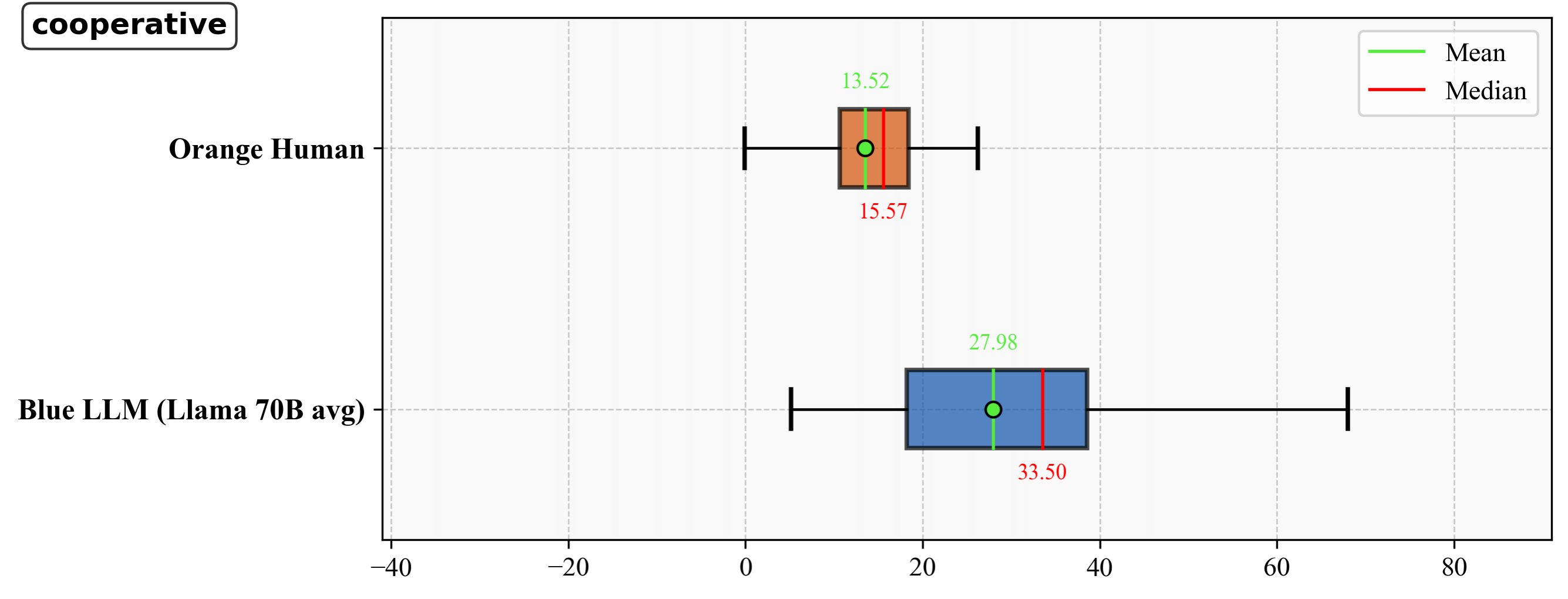} \\
\includegraphics[width=0.49\textwidth]{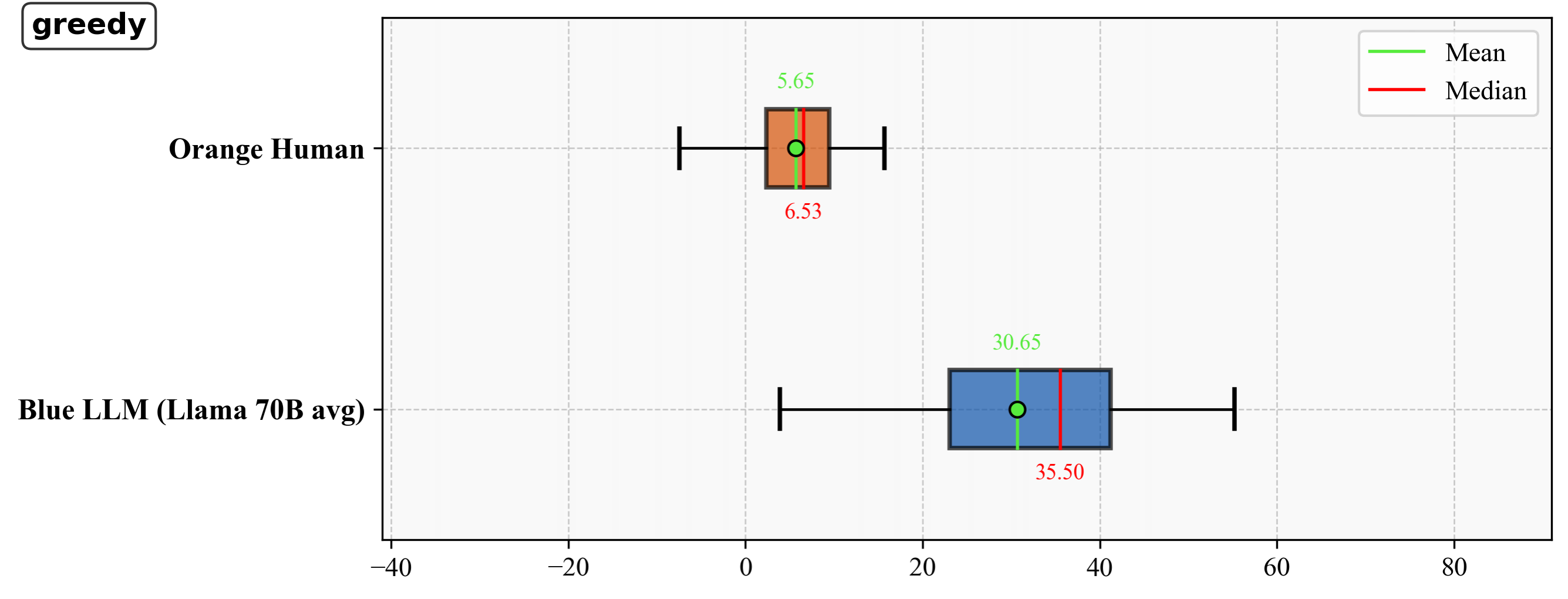} &
\includegraphics[width=0.49\textwidth]{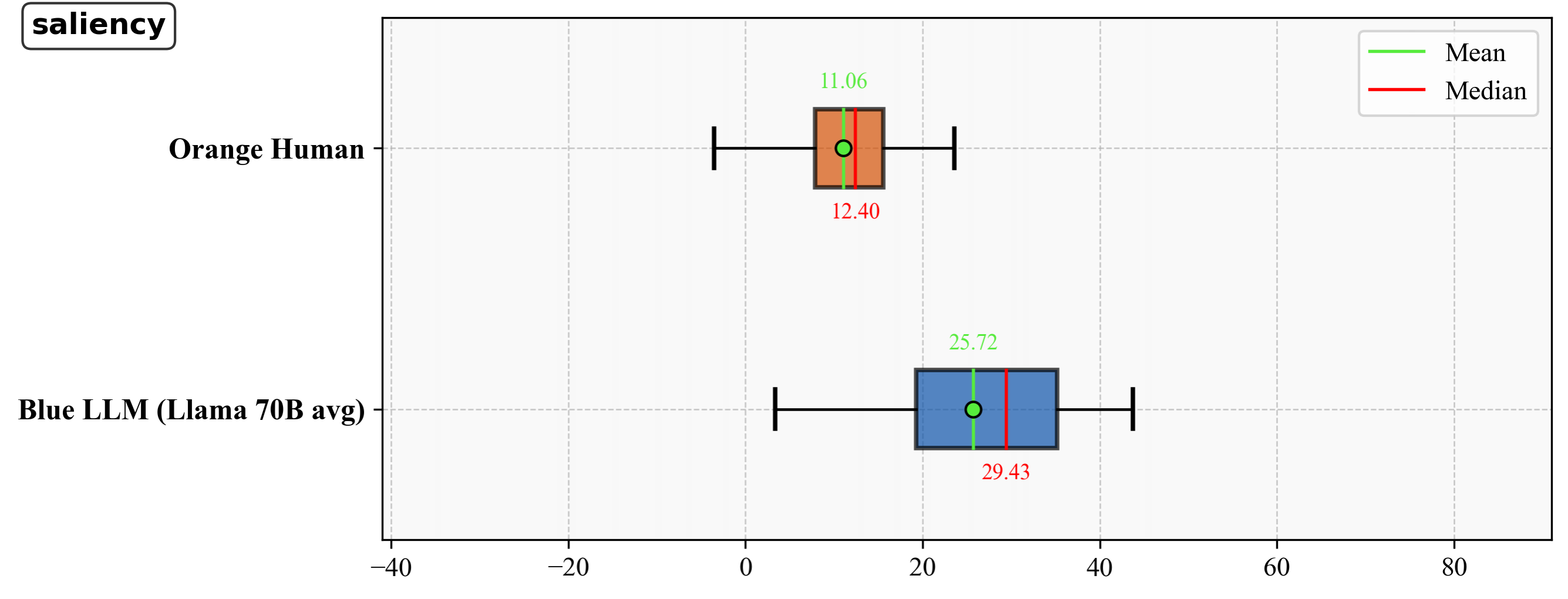} \\
\multicolumn{2}{c}{\includegraphics[width=0.49\textwidth]{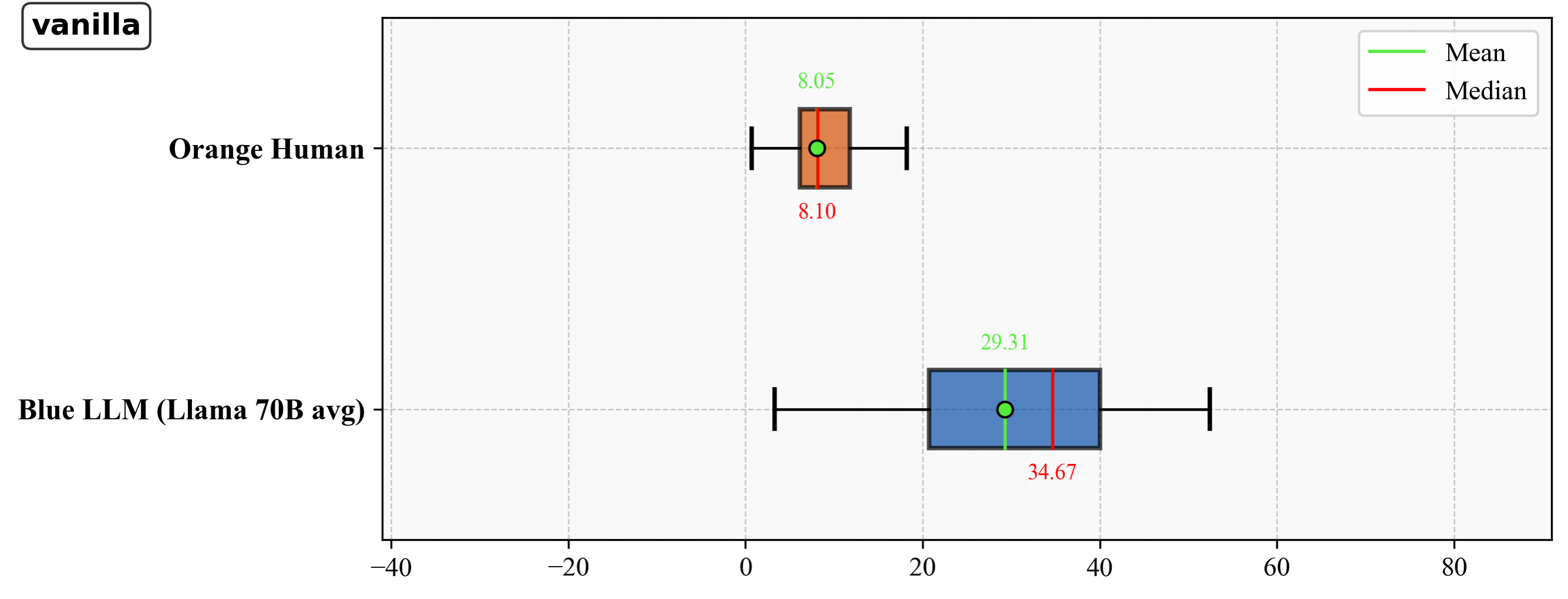}} \\
\end{tabular}
\vspace{-2mm}
\caption{Bargaining Table averages, averaged over the Llama-70B family: Llama-3.1-70B-Instruct, Llama-3.3-70B-Instruct, and Meta-Llama-3-70B-Instruct. Variants shown are all-features, cooperative, greedy, saliency, and vanilla. The LLM is the blue player.}
\label{fig:bt-avg-llama70b-p1-llm}
\end{figure*}

\begin{figure*}[t]
\centering
\setlength{\tabcolsep}{1pt}
\renewcommand{\arraystretch}{0}
\begin{tabular}{@{}cc@{}}
\includegraphics[width=0.49\textwidth]{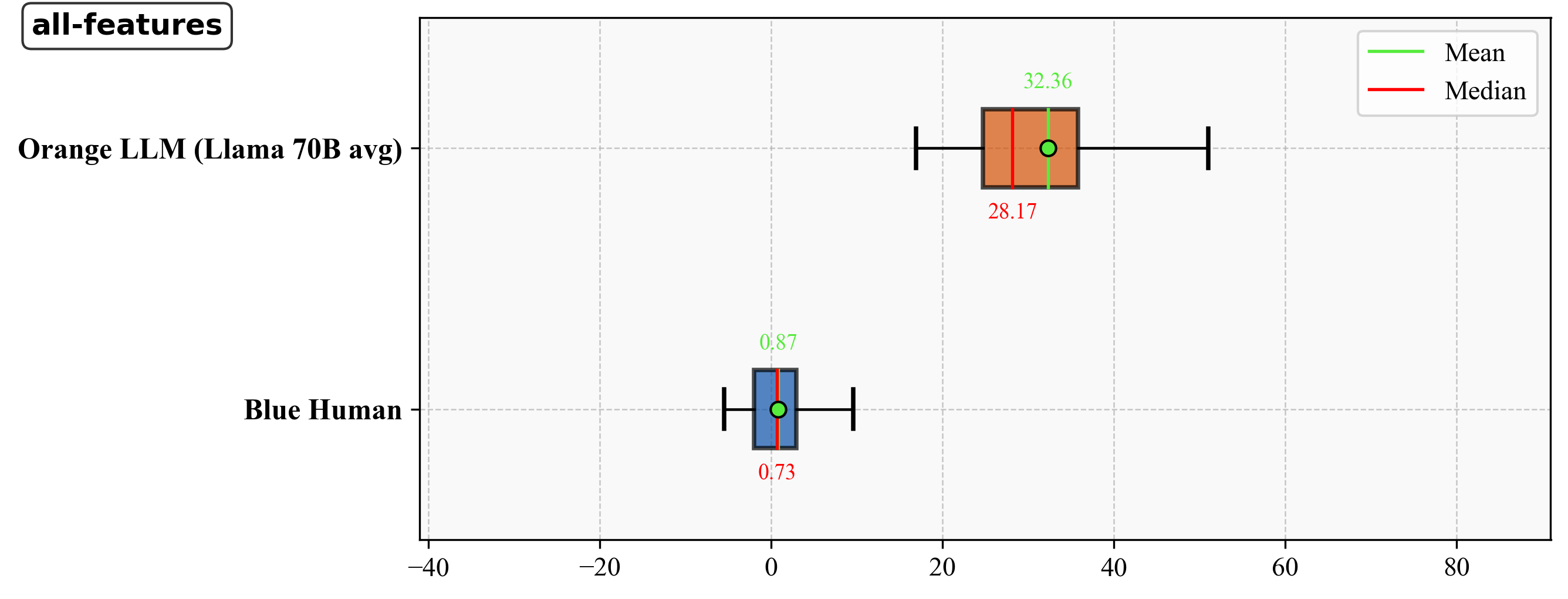} &
\includegraphics[width=0.49\textwidth]{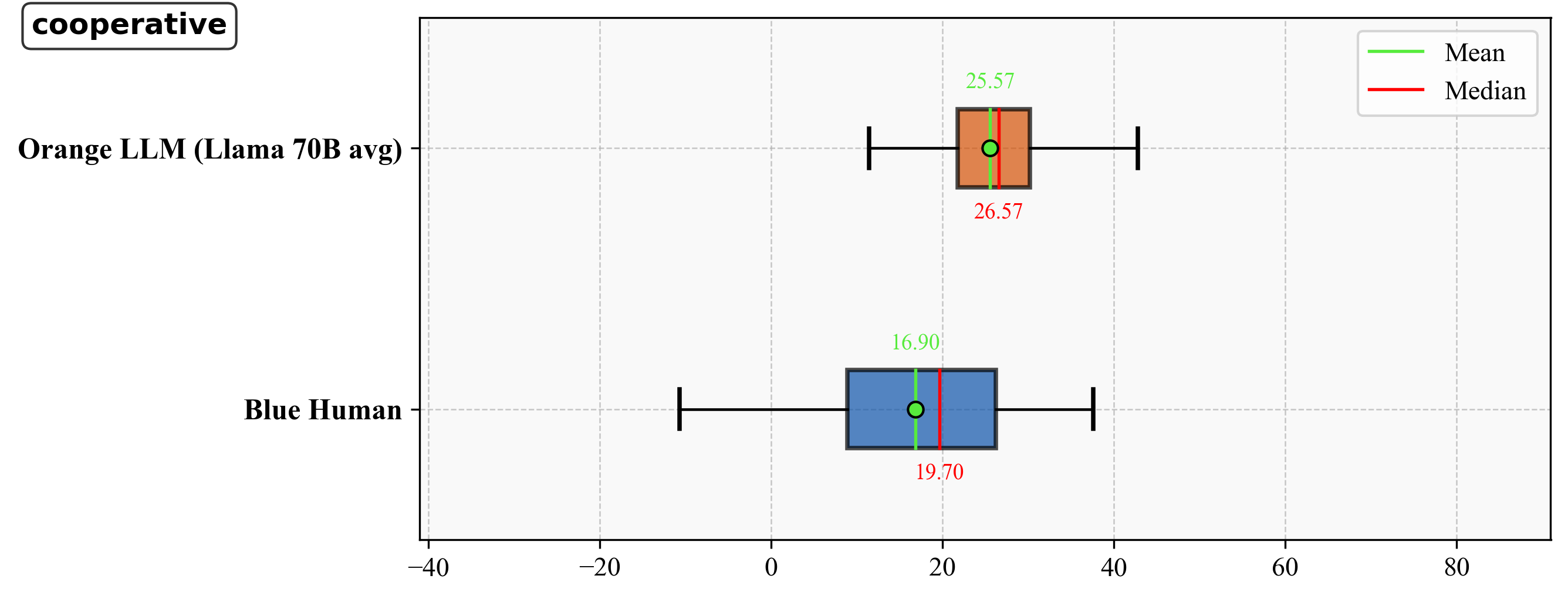} \\
\includegraphics[width=0.49\textwidth]{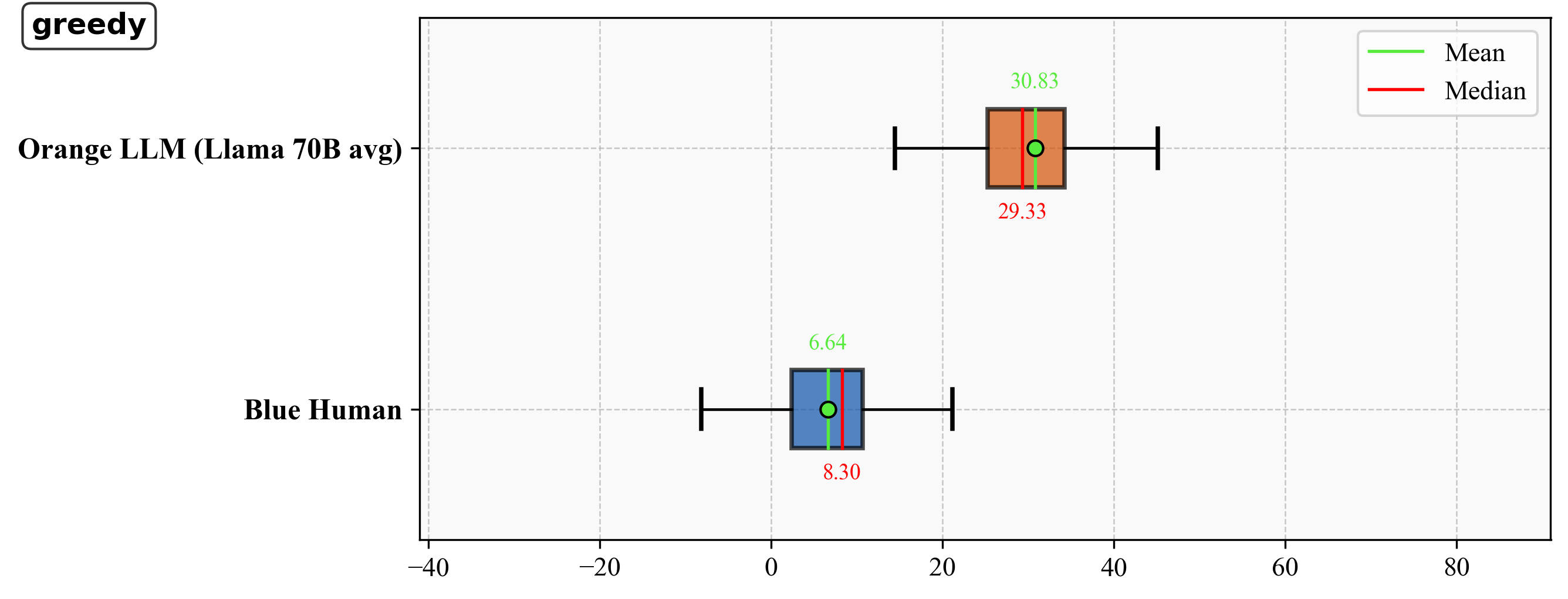} &
\includegraphics[width=0.49\textwidth]{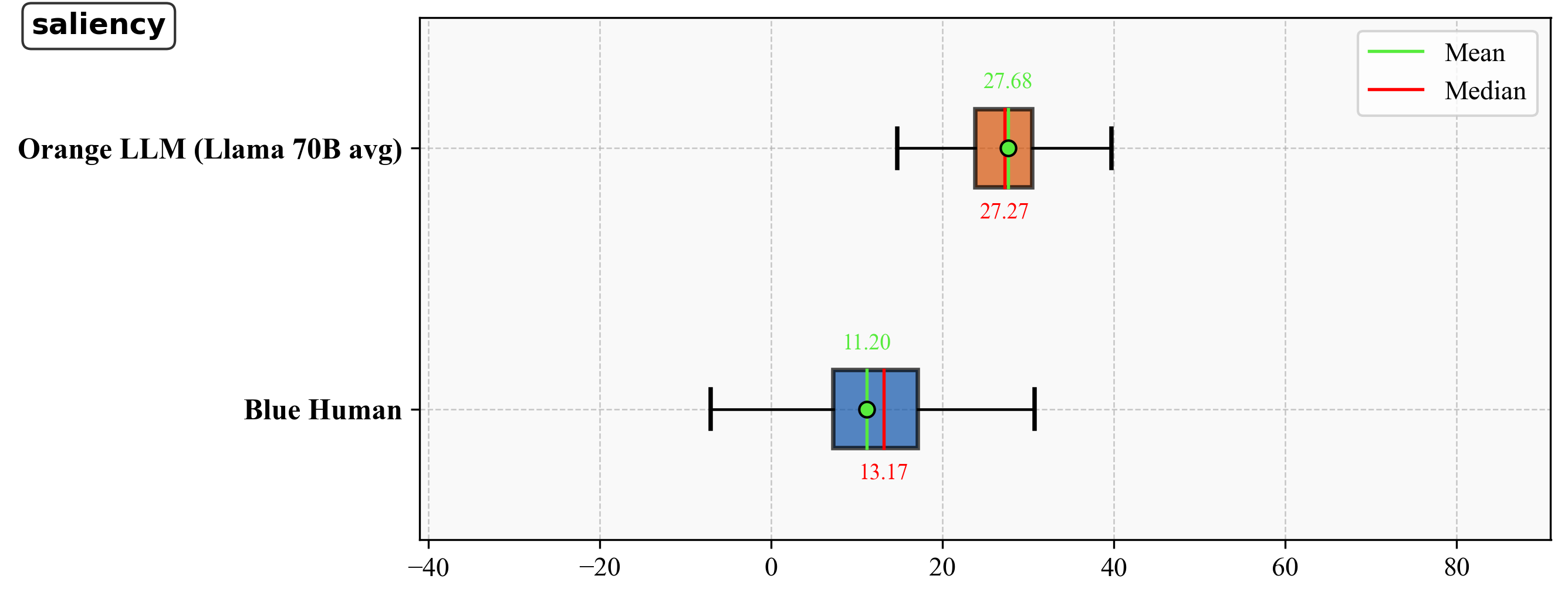} \\
\multicolumn{2}{c}{\includegraphics[width=0.49\textwidth]{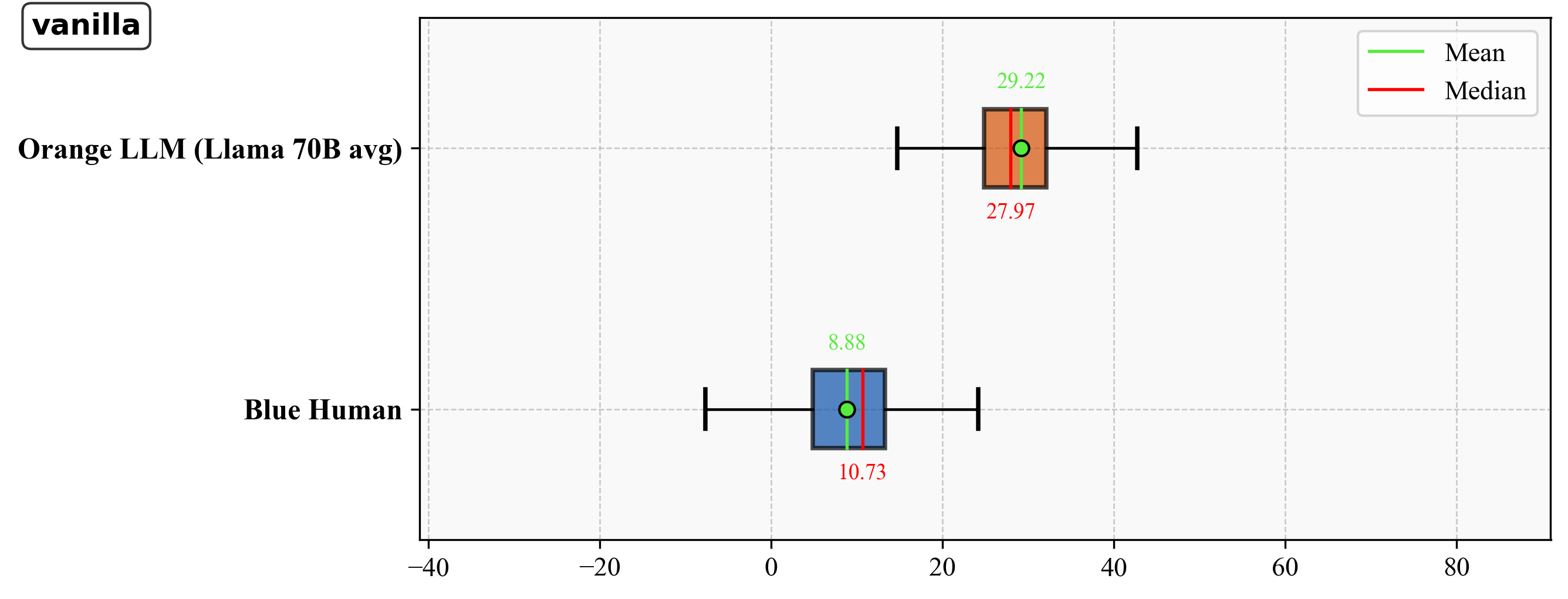}} \\
\end{tabular}
\vspace{-2mm}
\caption{Bargaining Table averages, averaged over the Llama-70B family: Llama-3.1-70B-Instruct, Llama-3.3-70B-Instruct, and Meta-Llama-3-70B-Instruct. Variants shown are all-features, cooperative, greedy, saliency, and vanilla. The LLM is the orange player.}
\label{fig:bt-avg-llama70b-p2-llm}
\end{figure*}

The Bargaining Table averaged results for the Llama-70B family (Llama-3.1-70B-Instruct, Llama-3.3-70B-Instruct, and Meta-Llama-3-70B-Instruct), for both blue and orange players as the LLM, are shown in Figures~\ref{fig:bt-avg-llama70b-p1-llm} and~\ref{fig:bt-avg-llama70b-p2-llm}.

Across both the Llama and Qwen settings, the GPT-OSS model family achieves higher scores for all variants except vanilla. Under vanilla prompting, GPT-OSS models exhibit more cooperative behaviour and consequently appropriate slightly fewer disks for themselves, whereas Llama and Qwen obtain marginally higher scores by behaving less cooperatively.

\subsubsection{Reproducibility of~\citet{mizrahi2020using}}\label{a:comparison-bargaining}
\begin{figure*}
    \centering
    \includegraphics[width=0.8\linewidth]{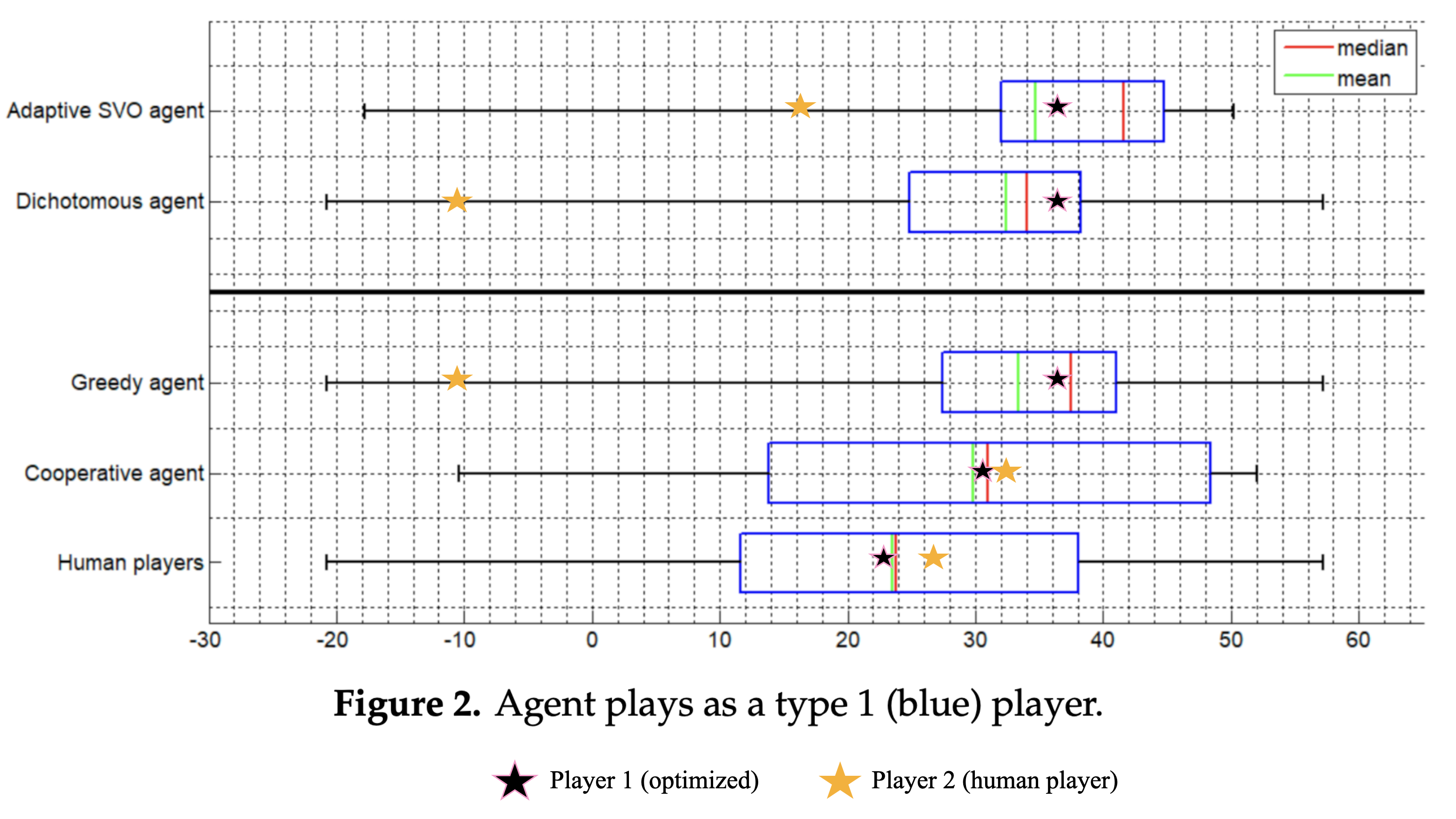}
    \caption{For reproducibility, we report the results (the stars) of our implementation of the techniques in~\citet{mizrahi2020using} on the Bargaining Table. The image and caption belong to~\cite{mizrahi2020using}.}
    \label{fig:appendix-comparison-dor}
\end{figure*}

Since our analysis of the Bargaining Table required us to recompute all the payoffs from the original data, we report how our experiments compare to those in~\citet{leland2018theory}. As reported in Figure~\ref{fig:appendix-comparison-dor}, our results are in line with those of the original paper.

\subsection{The Welfare Payoff of the Bargaining Table}
Table~\ref{tab:welfare-payoffs} reports the payoff of the Bargaining Table for each model, as well as the social welfare, i.e., the sum of the payoff of the two players.
Interesting additional analyses one can conduct are the Nash Social Welfare, which requires turning the payoffs to be strictly positive, and comparing the welfare in each game to the Price of Anarchy (i.e., the ratio between the optimal social welfare and the worst equilibrium welfare).

\clearpage 

\begin{figure*}
    \centering
    \includegraphics[width=1.1\linewidth]{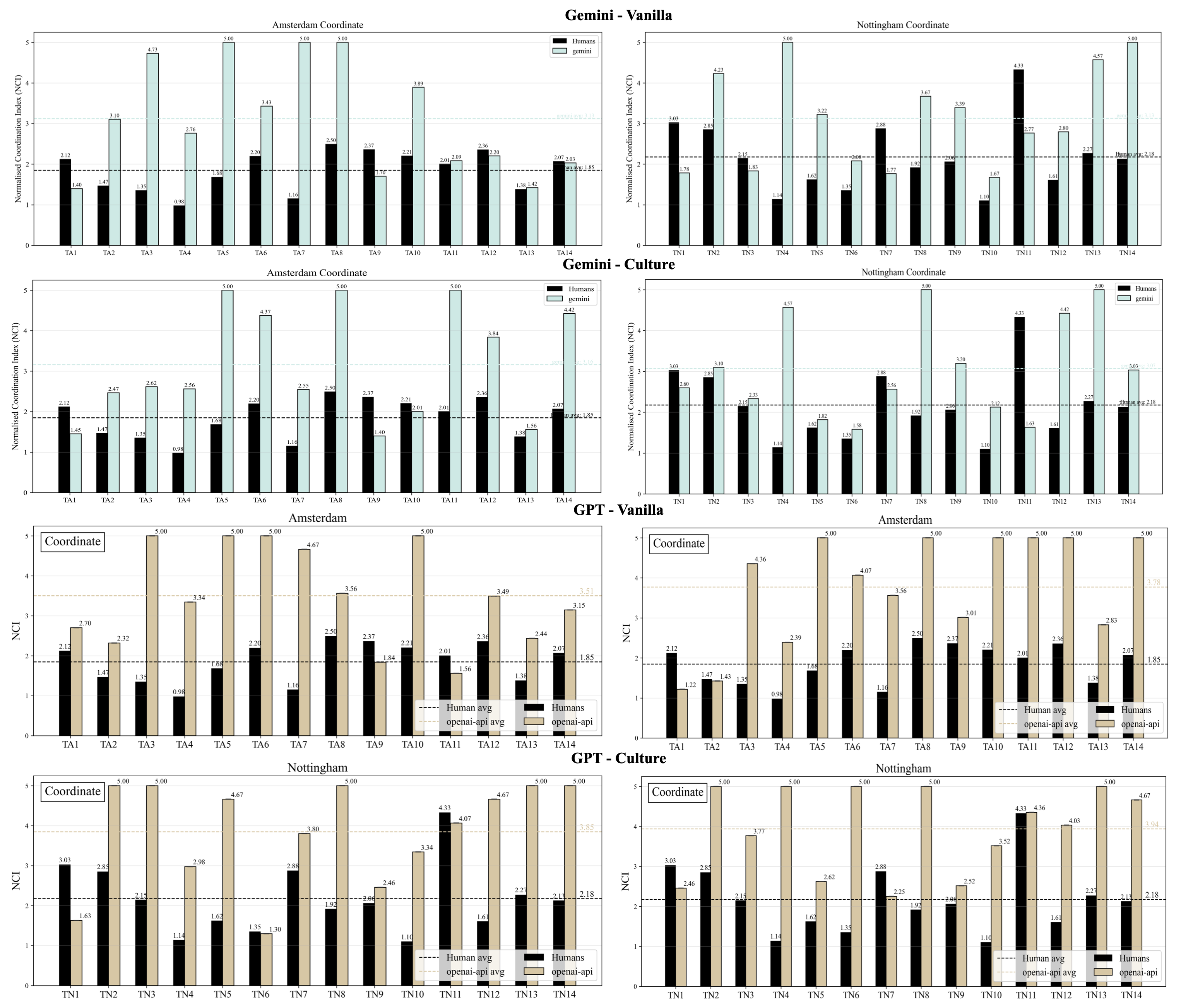}
    \caption{Normalised Coordination Index (NCI) of humans and all the GPT 5.4 and Gemini2.5-pro and Gemini3-pro models on the Amsterdam and Nottingham datasets. Full results in the code.}
    \label{fig:appendix-gpt-gemini}
\end{figure*}

\begin{table*}[ht]
\centering
\caption{Welfare payoffs: sum of Player 1 and Player 2 total payoffs (``Total payoff - across games'')}
\label{tab:welfare-payoffs}
\begin{tabular}{@{} l r r r @{}}
\toprule
Strategy & Player 1 total & Player 2 total & Welfare (sum) \\
\midrule
\texttt{Strategy: humans}            &  23.846 &  27.626 &  51.472 \\
\texttt{Strategy: p1\_greedy}       &  37.464 & -11.089 &  26.375 \\
\texttt{Strategy: p2\_greedy}       & -10.635 &  40.188 &  29.553 \\
\texttt{Strategy: both\_greedy}     & -20.800 & -20.800 & -41.600 \\
\texttt{Strategy: p1\_cooperative}  &  31.690 &  33.903 &  65.593 \\
\texttt{Strategy: p2\_cooperative}  &  38.493 &  37.250 &  75.743 \\
\texttt{Strategy: p1\_SVO}          &  36.668 & -11.222 &  25.446 \\
\texttt{Strategy: p2\_SVO}          & -10.622 &  40.268 &  29.646 \\
\texttt{Strategy: p1\_adaptive}     &  36.108 &  16.008 &  52.116 \\
\texttt{Strategy: p1\_llm}   &  29.117 &  31.707 &  60.824 \\
\texttt{Strategy: p2\_llm}          &  27.735 &  33.556 &  61.290 \\
\bottomrule
\end{tabular}
\end{table*}

\clearpage 

\subsection{Search \& Rescue: Task, Maps, Classification, and Evaluation}
\label{app:sar-details}
\label{a:additional-res-SAR}

This appendix provides the full methodological details for the search-and-rescue (SAR) experiment summarised in Section~\ref{sec:sar-main}. Wilderness SAR missions require teams to infer where a missing person may have moved after their last known location, often under uncertainty about route choice, terrain, physical condition, and decision-making. SAR is therefore a natural spatial analogue of focal-point reasoning: successful search depends on identifying terrain features, routes, landmarks, and movement affordances that might attract, constrain, or guide a missing person.

The experiment tests two related questions. First, can LLMs distinguish cases in which the missing person's realised movement ends near a salient terrain or rescue-relevant feature from cases more consistent with diffuse wandering? Second, when such focal structure exists, does explicitly prompting the model to attend to human movement, terrain salience, and rescue-relevant focal points improve localisation?

\subsubsection{SAR Location Prediction Task}
\label{app:sar-task}

We formulate SAR as a coordinate prediction task. Each instance consists of a real missing-hiker incident with an initial planning point (IPP), corresponding to the last known or initial search location, and a ground-truth find location. The model receives a terrain map centered on the IPP and predicts one coordinate at which the person is likely to be found. The find location is never shown during prediction.

The model output is constrained to a local metric coordinate $(x,y)$. The IPP is fixed at $(0,0)$, with $x$ denoting east--west displacement and $y$ denoting north--south displacement, both in meters. Thus, the model predicts a local offset rather than an absolute latitude and longitude.

We use the 65 missing-hiker incidents from \citet{hashimoto2022agent}. For each incident, the dataset provides latitude and longitude for both the IPP and the find location. We project these coordinates into an IPP-centered local metric frame to obtain the ground-truth displacement $(x_i^\ast,y_i^\ast)$ in meters. A model prediction $(\hat{x},\hat{y})$ is scored by its Euclidean distance from this projected target.

\subsubsection{Map Generation}
\label{app:sar-map-generation}

For each incident, we render a square terrain map centered on the IPP. The map combines elevation information with navigational and hydrological vector layers. Elevation cues are derived from USGS 3DEP data and rendered as grey contour and gradient features. Vector layers are obtained from OpenStreetMap via Overpass and include streams, riverbanks, roads, railways, hiking trails, lake interiors, river interiors, and lake and river shorelines. OpenStreetMap layers are \copyright{} OpenStreetMap contributors and are used under the Open Database License.\footnote{\url{https://www.openstreetmap.org/copyright}}

\begin{figure*}
    \centering
    \includegraphics[width=0.48\textwidth]{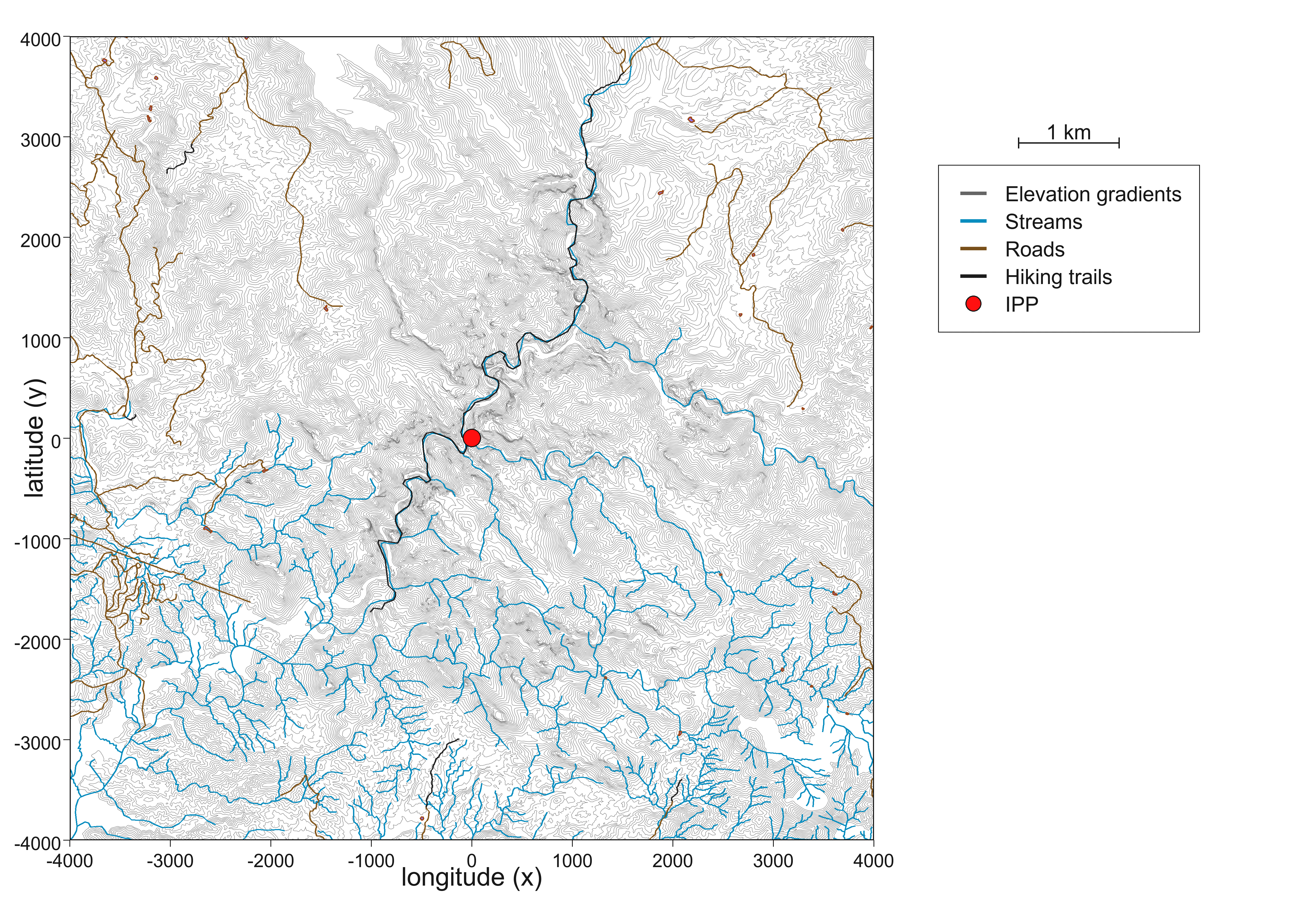}
    \hfill
    \includegraphics[width=0.48\textwidth]{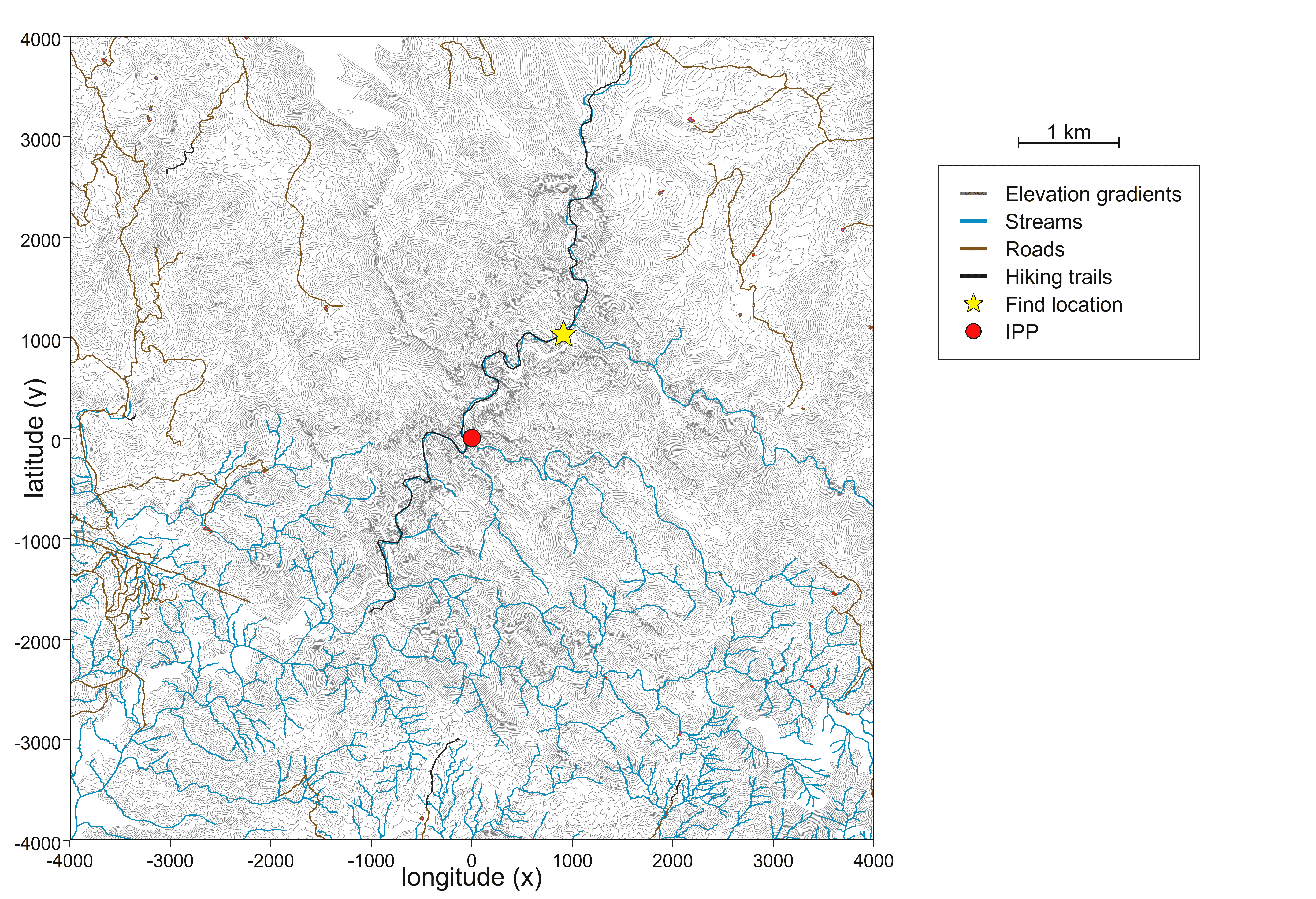}
    \caption{Two rendered versions of the same SAR incident. Left: the prediction map shown to the model, containing terrain and vector layers together with the red IPP marker. Right: the target-revealing map, which additionally overlays the yellow ground-truth find marker. The target-revealing version is used only for evaluation, classification, and visualisation.}
    \label{fig:sar-map-versions-app}
\end{figure*}

We generate two map variants for each incident. The \emph{prediction map}, used as model input, shows the terrain and vector layers together with a red IPP marker. The \emph{target-revealing map} additionally overlays the ground-truth find location as a yellow marker. Target-revealing maps are never used as prediction inputs; they are used only for focality classification, evaluation, and visualisation. Figure~\ref{fig:sar-map-versions-app} illustrates the two variants for one incident.

All maps include relative metric axes centered on the IPP, a scale bar, and a legend. The axes use the same local coordinate system as the model output and ground-truth displacement. Each axis has four labeled intervals per side plus a zero tick. Map width is selected separately for each incident so that the find location lies inside the frame with padding. Table~\ref{tab:sar-map-scales} reports the resulting map-width distribution.

\begin{table}
    \centering
    \small
    \begin{tabular}{lcc}
        \toprule
        Map width & Incidents & Share \\
        \midrule
        8 km  & 40 & 61.5\% \\
        12 km & 10 & 15.4\% \\
        16 km & 8  & 12.3\% \\
        20 km & 7  & 10.8\% \\
        \bottomrule
    \end{tabular}
    \caption{Distribution of map widths used in the SAR prediction task. Width is chosen per incident so that the ground-truth find location is inside the map frame with padding.}
    \label{tab:sar-map-scales}
\end{table}

\subsubsection{Focality Classification}
\label{app:sar-classification}

Before evaluating coordinate prediction, we classify incidents according to whether the observed find location is spatially focal. This classification separates two questions: whether the map contains salient search-relevant structure, and whether the actual find location lies at or near such structure.

Classification proceeds in two stages. In the first stage, the model sees the IPP-only prediction map and judges whether the terrain contains enough salient structure to support a non-random search hypothesis from the IPP. Relevant cues include trails, roads, streams, elevation changes, junctions, shorelines, drainage corridors, settlements, and other distinctive terrain or navigational features. This stage asks whether the map affords focal reasoning at all. GPT-5.5 judged all 65 maps informative under this criterion; Gemini 3.1 Pro Preview judged 64 of 65 informative.

In the second stage, the model sees the target-revealing map and judges whether the find location itself is salient enough to be a plausible search target from the map alone. We label a ``yes'' response as \emph{Focal} and a ``no'' response as \emph{Non-Focal}. Focal incidents are those in which the find location aligns with an affordance for movement or search, such as a trail junction, road, stream crossing, shoreline, drainage corridor, settlement, isolated lake, or distinctive terrain feature. Non-Focal incidents are those in which the find location is weakly tied to such features and appears more compatible with diffuse movement.

This two-stage design is important because almost all maps contain some salient terrain structure, but not every missing person is found at a salient endpoint. GPT-5.5 classified 36 incidents as Focal and 29 as Non-Focal. Gemini 3.1 Pro Preview classified 19 incidents as Focal and 46 as Non-Focal. The difference between these splits reflects a stricter notion of focality by Gemini, which labels fewer endpoints as sufficiently salient.

\begin{figure*}
    \centering
    \includegraphics[width=0.48\textwidth]{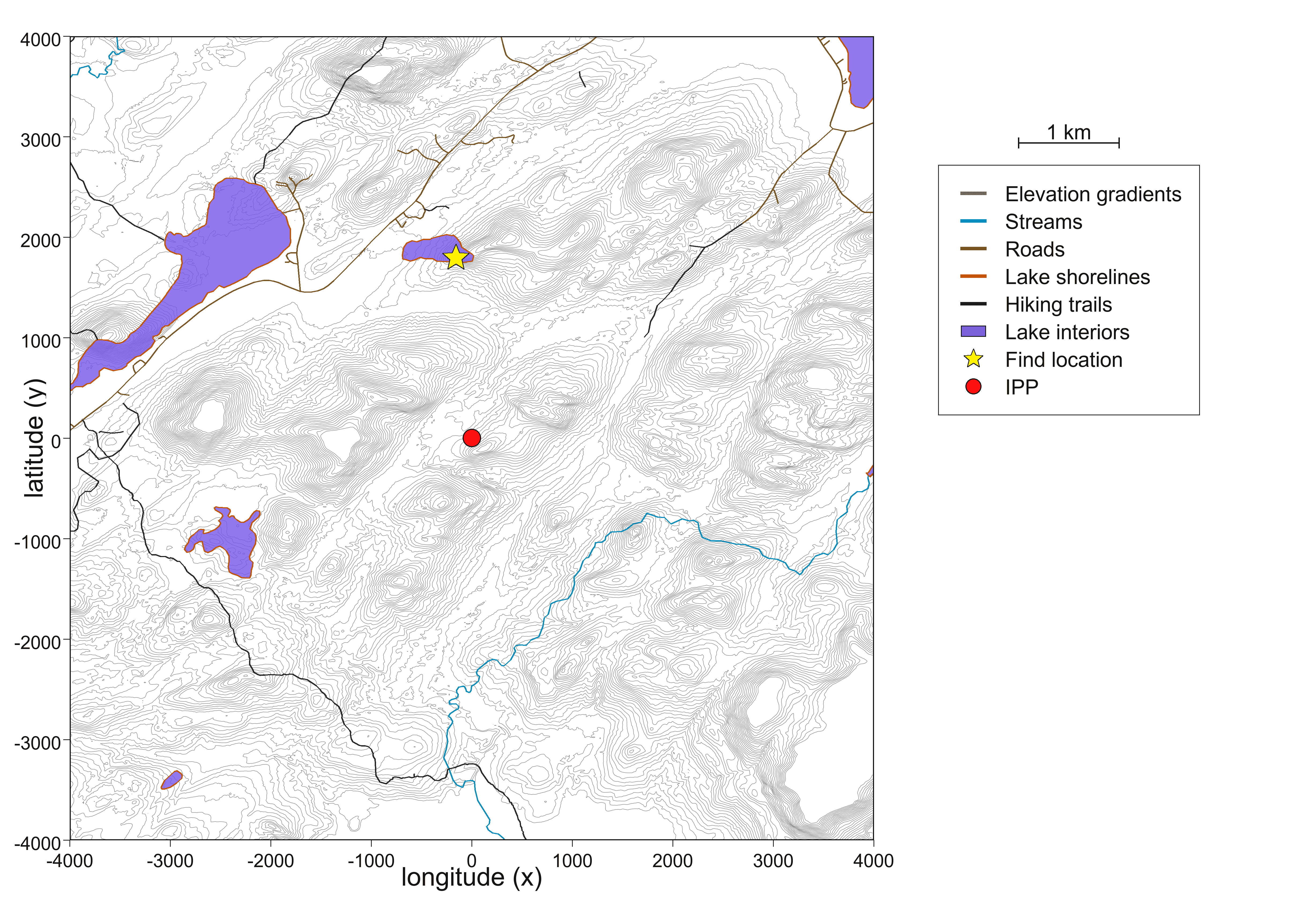}
    \hfill
    \includegraphics[width=0.48\textwidth]{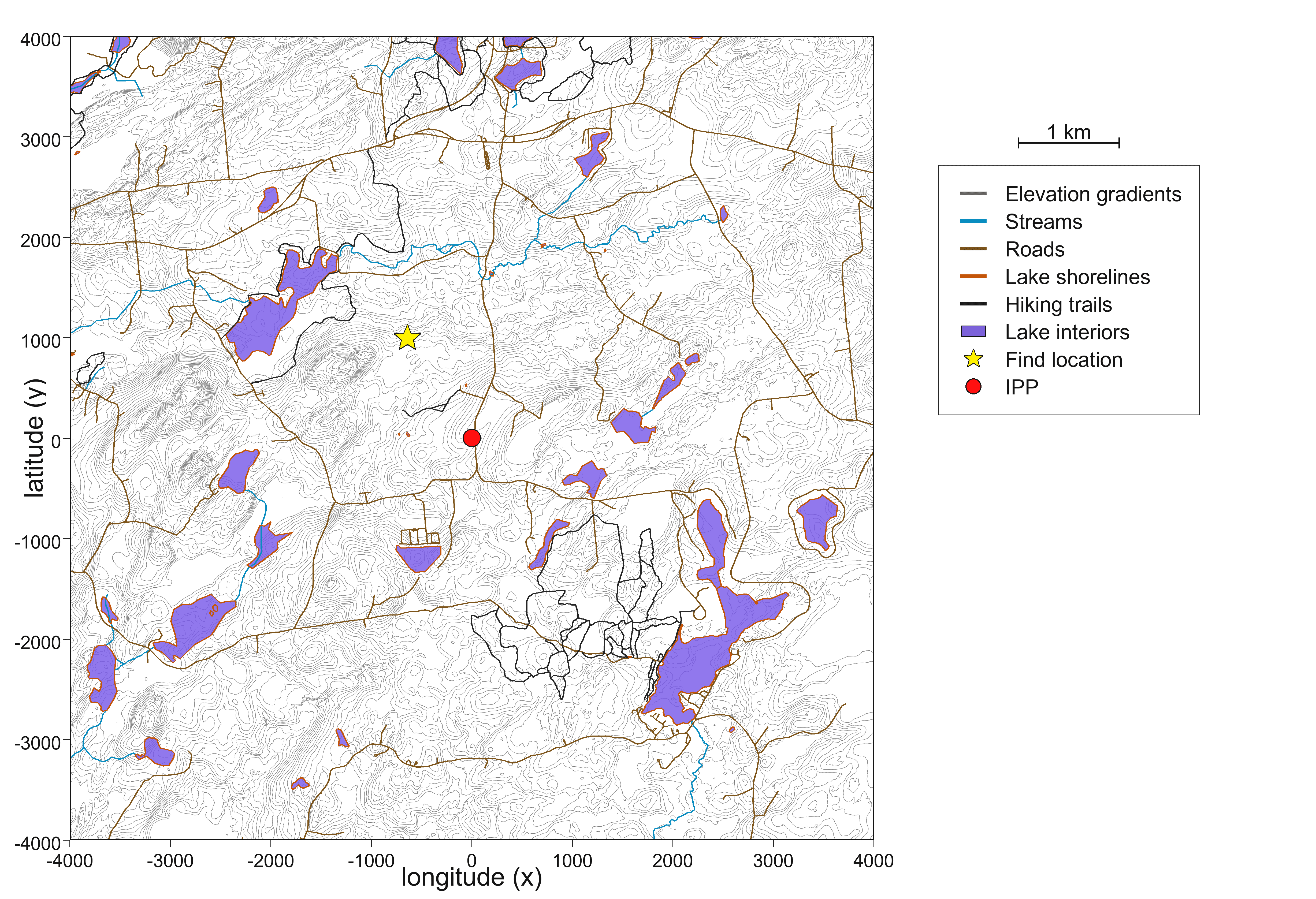}
    \caption{Examples from the focality classification task. Left: Incident 54, classified as Focal by GPT-5.5, where the find location lies beside a small isolated lake. Right: Incident 50, classified as Non-Focal by Gemini 3.1 Pro Preview, where the find location appears away from distinctive terrain or navigational features.}
    \label{fig:sar-classification-examples-app}
\end{figure*}

Figure~\ref{fig:sar-classification-examples-app} shows examples of both classes. In Incident 54, GPT-5.5 classifies the find location as Focal because the yellow marker lies beside a small isolated lake north of the IPP. The endpoint is therefore object-based and landmark-driven: the target is attached to a salient feature rather than lying in undifferentiated terrain. In Incident 50, Gemini 3.1 Pro Preview classifies the find location as Non-Focal because it lies away from roads, trails, streams, lakes, and distinctive topographic structure.

\subsubsection{Prediction Prompts and Evaluation}
\label{app:sar-prediction-eval}

For coordinate prediction, we compare a baseline prompt, denoted \emph{Vanilla}, with four behaviour-aware prompting variants, denoted \emph{S1--S4}. These variants test whether explicitly steering the model towards human movement, terrain salience, and rescue-relevant focal points improves prediction relative to the baseline. All prompts require the model to return a single coordinate $(\hat{x},\hat{y})$ in the IPP-centred local coordinate system.

For each prompt condition $c$, we run the model three times per incident. With 65 incidents, this yields 195 predictions per condition. This repeated-sampling design captures both variation across SAR incidents and within-incident variation across model outputs.

The primary evaluation metric is Euclidean distance error in metres. Let $(\hat{x}_{i,r}^{(c)}, \hat{y}_{i,r}^{(c)})$ denote the coordinate predicted for incident $i$ on repetition $r$ under prompt condition $c$, and let $(x_i^\ast,y_i^\ast)$ denote the corresponding ground-truth find coordinate. The prediction error is

\[
d_{i,r}^{(c)} =
\sqrt{
(\hat{x}_{i,r}^{(c)} - x_i^\ast)^2 +
(\hat{y}_{i,r}^{(c)} - y_i^\ast)^2
}.
\]

For each prompt condition, we report the mean and standard deviation of $d_{i,r}^{(c)}$ across all predictions. Lower mean distance indicates better localisation. The standard deviation reflects both incident-level difficulty and stochastic variation across repeated model outputs.

\subsubsection{Statistical Tests}
\label{app:sar-statistical-tests}

For statistical inference, incidents rather than individual repeated predictions are the unit of analysis. For each incident and prompt condition, we first average the three repeated distance errors. We then compute an incident-level paired improvement over the Vanilla baseline:

\[
\Delta_i^{(s)} =
\bar{d}_i^{(\text{Vanilla})}
-
\bar{d}_i^{(s)} ,
\]

where positive values indicate that prompt $s$ is closer to the true find location than Vanilla.

For each split and prompt variant, we test whether the mean paired improvement is greater than zero using a one-sided paired sign-flip permutation test. Because map widths vary across incidents, we also repeat the permutation test after normalising each improvement by the corresponding incident's map width. In addition, we report a one-sided sign test over the number of incidents improved versus worsened, and a bootstrap 95\% confidence interval for the mean improvement using incident-level resampling.

\subsubsection{Results}
\label{app:sar-results}

Across the full 65-incident evaluation, behaviour-aware prompting produces little overall improvement. The best full-set condition is S3, but its gain over Vanilla is only 11\,m on average. This suggests that focal-point prompting is not a general-purpose improvement for all SAR cases.

The picture changes when the evaluation is restricted to incidents classified as Focal. Under the GPT-5.5 focal split, S3 reduces mean error from 2580\,m to 2410\,m. This corresponds to a 170\,m average improvement over Vanilla, with bootstrap 95\% CI [48, 292]\,m, permutation $p=0.0048$, sign-test $p=0.0083$, and map-width-normalised permutation $p=0.0097$.

Under the Gemini 3.1 Pro Preview focal split, S3 reduces mean error from 3086\,m to 2882\,m. This corresponds to a 203\,m average improvement, with bootstrap 95\% CI [13, 388]\,m, permutation $p=0.0257$, sign-test $p=0.0318$, and map-width-normalised permutation $p=0.0435$.

No behaviour-aware prompt yields a reliable improvement on the full set or on the Non-Focal splits. In Non-Focal cases, behaviour-aware prompting is usually worse than Vanilla, consistent with the interpretation that imposing focal reasoning can mislead the model when the realised endpoint is not tied to a salient terrain cue.

\begin{figure*}[t]
    \centering
    \includegraphics[width=0.32\textwidth]{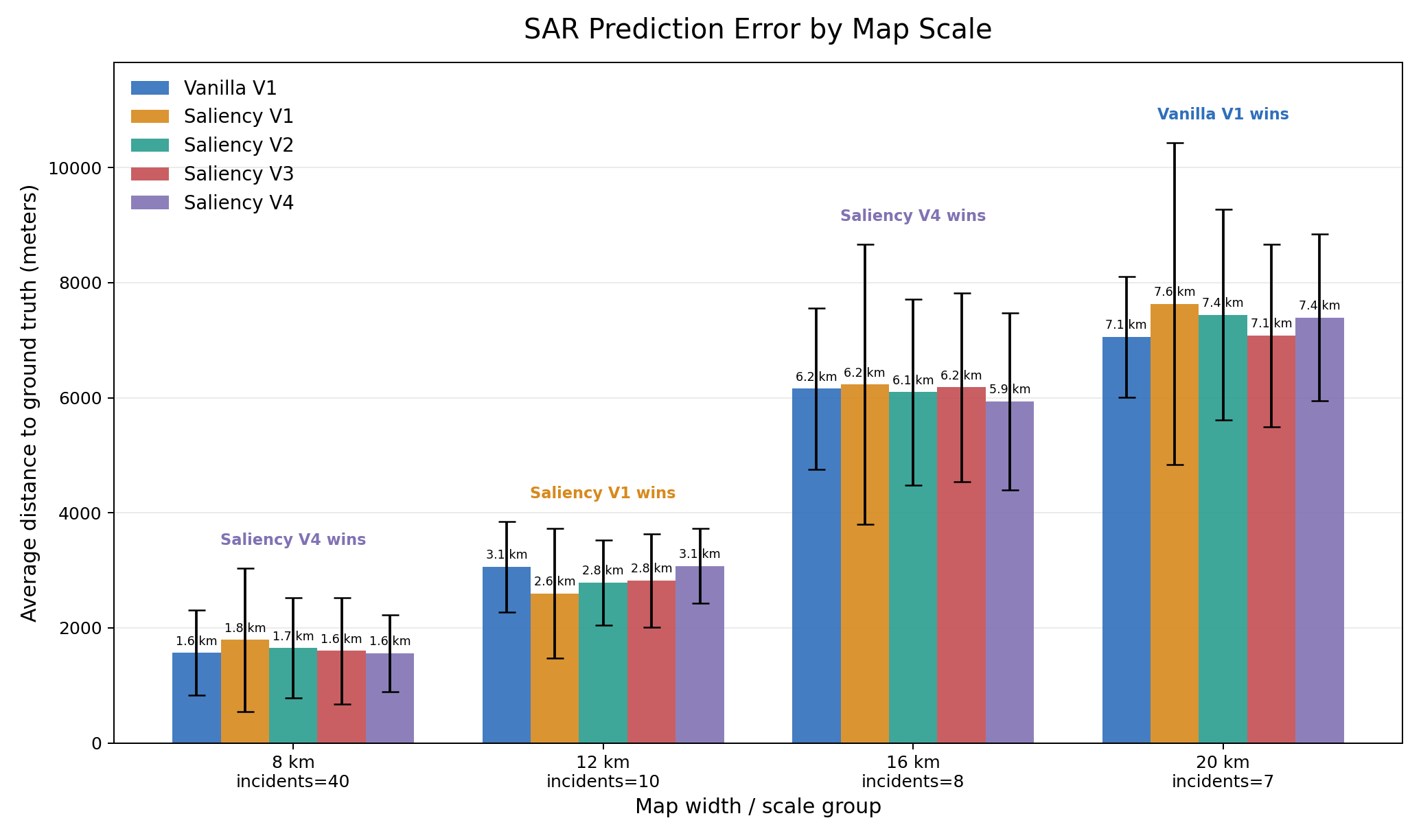}
    \hfill
    \includegraphics[width=0.32\textwidth]{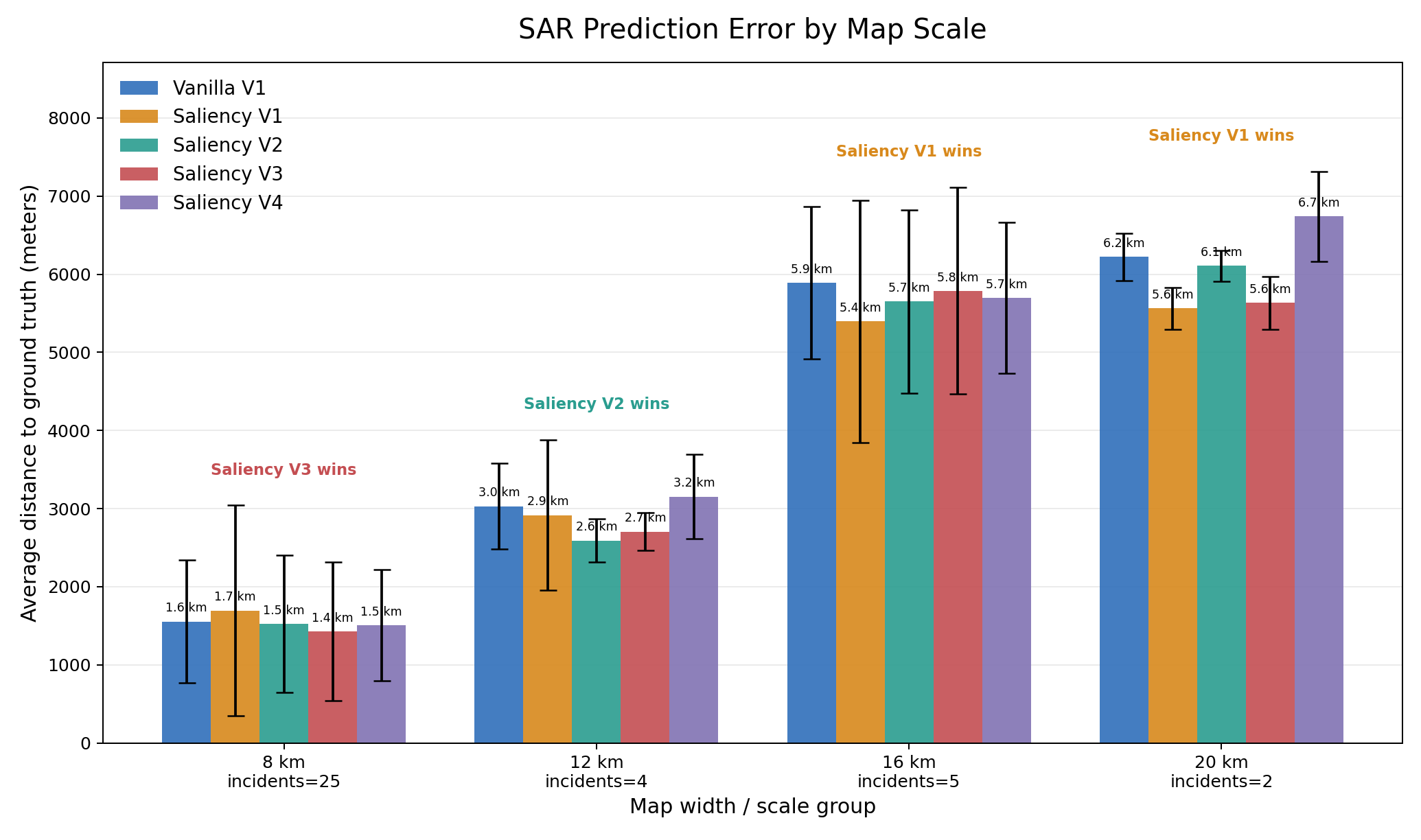}
    \hfill
    \includegraphics[width=0.32\textwidth]{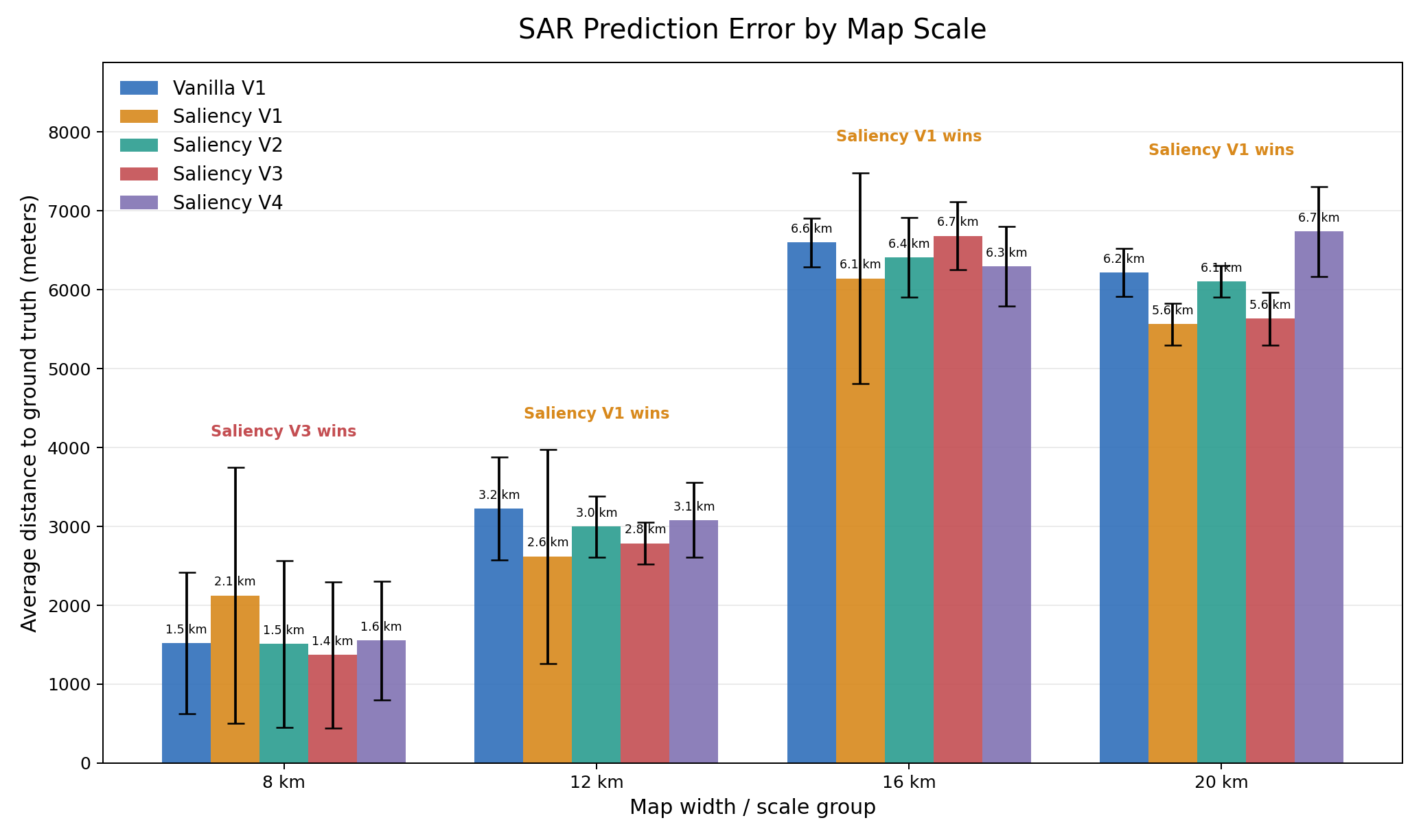}
    \caption{SAR prediction error by map scale. Bars show average Euclidean distance to the ground-truth find location for Vanilla and the four behaviour-aware prompts, shown as Saliency V1--V4. Left: all 65 incidents. Middle: incidents classified as Focal by GPT-5.5. Right: incidents classified as Focal by Gemini 3.1 Pro Preview. Error bars show variation across predictions.}
    \label{fig:sar-error-by-scale-app}
\end{figure*}

Figure~\ref{fig:sar-error-by-scale-app} provides a descriptive breakdown by map width for the full set and the two focal splits. As expected, larger maps generally produce larger absolute errors. More importantly, the improvement of S3 over Vanilla is most visible in the focal splits, matching the statistical results.

Overall, the SAR results support a conditional account of focal-point prompting. Behaviour-aware prompts do not make LLMs universally better at search localisation. They help when the missing person's realised endpoint is itself focal: that is, when it lies near a terrain feature, route, landmark, or movement affordance that can plausibly serve as a shared search target. When the endpoint is Non-Focal, the same prompts can over-regularise the model towards salient but incorrect locations.

\end{document}